\documentclass[useAMS,usenatbib]{mn2e}
\usepackage{graphicx}
\usepackage{times}

\newcommand{\photo}{{\it Photo}}
\newcommand{\lre}{$\log r_{\rm e}$}
\newcommand{\re}{$r_{\rm e}$}

\newcommand{\sn}{$n$}
\newcommand{\mie}{$<\! \mu\! >_{\rm e}$}
\newcommand{\ls}{$\log \sigma_0$}
\newcommand{\dt}{$\delta(\log t)$}
\newcommand{\dz}{$\delta(\log Z)$}

\newcommand{\ml}{$ M/L$}
\newcommand{\mls}{$M_{\ast}/L$}

\newcommand{\mr}{$^{0.07}M_{r}$}

\newcommand{\fja}{$\lambda_0$}
\newcommand{\fjb}{$\lambda_1$}
\newcommand{\papdata}{paper I}

\title[SPIDER -- II. The FP in $grizYJHK$]{SPIDER -- II. The Fundamental Plane of Early-type Galaxies in \MakeLowercase{griz}${\rm YJHK}$}
\author[F. La Barbera et al.]{F. La Barbera$^{1}$\thanks{E-mail: labarber@na.astro.it(FLB); rrdecarvalho2008@gmail.com (RRdC)}, R.R. de Carvalho$^{2}$, I.G. de la Rosa$^{3}$, P.A.A. Lopes$^{4}$\\
$^1$INAF -- Osservatorio Astronomico di Capodimonte, Napoli, Italy \\
$^2$Instituto Nacional de Pesquisas Espaciais/MCT, S. J. dos Campos, Brazil\\
$^3$Instituto de Astrofisica de Canarias, Tenerife, Spain\\
$^4$Observat\'orio do Valongo/UFRJ, Rio de Janeiro, Brazil}

\begin{document}
\date{Submitted on 2009 December 22}

\pagerange{\pageref{firstpage}--\pageref{lastpage}} \pubyear{2010}

\maketitle

\label{firstpage}
\begin{abstract}
We  present a  complete  analysis  of the  Fundamental  Plane (FP)  of
early-type  galaxies (ETGs)  in  the nearby  universe (z$<$0.1).   The
sample,  as defined  in  paper  I, comprises  39,993  ETGs located  in
environments covering  the entire domain in local  density (from field
to  cluster).  We derive  the FP  in the  $grizYJHK$ wavebands  with a
detailed  discussion  on  fitting  procedure, bias  due  to  selection
effects and bias due to correlated errors on the effective parameters,
$r_e$   and  \mie,  as   key  factors   in  obtaining   meaningful  FP
coefficients.  Studying  the Kormendy relation  (KR) we find  that its
slope  varies  from g  (3.44$\pm  0.04$)  through  K (3.80$\pm  0.02$)
implying that smaller size ETGs have  a larger ratio of optical to NIR
radii  than   galaxies  with  larger  $r_e$.   We   also  examine  the
Faber-Jackson (FJ) relation and find that its slope is similar for all
wavebands,   within  the   uncertainties,   with  a   mean  value   of
0.198$\pm$0.007.  {Writing  the FP equation as $\log  r_e = a \log
  \sigma_0 + b < \! \mu \!   >_e + c$, we find that the $''a''$ varies
  from $1.38 \pm 0.02$ in $g$, to $1.55 \pm 0.02$ in K, implying a $12
  \%$  variation  across  the  $grizYJHK$  wavelength  baseline.   The
  corresponding variation  of $''b''$ is negligible  ($b \sim 0.316$),
  while  $''c''$ varies  by $\sim  10\%$.  We  show that  the waveband
  dependence of  the FJ and KR  results from the  complex variation of
  the distribution of galaxies in  the face-on projection of the FP as
  well as  by the  change of FP  coefficients with waveband.   We find
  that $''a''$ and $''b'' $ become smaller for higher Sersic index and
  larger axis ratios, independent of the waveband.  This suggests that
  these  variations  are  likely  to  be  related  to  differences  in
  structural and dynamical (rather than stellar population) properties
  of  ETGs.  It  is noticeable  that galaxies  with bluer  colours and
  disc-like isophotes have smaller $''b''$, with the effect decreasing
  smoothly  from $g$  through $K$.   Considering a  power-law relation
  between  mass-to-light  ratio  and  (dynamical) mass,  $M/L  \propto
  M^{\gamma}$,  we   estimate  gamma  from  the   FP  coefficients  in
  $grizYJHK$.  The  $\gamma$ decreases from $0.224 \pm  0.008$ in $g$,
  to  $0.186 \pm 0.009$  in $K$  band. Using  the $\gamma$  values, we
  estimate  the  variation  of  age  and metallicity  of  the  stellar
  populations present in massive  galaxies per decade in stellar mass.
  This analysis shows that in the NIR the tilt of the FP is not due to
  stellar population's  variation, and  that ETGs have  coeval stellar
  populations with  an age  variation of a  few percent per  decade in
  mass, and  a corresponding  metallicity increase of  $\sim$23\%.  We
  also show  that current  semi-analytical models of  galaxy formation
  reproduce  very   well  these  amounts  of  variation   of  age  and
  metallicity with respect to stellar mass.}
\end{abstract}

\section{Introduction}
\label{sec:INTR}

One of  the most outstanding  and basic cosmological questions  is how
galaxies form and evolve.  Currently the favoured scenario assumes that
the assemblage of  baryonic matter is driven by  the evolution of dark
matter haloes  (Gott \& Rees~1975). Given the  difficulty of observing
dark  matter, we  rely  on the  luminous  counterpart to  be a  beacon
illuminating their  evolution.  The vast majority of  stars and metals
produced during the evolution of galaxies were formed and still reside
in them. Therefore, examining the star formation history and measuring
the metal  content of  galaxies may tell  us how these  systems evolve
through cosmic time.

The  study of  the global  properties  of elliptical  galaxies took  a
substantial  step   forward  with  the   application  of  multivariate
analysis, revealing potentially  meaningful scaling relations like the
Fundamental Plane  (FP, Brosche 1973). However, the  importance of the
technique pioneered by this paper  was not immediately realised by the
astronomical community. Determining which dimensions are statistically
significant  in a given  data set  is not  a simple  task, but  it can
reveal useful correlations involving the quantities defining a minimal
manifold  and  provide  insights  into  the physical  nature  of  such
correlations. Such is the case when the observed FP is associated with
the virial theorem (Djorgovski \& Davis~1987; Dressler et al.~1987).

Many studies  over the past twenty  years have tried  to interpret the
physical meaning of the FP  (e.g.  Faber et al.~1987; Djorgovski \& de
Carvalho~1990; Pahre et al.~1998b; \citealt{JFK96}; \citealt{JORG:99};
Dantas et al.~2003; \citealt{BER03a,BER03b,BER03c}; Nelan et al.~2005;
Cappellari  et al.~2006).   The  striking  feature of  the  FP is  its
narrowness,  implying  a regularity  among  the  global properties  of
early-type galaxies.  The quantities containing the entire variance of
the  data are:  effective radius,  $\rm r_{\rm  e}$,  central velocity
dispersion,  $\sigma_{\circ}$, and  mean  surface brightness  measured
within the effective radius, $\mu_{\rm e}$. The best representation of
the FP is $\rm r_{\rm e}$ $\sim$ $ \sigma_{\circ}^{\rm A} \rm I_e^{\rm
  B}$, { where $I_e$  is mean surface  brightness in  flux units}.
Bernardi et  al.~(2003c) show a  comprehensive table listing  the most
important papers  presenting values  of A and  B and  their respective
errors.  A  seems to  vary with the  passband used in  the photometric
observation, while B does  not { (see e.g.~\citealt{PDdC98b, SCO98,
    MGA99})}.   However,  \citealt{BER03c},  found only  a  marginally
significant variation  of A  in the SDSS  optical passbands  (see also
Hyde \&  Bernardi~2009).  La  Barbera et al.~(2008)  (hereafter LBM08)
also  found a  small difference  in A  when measured  between r  and K
bands.

Assuming that early-type galaxies  are homologous systems in dynamical
equilibrium  and that velocity  dispersion is  related to  the kinetic
energy per  unit mass we can  write down expressions for  mass (M) and
luminosity (L), namely (M/L) $\sim \sigma^{\rm 2-A} \rm I^{\rm -1-B}$.
In the case of a fully virialized system, A = 2 and B = -1, implying a
constant mass-to-light  ratio. However,  A and B  are found  to differ
significantly from the virial  values, resulting in the so-called $\sl
tilt$ of the FP. In this  case, M/L $\sim \rm M^{\gamma}$, where gamma
is   $\sim$0.25  (Faber   et  al.~1987).    This  dependence   of  the
mass-to-light  ratio on galaxy  mass has  been interpreted  as arising
either from differences in the  stellar populations or the dark matter
fractions  among ETGs.   It  is important  to  emphasize that  another
option to explain the tilt is  related to the assumption that ETGs are
truly virialized systems - in which case they should have self-similar
density distributions and similar orbital distributions. Any departure
from either or both of these conditions may well explain the tilt, and
several studies have tried to disentangle these effects. For instance,
non-homology  seems  to  contribute  to  at least  part  of  the  tilt
(\citealt{HM95},  Capelato,  de  Carvalho  \&  Carlberg~1995;  Ciotti,
Lanzoni   \&   Renzini~1996;  Ciotti   \&   Lanzoni~1997;  Graham   \&
Colless~1997;   Busarello  et   al.~1997;  Bertin,   Ciotti,   \&  del
Principe~2002; Trujillo, Burkert \& Bell~2004). Even studies trying to
explain  the  tilt  as  a  stellar population  effect  concluded  that
non-homology may  play a significant  role in determining the  tilt of
the  FP (e.g.   Pahre,  Djorgovski  \& de  Carvalho  1998b; Forbes  \&
Ponman~1999).  Another  interesting finding  from  the simulations  of
Capelato  et   al.~(1995)  is  that  when   measuring  the  structural
parameters defining  the FP  inside larger apertures,  of order  a few
r$_{\rm  e}$, the  coefficients are  similar to  those implied  by the
virial theorem.   More recently, Bolton  et al.~(2008) find  a similar
result  when  using the  surface  density  term  defined by  the  mass
measured through strong lensing, and  conclude that the tilt of the FP
is due to the fraction of dark matter inside one effective radius (see
also Tortora et al.~2009).

This is the second paper of  a series analysing the properties and the
scaling relations of ETGs as  a function of the environment where they
reside.   The   Spheroids  Panchromatic  Investigation   in  Different
Environmental  Regions  (SPIDER)  utilises optical  and  Near-Infrared
(NIR)  photometry  in  the  $grizYJHK$  wavebands,  along  with
spectroscopic  data,  taken from  the  {  UKIRT  Infrared Deep  Sky
  Survey}-Large  Area Survey  (UKIDSS-LAS) and  the Sloan  Digital Sky
Survey (SDSS). The  selection of ETGs for this  project is detailed in
Paper I, and we refer the reader  to that paper for all the details of
sample  selection  and  the  procedures  used  to  derive  the  galaxy
parameters.

In this  work we focus on the  derivation of the FP  in the $grizYJHK$
wavebands  for the  entire SPIDER  sample.  {  Although  our sample
  contains ETGs over the entire domain of local density (from field to
  clusters), we postpone the  study of the environmental dependence of
  the FP  to another paper in  the SPIDER series (paper  III).  In the
  present  work,  we discuss  the  main  pitfalls  of the  FP  fitting
  procedure  and how to  account for  selection effects  and different
  sources of  biases.  We analyse the edge-on  and face-on projections
  of the FP, as well as the two other projections of the FP, i.e.  the
  Kormendy  and Faber-Jackson  relations.  The  analysis of  these two
  scaling  relations  serves as  a  reference  at  the local  Universe
  (z$<$0.1),  and at  different wavebands,  for other  studies lacking
  data for a full FP analysis. We find a consistent picture connecting
  the waveband variation  of the edge- and face-on  projections of the
  FP with  that of the Kormendy and  Faber-Jackson relations. Finally,
  we show how the optical  and NIR FPs can constrain various scenarios
  for  galaxy  formation  and   evolution,  by  using  the  wavelength
  dependence of  the FP to  infer the variation of  stellar population
  parameters along the ETG's sequence}.

The layout  of the paper is  as follows. Sec.~2  shortly describes the
SPIDER dataset. Sec.~3 presents  the different subsamples of ETGs used
to derive the FP in $grizYJHK$  and to analyse the impact of different
biases  on the  FP.  Sec.~4  details the  FP fitting  procedure.  {
  Secs.~5  and~6  analyse the  Kormendy  and Faber-Jackson  relations,
  respectively.  Sec.~7  presents one main result of  this study, i.e.
  the dependence of  FP slopes on waveband, from  g through K.  Sec.~8
  analyses the waveband variation of the edge- and face-on projections
  of  the  FP.  Sec.~9  describes  how  the  optical and  NIR  scaling
  relations  of ETGs  constrain  the variation  of stellar  population
  properties along the FP.  Discussion follows in Sec.  10.  A summary
  is provided in Sec.~11}.

Throughout the paper,  we adopt a cosmology with $\rm H_0  \! = \!  75
\,  km \,  s^{-1} \,  Mpc^{-1}$, $\Omega_{\rm  m} \!   = \!  0.3$, and
$\Omega_{\Lambda} \! = \! 0.7$.

\section{Data}
\label{sec:data}
The  SPIDER   data-set  is  based   on  a  sample  of   $39,993$  ETGs
(see~\papdata \,  for details),  with available $griz$  photometry and
spectroscopy from SDSS-DR6. Out of these galaxies, $5,080$ objects have
also photometry available in the $YJHK$ wavebands from UKIDSS-LAS. All
galaxies have  two estimates of  the central velocity  dispersion, one
from SDSS-DR6 and an  alternative measurement obtained by fitting SDSS
spectra  with  the software  STARLIGHT~\citep{CID05},  using a  linear
combination of  simple stellar population models  (rather than single
templates as in  SDSS) with different ages and  metallicities. { In
  both   cases,    STARLIGHT   and SDSS-DR6,  the $\sigma_0$'s   are
  aperture-corrected      to     an      aperture      of     $r_e/8$,
  following~\citet{JFK95}  . In  order to  make proper  comparisons to
  earlier studies (e.g. Bernardi et al.~2003a), we use SDSS
  velocity dispersion measurements to examine the scaling relations presented in this paper. 
  In paper  I, we find that the mean
  difference  between  $\sigma_{0}$(SDSS-DR6) and  $\sigma_{0}$(STARLIGHT)
  does not  change significantly  with $\sigma_{0}$. Therefore,  we do
  not  expect   that  the  choice  of  a   given  velocity  dispersion
  measurement  might have  a dramatic  impact on  the  FP relation.
  This is further discussed in Secs.~6,~7, and~9}.
 
In all wavebands, structural  parameters -- i.e. the effective radius,
\re, the  mean surface  brightness within that  radius, \mie,  and the
Sersic  index,  \sn  --   have  been  all  homogeneously  measured  by
2DPHOT~\citep{LBdC08}.  In the optical ($griz$), alternative estimates
of the effective parameters, \re  \, and \mie, are also available from
the SDSS-DR6 \photo \, pipeline. In~\papdata, we compare the different
estimates of  photometric and spectroscopic  parameters, deriving also
an  estimate of the  $95\%$ completeness  limit of  the sample  in all
wavebands. We find that 2DPHOT total magnitudes are brighter than SDSS
model magnitudes,  with the difference amounting to  $\sim 0.2$~mag in
r-band, for  the faintest galaxies  in the sample. This  difference is
due to the use of  Sersic (2DPHOT) rather than de Vaucouleurs (\photo)
models to  fit the light distribution of  ETGs, as well as  to the sky
estimate  bias affecting  SDSS effective  parameters~\citep{DR6, DR7}.
Hence, the completeness  limit of the sample is  also dependent on the
source of effective  parameters (2DPHOT vs.  \photo).  In  r band, the
sample is  $95\%$ complete at $-20.32$  and $-20.55$ for  the SDSS and
2DPHOT parameters, respectively.   In the following, unless explicitly
said, we refer to 2DPHOT total magnitudes.

\section{The samples}
\label{sec:samples}
The  waveband  dependence  of  the  FP  is  analysed  using  different
subsamples of ETGs, extracted from the SPIDER sample.  Details on each
subsample  are  provided  in  Sec.~\ref{sec:FPsamples}.  In  order  to
analyse the effect of different sources of bias on the FP relation, we
also utilise several  samples of ETGs, with effective  parameters in r
band.   We  describe   the   characteristics  of   these  samples   in
Sec.~\ref{sec:FPcontrolsamples}, referring to  them, hereafter, as the
control samples of ETGs.

\subsection{The $grizYJHK$ (SDSS+UKIDSS) samples of ETGs}
\label{sec:FPsamples}

\begin{table}
\centering
\small
\begin{minipage}{80mm}
\caption{Magnitude limits in $grizYJHK$ adopted to derive the FP.}
\begin{tabular}{c|c|c|c|c}
\hline
waveband       &   $X$\footnote{X is the equivalent magnitude limit as used in the colour-selected samples}
 & $^{0.07}M_{X}$ limit  & $N_{X}$  \\
\hline
 g & -19.75 & -19.71 & 4467\\
 r & -20.60 & -20.55 & 4478\\
 i & -21.02 & -20.99 & 4455\\
 z & -21.34 & -21.22 & 4319\\
 Y & -22.03 & -21.95 & 4404\\
 J & -22.55 & -22.54 & 4317\\
 H & -23.22 & -23.21 & 4376\\
 K & -23.60 & -23.60 & 4350\\
\hline
\end{tabular}
\label{tab:FP_samples}
\end{minipage}
\end{table}

In order  to analyse how  different selection procedures  might affect
the dependence of the FP relation on waveband, we derive the FP in the
$grizYJHK$  wavebands  for  ETG's  SPIDER subsamples  defined  by  two
different selection procedures. In case  (i), we derive the FP for the
same  sample of  ETGs in  all wavebands,  by selecting  those galaxies
brighter  than  the  r-band  completeness  limit  (\mr$=-20.55$).   We
exclude galaxies whose Sersic fit,  in one of the available wavebands,
has an high reduced $\chi^2$  value ($\ge 3$). This cut removes less than
$2\%$ of galaxies, resulting in a  sample of $4,589$ ETGs. In case (ii),
we select  different samples of  ETGs in the different  wavebands, but
according  to  {\it  equivalent}  magnitude  limits.   The  equivalent
magnitude limits are derived  by using the optical-NIR colour-magnitude
relations (see sec.~4 of~\papdata).  To  this effect, we first fix the
r-band magnitude limit to $-20.6$  and then translate it into the other
wavebands using  the colour-magnitude relations. The  value of $-20.6$
is chosen  so that, for each  band, the equivalent  magnitude limit is
brighter  than the  completeness magnitude  in that  band,  as defined
in~\papdata.   This  makes  the  samples  magnitude  complete  in  all
wavebands.   The {\it  equivalent}  magnitude limits  are reported  in
column   (2)  of  Tab.~\ref{tab:FP_samples},   along  with   the  95\%
completeness magnitude  limits, from~\papdata, in column  (3), as well
as the number of ETGs selected in each band in column (4).

In  the following,  we refer  to the  ETG sample  of case  (i)  as the
(r-band) magnitude-selected sample of  ETGs, while the samples of case
(ii) are referred to as the colour-selected samples of ETGs.

\subsection{Control samples of ETGs}
\label{sec:FPcontrolsamples}
We use five control samples of
ETGs selected from SDSS-DR6, with photometry available in r band. The
control samples consist of ETGs selected in different redshift ranges,
with effective parameters and central velocity dispersions measured with
different methods. In all cases, velocity dispersions are corrected to
an aperture of $r_e/8$,
following~\citet{JFK95}. Each control sample is named with a letter, as
shown in Tab.~\ref{tab:opt_samples}, where we summarise the basic
characteristics of the five samples. 
\begin{description}
\item[--] Sample $A$  is obtained from  the
sample of $39,993$ ETGs defined in paper I. We select all galaxies with an
r-band model magnitude brighter than $-20.32$. This
magnitude cut corresponds to the $95 \%$
completeness limit in r band, as defined in paper I, when using SDSS model
magnitudes (see Sec.~\ref{sec:data}). Effective parameters are obtained
from SDSS, as in~\citet{BER03a}. 
\item[--] Sample $B$ is a subsample of sample $A$,
consisting of all the ETGs that also have photometry available in the
$YJHK$ wavebands (see~\papdata). Such sample is used to estimate the impact
of matching SDSS to UKIDSS data on the FP relation. 
\item[--] Sample $C$ is defined to explore a wider
magnitude range than that of samples $A$ and $B$. We query the  SDSS-DR6
database for ETGs in a redshift range of $z=0.02$ to 
$0.03$. ETGs are defined according to the  same criteria  as in~\papdata,
i.e.  $zwarning=0$, $eclass < 0$  and $fracDev_r  \! > \!   0.8$. No
requirement  is done  for
the  galaxy velocity  dispersion.  The query results into  a list  of 3732
galaxies,  that hereafter we  refer to as the low-redshift  sample
of ETGs. All galaxies have effective parameters from SDSS. Using the
same procedure  as in~\papdata, we  estimate a $95 \%$
completeness limit of $^{0.07}M_r=-17.64$ (model magnitude).  Since
velocity dispersions from SDSS are not available for all galaxies in
this sample, we assign fake  $\sigma_0$ values to each pair
of \re \,  and \mie \, values, as described in
Sec.~\ref{sec:sel_effects}.
\item[--] Samples  $D$ and $E$ are defined in the same way as samples
$A$ and $B$, respectively, but using 2DPHOT rather than SDSS effective
parameters. For both samples, we select all the ETGs in the SPIDER sample 
with total magnitude brighter than \mr$=-20.55$ (corresponding to
the 2DPHOT completeness magnitude). Sample $E$ is obtained from sample
$D$ by selecting only those objects with matched photometry in UKIDSS.
Sample $E$ coincides with the magnitude-selected sample of
ETGs in r band (Sec.~\ref{sec:FPsamples}).
\end{description}

\begin{table*}
\centering
\small
\begin{minipage}{160mm}
\caption{Control samples of ETGs.}
\begin{tabular}{c|c|c|c|c|c}
\hline
                   & A                      &             B     
     &          C            &            D            &            E      \\
\hline
 Number of galaxies        & 37273                  &           4796       
  &      3690              &  36205                  &           4589      
   \\
 Redshift  range           & $0.05 \le z \le 0.095$ &  $0.05 \le z \le 0.095$ & $0.02 \le z \le 0.03$  & 
$0.05 \le z \le 0.095$  &  $0.05 \le z \le 0.095$ \\
 Limiting $^{0.07}M_{r}$   & $-20.32$               &         $-20.32$     
  &        $-17.64$        &  $-20.55$               &         $-20.55$   
   \\
 Source of \re \, and \mie & SDSS                   & SDSS                    &        SDSS           &  
2DPHOT                 & 2DPHOT                  \\
 Available wavebands       & $r$                    & $grizYJHK$              &        $r$            &  $r$                    
& $grizYJHK$              \\
\hline
\end{tabular}
\label{tab:opt_samples}
\end{minipage}
\end{table*}

\section{Deriving the FP}
\label{sec:FP_FIT}

We write the FP relation as:
\begin{equation}
\log r_{\rm e} = a \log \sigma_{\rm 0} + b < \! \mu \! >_{\rm e} + c,
\label{eq:FP}
\end{equation}
where $''a''$ and  $''b''$ are the slopes, and  $''c''$ is the offset.
We denote the  rms of residuals around the FP with  respect to \lre \,
as $s_{r_e}$, referring to $''a''$, $''b''$, $''c''$, and $s_{r_e}$ as
the  coefficients of the  FP.  We  estimate the  FP coefficients  by a
procedure consisting  of three steps.  First, we derive the  values of
$''a''$ and  $''b''$, as  described in Sec.~\ref{sec:fit}.  The slopes
are  then  corrected  for   different  sources  of  biases,  including
selection  effects  (Sec.~\ref{sec:sel_effects})  and  the  effect  of
correlated     uncertainties     on      \lre     \,     and     \mie,
(Sec.~\ref{sec:errors_bias}). The bias-corrected values of $''a''$ and
$''b''$   are   then   used   to  estimate   $''c''$   and   $s_{r_e}$
(Sec.~\ref{sec:fit}).  This  procedure  is  tested through  the  ETG's
control samples, as discussed in Sec.~\ref{sec:test_bias_corr}.

\subsection{Fitting procedure}
\label{sec:fit}
We obtain  a first estimate of  $''a''$ and $''b''$  by minimising the
sum of absolute residuals around the FP. When compared to the ordinary
least-squares fitting  method, where one minimises the  sum of squared
residuals,  this procedure  is more  robust, being  less  sensitive to
outliers   in    the   distribution   of    data-points   around   the
plane~\citep{JFK96} (hereafter JFK96).   We adopt two different fitting
methods,  by minimising  the residuals  in \ls  \, and  the orthogonal
residuals about  the plane.  The orthogonal  fit - adopted  in most of
previous works - has the main advantages of treating all the variables
symmetrically, while the \ls  \, regression is essentially independent
of selections  effects in the  plane of effective parameters,  such as
the magnitude  limit (see~\citealt{LAB00}, hereafter LBC00). The values  of $''a''$ and
$''b''$ are  corrected for selection effects and  correlated errors on
effective       parameters       (see      Secs.~\ref{sec:sel_effects}
and~\ref{sec:errors_bias}).  The  value of $''c''$ is  then derived as
the median value  of the quantity $log r_{\rm e}  - a \log \sigma_{\rm
  0} -  b < \!  \mu  \! >_{\rm e}$, over  all the galaxies  of a given
sample, with  $''a''$ and $''b''$  being the bias-corrected  values of
FP slopes.  As shown in Sec.~4.2, when compared to the more common
practise  of estimating  $''c''$ through  the  least-squares procedure
itself, the above estimate has  the advantage of providing an unbiased
value  of $''c''$  ,  regardless  of the  magnitude  selection of  the
sample. For both fitting methods,  we calculate the scatter of the FP,
$s_{r_e}$, from  the mean value of  the absolute residuals  in \lre \,
around  the plane, using  the bias-corrected  slopes. As  for $''c''$,
this procedure  provides an unbiased  estimate of the FP  scatter (see
Sec.~\ref{sec:sel_effects}).

\subsection{Bias due to selection effects}
\label{sec:sel_effects}

\begin{figure}
\includegraphics[height=80mm]{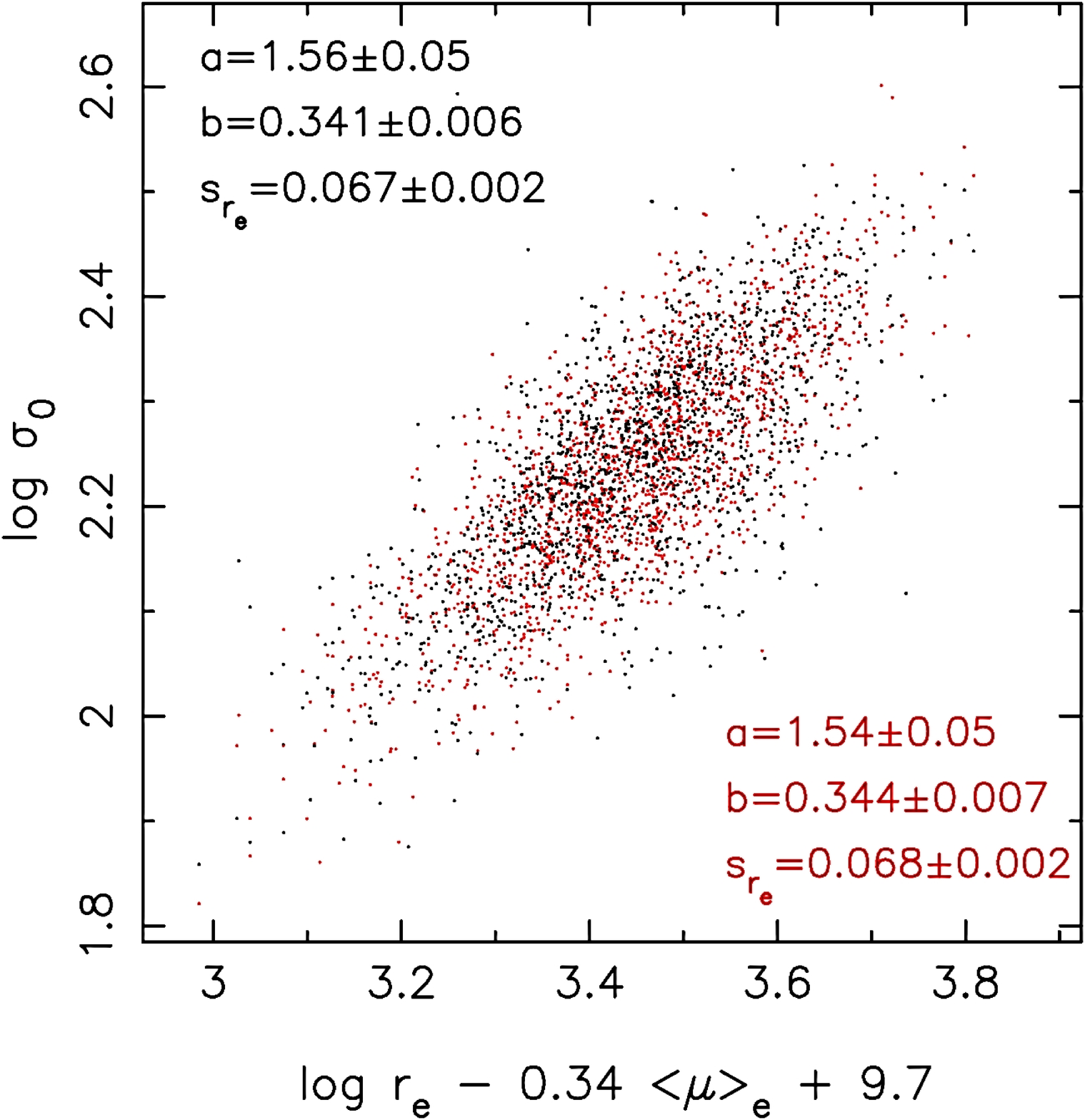}
\caption{Comparison of  the {\it short}  edge-on projection of  the FP
  for galaxies in sample C with available $\sigma_0$ from SDSS (black)
  and  one toy  sample  (red).   For both  samples,  only points  with
  $^{0.07}M_r \le -20.28$ and $ 70  \le \sigma_0 \le 420 \, km/s$ have
  been selected. The values of \ls \, are plotted against the variable
  \lre$-b$\mie,  i.e.   the   combination  of  photometric  parameters
  entering  the FP.   The  slopes ($''a''$  and  $''b''$) and  scatter
  ($s_{r_e}$)  of the relation,  obtained from  the \ls  \, regression
  procedure, are reported in the upper-left and lower-right corners of
  the plot for the observed  and toy samples, respectively. Notice the
  negligible difference between the two sets of coefficients.}
\label{fig:simfp}
\end{figure}

To estimate  how selection criteria (e.g. the  magnitude limit) affect
the FP coefficients,  we use a simulated sample  of data-points in the
space of \lre,  \mie, and \ls, resembling the  distribution of ETGs in
that space.  The  simulated sample is created from  the control sample
$C$, namely all ETGs from SDSS-DR6  in the redshift range of $0.02$ to
$0.03$, brighter than an r-band model magnitude of $^{0.07}M_r=-17.64$
(Sec.~\ref{sec:FPcontrolsamples}).  Since  galaxies in this  sample do
not  have available  velocity dispersions,  we assign  fake $\sigma_0$
values.  For each galaxy,  we use its \lre \, and \mie  \, to obtain a
value of $\sigma_0$ from the FP relation (Eq.~\ref{eq:FP}). That value
is then shifted  according to a random Gaussian  deviate, with a given
width value $s_{\sigma_0}$, that describes the scatter of the FP along
the $\sigma_0$  axis.  The slopes, offset, and  scatter parameters are
chosen with an iterative procedure.
\begin{description}
\item[--] First, we  select all galaxies in sample  $C$ with available
  $\sigma_0$  from SDSS-DR6,  applying similar  cuts in  magnitude and
  velocity  dispersion  as  those  for the  r-band  magnitude-selected
  sample  of ETGs  (sec  Sec.~\ref{sec:FPsamples}).  This  is done  by
  selecting   all  galaxies   with  model   magnitude   brighter  than
  $-20.28$~\footnote{Notice  that the  value of  $-20.28$  is 0.04~mag
    fainter   than   the  r-band   model   magnitude   limit  of   the
    magnitude-selected    ETG    sample   ($^{0.07}M_r=-20.32$,    see
    Sec.~\ref{sec:data}) We obtain $-20.28$  by adding to $-20.32$ the
    difference of evolutionary  correction between the median redshift
    of the  ETG's sample of~\papdata  ($z=0.0725$) and that  of sample
    $C$ ($z=0.025$).  To this aim, following~\citet{BER03b} (hereafter
    BER03b),  we parametrize  the evolutionary  correction as  $-2.5 Q
    \log (1+z)$, where the coefficient  $Q$ is equal to $\sim 0.85$ in
    r-band , at redshift $z<0.3$.  }  and $ 70 \le \sigma_0 \le 420 \,
  km \, s^{-1}$. This subsample  consists of $1682$ ETGs out of $3690$
  galaxies in sample $C$.   We derive the best-fitting FP coefficients
  for  this  subsample,  referring  to  them as  the  {\it  reference}
  coefficients of the FP.
\item[--] We assign fake $\sigma_0$  values to sample C by using guess
  values of $''a''$, $''b''$, $''c''$ and $s_{\sigma_0}$. Applying the
  same cuts in magnitude and velocity dispersion as in the above step,
  we derive the best-fitting FP  coefficients and compared them to the
  {\it  reference}  FP coefficients.   The  guess  values of  $''a''$,
  $''b''$,   $''c''$  and   $s_{\sigma_0}$  are   changed   until  the
  best-fitting simulated FP matches  the {\it reference} relation.  In
  practise, we  are able  to match the  simulated and  {\it reference}
  coefficients at better than $2\%$ for both the \ls \, and orthogonal
  regressions.
\end{description}
Fig.~\ref{fig:simfp} compares the distribution of the $1682$ ETGs with
available $\sigma_0$'s from SDSS in  sample C with that of data-points
for  one  of  the toy  samples,  showing  the  similarity of  the  two
distributions.   The above  procedure  allows us  to create  simulated
samples  in the space  of \lre,  \mie, and  \ls \,  down to  a (model)
magnitude  limit of  $-17.64$,  which is  more  than $2.5$  magnitudes
fainter  than the $r$-band  completeness limit  ($-20.32$) of  the ETG
samples of  Sec.~\ref{sec:FPsamples}. The effect of  any selection cut
on the FP can then be estimated by computing the relative variation of
FP coefficients as one applies that selection to the toy samples.

\begin{figure}
\includegraphics[height=80mm]{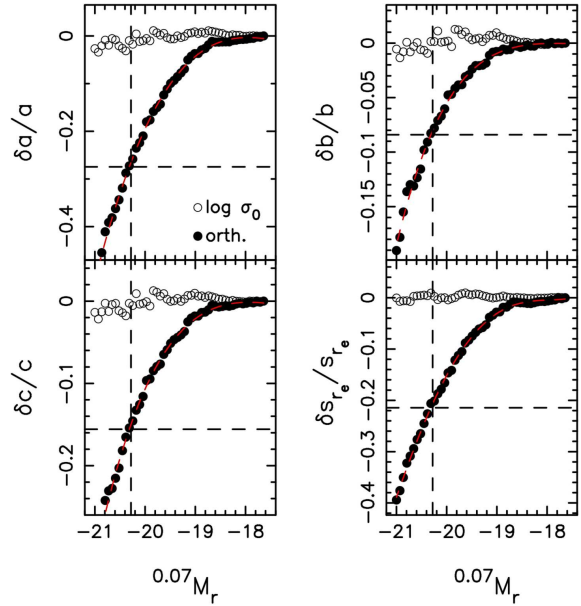}
\caption{Relative variation  of FP coefficients  as a function  of the
  magnitude cut (see the text).  The variation is computed between the
  magnitude selected and entire toy samples.  Empty and filled circles
  correspond to  the results  obtained for the  \ls \,  and orthogonal
  fits, respectively, as shown in lower-right corner of the upper-left
  panel.  From left  to right and top to bottom,  the four panels show
  the  relation variation  (bias)  in $''a''$,  $''b''$, $''c''$,  and
  $s_{r_e}$, respectively.   The vertical and  horizontal dashed lines
  mark the  completeness of the magnitude-selected sample  of ETGs and
  the corresponding  bias values, respectively.  The  red dashed curve
  in each panel is the  fourth order polynomial fit performed to model
  the bias as a function of \mr.  }
\label{fig:bias}
\end{figure}

Fig.~\ref{fig:bias} plots the relative variation of FP coefficients as
a function  of the magnitude cut.   The relative variation  of a given
quantity,  $x$, out of  $''a''$, $''b''$,  $''c''$, and  $s_{r_e}$, is
computed as $(x_{cut}-x) / x$,  where $x_{cut}$ is the value estimated
for that quantity when the cut  is applied. Here, instead of using the
procedure  of Sec.~\ref{sec:fit},  the  value of  $''c''$ is  directly
derived  from the  fit,  and the  $s_{r_e}$  is obtained  as the  mean
absolute  deviation  of residuals  around  the  plane,  using the  (no
bias-corrected)   best-fitting  coefficients  $''a''$,   $''b''$,  and
$''c''$. For  the orthogonal fit,  we see that brighter  the magnitude
cut, more the FP coefficients  tend to be underestimated. This finding
is  consistent  with  that  of previous  studies  (see~LBC00,~\citealt{HB09}).  For  the  \ls  \,   fit,  the  FP  coefficients  are  very
insensitive, as somewhat expected, to the selection in magnitude.
The vertical  lines in Fig.~\ref{fig:bias} correspond  to the (r-band)
model magnitude limit  (\mr$=-20.28$) of the magnitude-selected sample
of  ETGs (see  Sec.~\ref{sec:FPsamples}),  after the  small amount  of
luminosity evolution between $z=0.025$  and $z=0.075$ has been removed
(see above). For that magnitude limit, the amounts of bias in $''a''$,
$''b''$,  $''c''$ and  $s_{r_e}$ (horizontal  lines)  are significant,
amounting   to   about  $27   \%$,   $8\%$,   $16   \%$  and   $22\%$,
respectively. The same amounts of bias are also expected to affect the
colour--selected samples  of ETGs (see  Sec.~\ref{sec:FPsamples}), whose
magnitude  limit   in  r-band   is  very  similar   to  that   of  the
magnitude-selected  sample.  We  also  used the  simulated samples  to
estimate the impact  of the $\sigma_0$ cut of the  ETG's sample on the
FP  relation. To  this aim,  we  selected only  simulated points  with
magnitudes  brighter  than   \mr$=-20.28$.   Applying  the  $\sigma_0$
selection ($70 \le \sigma_0 \le  420 \, km/s$), we found that relative
variation of  FP slopes is  completely negligible ($<$1$\%$).  This is
due to the fact that,  for the magnitude range considered here, almost
all  galaxies  have  $\sigma_0>70  \,  km/s$,  making  the  $\sigma_0$
selection unimportant.

\begin{figure}
\begin{center}
\includegraphics[height=100mm]{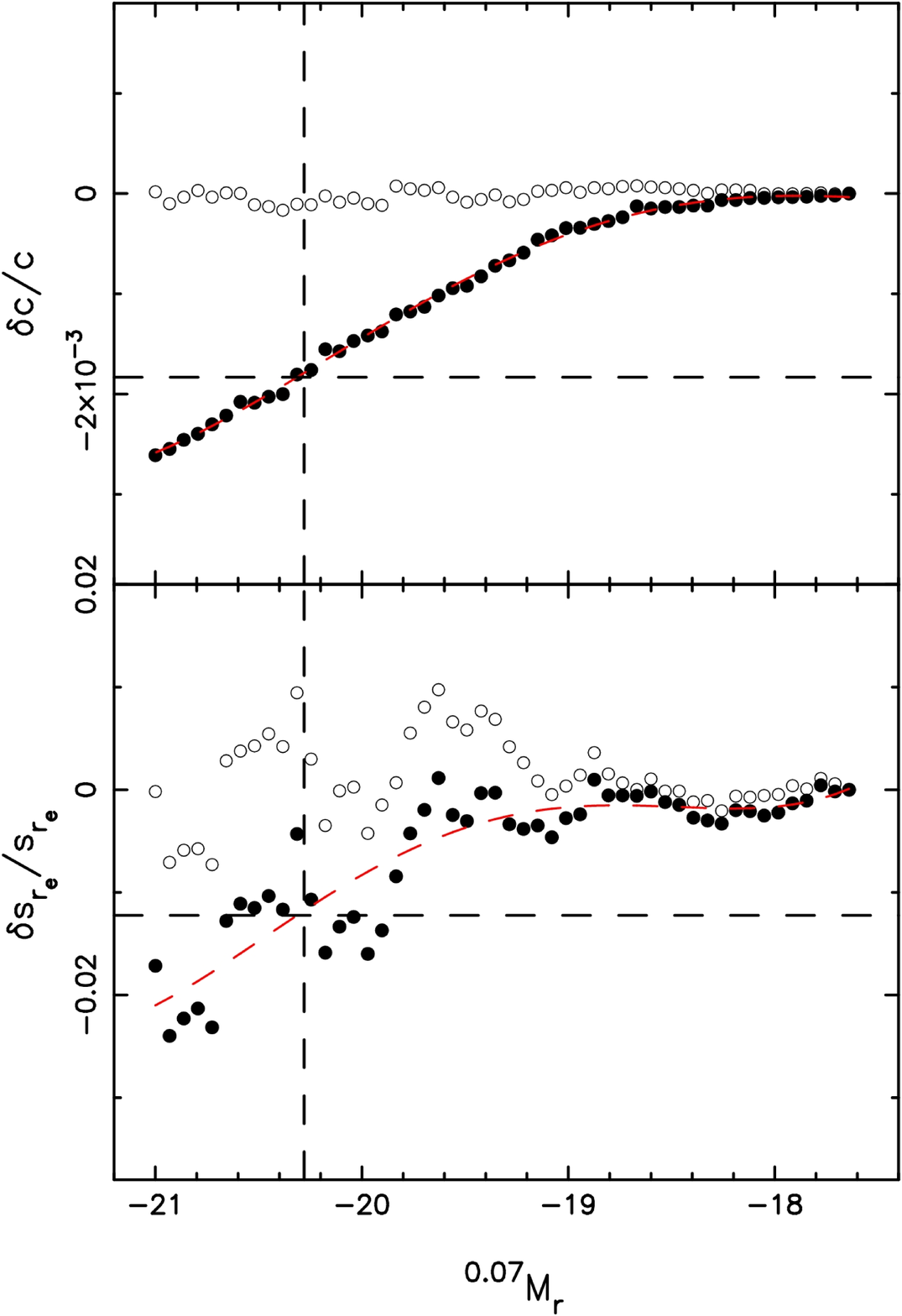}
\caption{Relative variation of the FP offset (upper panel) and scatter
  (lower panel) as a function  of the magnitude cut, as estimated from
  the toy FP samples. The offset, $''c''$, and scatter, $s_{r_e}$, are
  estimated  from the  bias-corrected FP  slopes.   Empty and  filled
  circles  correspond  to  the  \ls  \,  and  orthogonal  fit  values,
  respectively. The  red dashed lines  show a fourth  order polynomial
  fit of  the filled circles. The  vertical and dashed  lines mark the
  completeness   of  the   magnitude-selected  ETG   sample   and  the
  corresponding  expected  bias  values.   Notice  that  the  bias  is
  negligible for both quantities, being smaller than $\sim 2\%$.
\label{fig:bias_slopes}}
\end{center}
\end{figure}

For each  sample of ETGs,  as defined in  Sec.~\ref{sec:FPsamples}, we
consider  the corresponding  2DPHOT  r-band magnitude  limit. For  the
magnitude-  and  colour-selected  subsamples,  these limits  amount  to
$-20.55$  and $-20.6$,  respectively.  The  2DPHOT magnitude  limit is
translated to a  (model) magnitude limit by adding  the term $0.23$~mag
which  is the difference  of 2DPHOT  and SDSS  completeness magnitudes
(Sec.~\ref{sec:data}).   For a  given sample,  the amount  of  bias on
$''a''$  and  $''b''$  is  then  estimated evaluating  the  trends  in
Fig.~\ref{fig:bias}  for  the r-band  model  magnitude  limit of  that
sample. This is done only  for the orthogonal regression procedure, by
modelling   the  trends  in   Fig.~\ref{fig:bias}  with   fourth  order
polynomials.  The biased values  of $''a''$ and $''b''$ are multiplied
by the estimated $x/x_{cut}$ factors.  Notice that the same correction
factor  is   applied  to  all  the  $grizYJHK$   wavebands  (see  also
Sec.~\ref{sec:fp_slopes}).  The  bias-corrected values of  $''a''$ and
$''b''$    are    used    to    estimate   $''c''$    and    $s_{r_e}$
(Sec.~\ref{sec:fit}). Fig.~\ref{fig:bias_slopes}  shows how the values
of $''c''$  and $s_{r_e}$ vary as  a function of  the magnitude limit,
when  this  procedure  is  applied,  rather than  estimating  $c$  and
$s_{r_e}$  from the  fit,  as in  Fig.~\ref{fig:bias}.   As stated  in
Sec.~\ref{sec:fit}, the  estimates of  $''c''$ and $s_{r_e}$  from the
bias-corrected values  of $''a''$ and $''b''$  are almost insensitive,
within $\sim 2\%$, to the magnitude selection.

\subsection{Bias due to correlated errors on $r_e$ and \mie}
\label{sec:errors_bias}
Another possible source of bias  on FP coefficients is the correlation
of uncertainties on \lre \, and \mie. As shown in \papdata, the errors
on effective parameters mainly depend on the signal-to-noise per pixel
of  galaxy images,  and are  slightly larger  in the  NIR than  in the
optical  wavebands. For  instance, the  median  value of  the \lre  \,
uncertainties increases from $\sim 0.09$  in g-band, to $\sim 0.14$ in
K-band.  This  variation  might  imply  a spurious  dependence  of  FP
coefficients on  waveband, and thus we  have to correct  the FP slopes
separately  in each  band. \\  The  corrections are  estimated by  (1)
constructing simulated  samples of data-points  in the space  of \lre,
\mie, and  \ls, resembling  the distribution that  galaxy's parameters
would have in that space if no correlated errors on \re \, and \mie \,
would be present  (Sec.~\ref{sec:sim_samples}), and (2) estimating how
the  FP  slopes  change  by  adding correlated  uncertainties  on  the
effective      parameters      of      such     simulated      samples
(Sec.~\ref{sec:corr_errors}).    Notice  that   the  toy   samples  of
Sec.~\ref{sec:sel_effects}  are  not   suitable  to  apply  the  above
procedure,  since  the   corresponding  effective  parameters  already
include the effect of correlated errors on the effective parameters.

\subsubsection{Simulated samples with no correlated errors}
\label{sec:sim_samples}
Each simulated sample is generated as follows.  First, we extract \lre
\,  values from a  random deviate  whose centre  and width  values are
given by the mean ($0.27$~dex) and standard deviation ($s_e=0.25$~dex)
of the \lre \, distribution of sample C. For a given \lre, we assign a
\mie \, value by the Kormendy relation (hereafter KR)
\begin{equation}
 < \! \mu \! >_e = p_1 + p_2 \log r_e,
\label{eq:KR}
\end{equation}
where  $p_1$ and  $p_2$ are  the offset  and slope,  respectively. The
values  of $p_1$  and  $p_2$  are derived  by  a robust  least-squares
fitting procedure for galaxies in sample C, by minimising the absolute
sum  of  residuals   in  \mie  \,  around  the   relation.   As  shown
by~\citet{LBM03} (hereafter LBM03), the KR fit is quite insensitive to
the correlated  errors on \lre  \, and \mie.  The fit gives  $p_1 \sim
18.969$ and $p_2 \sim 1.95 $, respectively.  Then, we shift the values
of \mie  \, according to  a normal Gaussian  deviate of width  $0.4 \,
mag/arcsec^2$, corresponding to the intrinsic dispersion in \mie \, of
the KR  (LBM03). For a given  pair of \lre  \, and \mie \,  values, we
assign a \ls  \, value by the FP  relation (Eq.~\ref{eq:FP}).  The \ls
\, values are shifted according  to a random deviate with given width,
$s_0$. The free parameters of  this procedure, i.e.  the FP slopes and
offset, and the value of $s_0$, are chosen so that, on average, the FP
coefficients    of   simulated    samples   match    those    of   the
magnitude-selected sample  of ETGs, with the  same iterative procedure
as in Sec.~\ref{sec:sel_effects}.

\subsubsection{The effect of correlated uncertainties}
\label{sec:corr_errors}
The \lre  \, and  \mie \,  of the simulated  samples are  then shifted
according to a two-dimensional random deviate, whose covariance matrix
terms are given by the median uncertainties on \lre \, and \mie \, for
galaxies in the magnitude-selected  samples of ETGs.  The procedure is
repeated  for  each  waveband,   by  using  the  corresponding  median
covariance matrix of uncertainties  on effective parameters. We derive
the FP slopes by (i)  applying the correlated errors, and (ii) without
applying  any simulated  uncertainty on  the effective  parameters. We
indicate as $\delta_a$ and $\delta_b$  the ratios of FP slopes of case
(ii)  with respect to  those obtained  in case  (i).  Each  toy sample
includes  $N=2000$  data-points,  and  the values  of  $\delta_a$  and
$\delta_b$  are  averaged  over  $300$ realisations.   The  values  of
$\delta_a$ and $\delta_b$ in $grizYJHK$ bands, for both the orthogonal
and    \ls    \,    regression    procedures,    are    reported    in
Tab~\ref{tab:corr_bias_errors}.    The  correlated   uncertainties  on
effective parameters tend to increase the value of the \ls \, slope of
the FP, and  decrease the coefficient of the \mie  \, term. The effect
is quite  small, in particular for the  coefficient $''a''$, amounting
to  less than  a few  percent.  The  bias is  larger for  $''b''$, and
varies  almost  by  a factor  of  two  from  the  optical to  the  NIR
wavebands.   Moreover, unlike the  bias due  to selection  effects, it
affects both the  orthogonal and \ls \, regression  procedures. Due to
the  large number of  galaxies in  the SPIDER  sample, the  factors in
Tab.~\ref{tab:corr_bias_errors} are not negligible with respect to the
typical errors on FP  slopes (see Sec.~\ref{sec:fp_slopes}). Hence, we
correct the  slopes of  the FP  in each band  multiplying them  by the
corresponding     $\delta_a$     and     $\delta_b$     factors     in
Tab~\ref{tab:corr_bias_errors}. We  have also performed  some tests to
check  how robust  the values  of $\delta_a$  and $\delta_b$  are with
respect to the  procedure outlined above.  First, one  can notice that
the adopted slope of the KR  ($p_2=1.95$) is smaller than that of $p_2
\sim 3$ found by other  studies (see~LBM03 and references therein) and
by that  reported for the SPIDER samples  in Sec.~\ref{sec:kr}. Hence,
we derived  the offset of  the KR by  fixing $p_2=3$ and  repeated the
above   procedure  with   the  corresponding   values  of   $p_1$  and
$p_2$. Second, one may notice that  the width value itself of the \lre
\,  distribution, $s{_e}=0.25$~dex  (see above),  is broadened  by the
measurement  errors on  \lre, and  hence  does not  correspond to  the
intrinsic  width of  the \lre  \,  distribution. To  account for  this
effect, we  repeated the above procedure by  subtracting in quadrature
$0.1$~dex (the typical uncertainty on  \lre \, in r-band) to the value
of  $s_{e}$.  For  both  tests, we  found  that the  variation of  the
$\delta_a$ and $\delta_b$ estimates in Tab.~\ref{tab:corr_bias_errors}
is completely negligible, being smaller than $0.5\%$.

\begin{table}
\centering
\small
\begin{minipage}{80mm}
\caption{Effect   of  the   correlated   uncertainties  of   effective
  parameters on the slopes of the FP in different wavebands.}
\begin{tabular}{c|c|c|c|c}
\hline
& \multicolumn{2}{c|}{$orthogonal \, fit$} &  \multicolumn{2}{c}{$\log \sigma_0 \, fit$} \\
waveband & $\delta_a$ & $\delta_b$ & $\delta_a$ & $\delta_b$ \\
\hline
$g$ & $0.995$ & $1.038$ & $0.984$ & $1.039$ \\
$r$ & $0.990$ & $1.035$ & $0.980$ & $1.035$ \\
$i$ & $0.996$ & $1.031$ & $0.992$ & $1.031$ \\
$z$ & $0.986$ & $1.043$ & $0.975$ & $1.043$ \\
$Y$ & $0.992$ & $1.040$ & $0.980$ & $1.039$ \\
$J$ & $0.985$ & $1.060$ & $0.977$ & $1.062$ \\
$H$ & $0.980$ & $1.066$ & $0.964$ & $1.065$ \\
$K$ & $0.975$ & $1.070$ & $0.956$ & $1.069$ \\
\hline
  \end{tabular}
\label{tab:corr_bias_errors}
\end{minipage}
\end{table}

\subsection{Comparison of bias-corrected FP coefficients in r-band}
\label{sec:test_bias_corr}
\begin{figure}
\begin{center}
\includegraphics[height=80mm]{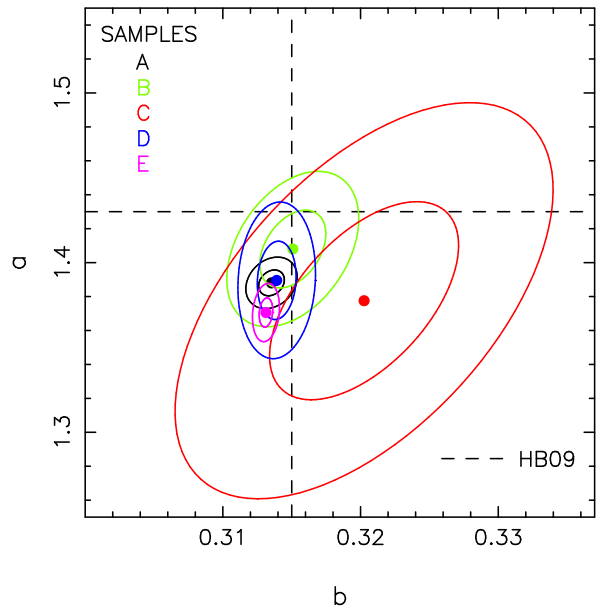}
\caption{Slopes of the r-band  FP, corrected for selection effects and
  correlated errors  on effective parameters, for  the control samples
  of ETGs (Tab.~\ref{tab:opt_samples}). Each  sample is plotted with a
  different colour, as shown in  the upper-left corner. For each point,
  the  corresponding  concentric  ellipses  denote  the  one  and  two
  $\sigma$ confidence  contours for a  two-dimensional normal Gaussian
  deviate. The  dashed lines mark the  values of $''a''$  and $''b''$ obtained
  from~\citet{HB09}.~\label{fig:COFF_BIAS_a_b}}
\end{center}
\end{figure}

In order to test the  above procedure for deriving the coefficients of
the FP and correct them for  the different sources of biases, we apply
it to  the control samples  of ETGs (Sec.~\ref{sec:FPcontrolsamples}).
In Fig.~\ref{fig:COFF_BIAS_a_b},  we plot the corrected  slopes of the
r-band FP for the five control samples. For sample $C$, we select only
those  $1682$  galaxies with  available  $\sigma_0$'s  from SDSS,  and
(model)      magnitudes      brighter      than     $-20.28$      (see
Sec.~\ref{sec:errors_bias}).   The values of  $''a''$ and  $''b''$ are
also compared  to those recently obtained  from~\citet{HB09}, who took
into account  selection effects in the fitting  procedure, rather than
applying correction  factors as we do  here.  For all  samples, the FP
slopes are corrected for  the magnitude bias evaluating the polynomial
curves in Fig.~\ref{fig:bias} at  a (model) magnitude of \mr$=-20.28$.
For  samples  $D$  and $E$,  this  \mr  \,  value corresponds  to  the
magnitude limit of $-20.55$, after difference between model and 2DPHOT
total  magnitudes is taken  into account  (see~\papdata). In  order to
remove the  effect of correlated  errors on effective  parameters, the
slopes of  samples $D$ and  $E$ have also  been divided by  the r-band
correction               factors              reported              in
Tab.~\ref{tab:corr_bias_errors}~\footnote{We    also   estimated   the
  uncertainties on  SDSS Photo parameters in  the same way  as for the
  2DPHOT effective parameters, i.e. by  comparing the values of \re \,
  and \mie  \, from SDSS in  r and i bands  (see~\papdata).  For these
  uncertainties,  we found that  the r-band  correction factors  on FP
  slopes    are    even     smaller    than    those    reported    in
  Tab.~\ref{tab:corr_bias_errors}. Hence, we  decided not to apply any
  further correction factor to samples $A$, $B$, and $C$.}.

Fig.~\ref{fig:COFF_BIAS_a_b} shows  that the FP slopes  of all control
samples  are remarkably  consistent  within the  2$\sigma$ level,  and
differ  by less  than $\sim  3  \%$ from  the values  of~\citet{HB09},
proving  the robustness  of  the procedure  outlined  above to  derive
bias-corrected FP coefficients.  The  consistency of FP slopes between
samples $A$ and $B$ ($D$ and $E$) shows that matching the ETG's sample
with UKIDSS does  not lead to any significant bias  in the estimate of
FP coefficients, in agreement  with~LBM08. One can also notice
that,  although  the  SDSS  and  2DPHOT  effective  parameters  differ
significantly (see~\papdata), the  corresponding FP relations are very
consistent, as  shown by the  consistency of FP slopes  between sample
$A$  and  $D$ ($B$  and  $E$).   This is  due  to  the  fact that  the
combination of  \re \, and  \mie \, that  enters the FP  is determined
with  much  better  accuracy  that  \re  \,  and  \mie  \,  themselves
(see~\citealt{Kelson00}), making the FP relation very stable.

\section{The Kormendy relation}
\label{sec:kr}
\begin{figure*}
\begin{center}
\includegraphics[width=170mm]{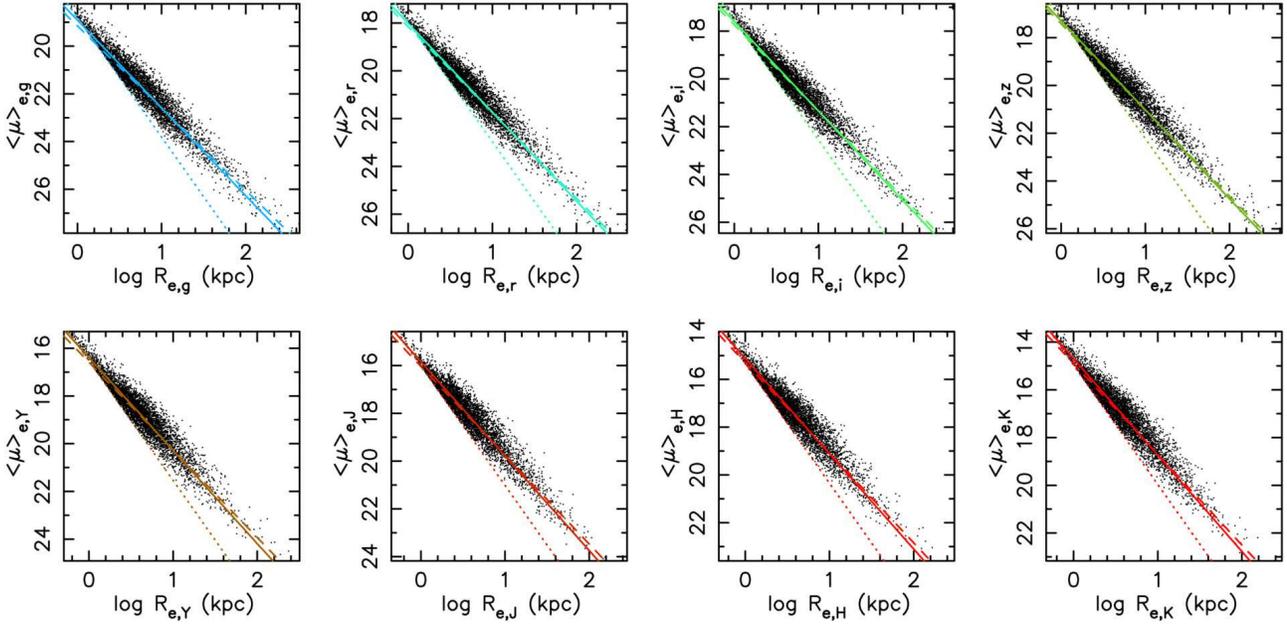}
\caption{Kormendy relation  of ETGs in the  $grizYJHK$ wavebands (from
  left to  right and top to  bottom). For each panel,  the dotted line
  mark the  completeness magnitude in the  corresponding waveband. The
  solid and  dashed lines are the best-fitting  relations, obtained by
  the orthogonal  and bisector fitting methods,  respectively (see the
  text).~\label{fig:KR_grizYJHK}}
\end{center}
\end{figure*} 
Fig.~\ref{fig:KR_grizYJHK}  plots  the \re--\mie  \,  diagram for  the
colour-selected  samples of  ETGs (Sec.~\ref{sec:FPsamples}),  from $g$
through $K$. For each band,  the figure also exhibits the completeness
limit  of the  sample  in that  band, from  Tab.~\ref{tab:FP_samples}.
Galaxies follow a  well-defined KR in all wavebands.   We write the KR
as in  Eq.~\ref{eq:KR}.  In order  to characterise the  offset, $p_1$,
the slope,  $p_2$, and the scatter,  $s_{_{KR}}$, of the  KR, we apply
the  {\it modified}  least-squares (hereafter  MLS)  fitting procedure
of~LBM03.  The MLS fit allows the coefficients of the KR to be derived
by accounting for selection cuts  in the \re--\mie \, diagram, such as
the magnitude  limit.  LBM03 applied  three MLS fits.   The MLS$_{\log
  r_e}$  and MLS$_{<  \!  \mu  \!  >_e}$  regressions are  obtained by
minimising the residuals  around the relation with respect  to \lre \,
and \mie, respectively. The MLSB  fit corresponds to the bisector line
of the  MLS$_{\log r_e}$ and MLS$_{<  \! \mu \! >_e}$  fits.  The MLSB
method  is more  robust and  effective (i.e.   lower  uncertainties on
fitting coefficients)  with respect to  the other MLS fits.   For this
reason, we apply here only  the MLSB fit.  Moreover, we generalise the
MLS method to the case  where orthogonal residuals around the relation
are minimised.  This orthogonal  MLS fit (hereafter MLSO) is described
in  App.~\ref{app:mlso}. For  both  the  MLSB and  MLSO  fits, the  KR
coefficients  are derived accounting  for the  magnitude limit  of the
sample  in the  corresponding  waveband.   The scatter  of  the KR  is
obtained by the standard deviation  of the \lre \, residuals about the
line,   accounting   for   the    magnitude   cut   as   detailed   in
App.~\ref{app:mlso}.   Fig.~\ref{fig:KR_grizYJHK} also plots  the MLSB
and MLSO  lines.  The corresponding fitting  coefficients are reported
in Tab.~\ref{tab:KR_cof}.

\begin{figure}
\begin{center}
\includegraphics[width=80mm]{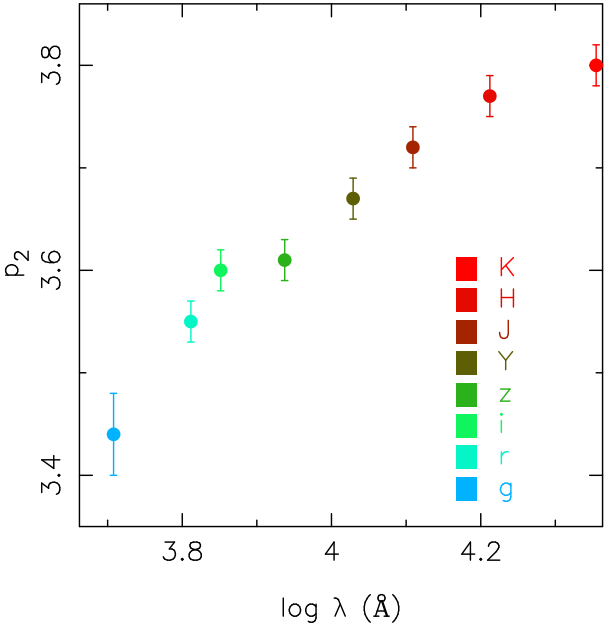}
\caption{The slope  of the KR, obtained  with the MLSB  fit, $p_2$, is
  plotted as a function of the logarithmic effective wavelength, $\log
  \lambda$,   of  the   passbands  where   effective   parameters  are
  measured.~\label{fig:KR_slopes}}
\end{center}
\end{figure} 

From Tab.~\ref{tab:KR_cof} one  sees that the MLSO fit  gives a larger
value  of the  slope, $p_2$,  with respect  to the  bisector  fit. The
scatter around the KR is independent of waveband, and larger, by $\sim
0.01$~dex,  for the  MLSB  than for  the  MLSO fit.   The KR  smoothly
steepens  from  the  $g$ through  the  $K$  band.   This is  shown  in
Fig.~\ref{fig:KR_slopes}, where we plot the  MLSB slope of the KR as a
function of  the logarithmic effective wavelength of  each filter. The
$p_2$ smoothly increase from a value  of $~\sim 3.44$ in $g$ to $~3.8$
in $K$.  A similar trend is  also observed for the results of the MLSO
fit.

\begin{figure}
\begin{center}
\includegraphics[width=80mm]{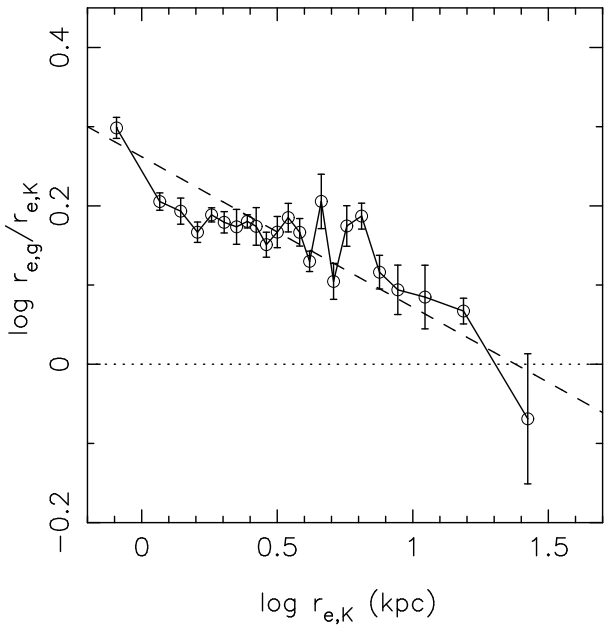}
\caption{Logarithmic  ratio of $g$  to $K$-band  effective radii  as a
  function  of  \lre  \,  in  K-band.  The  solid  line  connects  the
  data-points  obtained by  median binning  the distribution  of $\log
  r_{e,g}/r_{e,K}$  with  respect to  $\log  r_{e,K}$,  with each  bin
  including the  same number  (N=200) of points.   The trend  is fully
  consistent with that expected from  the variation of KR slope from g
  to K (dashed line).  The dotted  line marks the value of zero. Error
  bars  denote   $1\sigma$  errors  on  median   values  in  different
  bins.~\label{fig:reg_rek}}
\end{center}
\end{figure}

In order  to analyse the trend  of $p_2$ with waveband,  we follow the
same approach  as in~\citet{LMB04}. Given  two wavebands, $X$  and $W$
($X,W=grizYJHK$), one  can relate the corresponding slopes  of the KR,
$p_{2,X}$ and $p_{2,W}$, through the following equation:
\begin{equation}
 p_{2,X} = \frac{(p_{2,W}+5 \Delta_{XW})-\zeta
(5-p_{2,W})}{1+\Delta_{XW}}, 
\label{eq:p2XW}
\end{equation}
where $\Delta_{XW}$ is the slope  of the $X-W$ vs. $W$ colour-magnitude
relation,  and   $\zeta$  parametrizes  the  variation   of  the  mean
logarithmic   ratio   of   $X$   to   $W$   effective   radii,   $\log
\frac{r_{e,W}}{r_{e,X}}$, as a function of \re:
\begin{equation}
 \log \frac{r_{e,W}}{r_{e,X}} \propto \zeta \log {r_{e,X}}.
\label{eq:re_var}
\end{equation}
We first  consider the  case where the  mean ratio of  effective radii
does not change  along the sequence of ETGs,  i.e. $\zeta=0$.  Setting
$X=K$  and $W=g$, using  the value  of $\Delta_{gK}$  from~\papdata \,
($0.034  \pm 0.016$)  and  the value  of  $p_{2,g}$ of  the MLSB  fit,
Eq.~\ref{eq:p2XW} would imply $p_{2,K}=3.53  \pm 0.02$.  This value is
significantly  smaller  than  that reported  in  Tab.~\ref{tab:KR_cof}
($3.80  \pm   0.02$),  implying  that  the   assumption  $\zeta=0$  is
incorrect.   Indeed, inverting  Eq.~\ref{eq:p2XW} and  using  the MLSB
values  of  $p_{2,g}$  and  $p_{2,K}$, one  obtains  $\zeta=-0.19  \pm
0.02$. The negative sign of $\zeta$ implies that galaxies with smaller
$r_{e,K}$    tend     to    have,    on     average,    also    larger
$\frac{r_{e,g}}{r_{e,K}}$  value.   In  other  terms,  the  NIR  light
profile of ETGs is more concentrated in the centre with respect to the
optical for  small (relative to  larger) galaxies.  The  dependence of
$\frac{r_{e,g}}{r_{e,K}}$ on $r_{e,K}$ \,  can be directly analysed by
binning the SPIDER sample with  respect to $r_{e,K}$ and computing the
median value of $\frac{r_{e,g}}{r_{e,K}}$  in each bin.  The result of
this  procedure is  shown in  Fig.~\ref{fig:reg_rek}.  We  clearly see
that  the  median  value  of  $\frac{r_{e,g}}{r_{e,K}}$  decreases  as
$r_{e,K}$ increases, and that the  trend is fully consistent with what
expected   from    the   waveband   variation   of    the   KR   slope
(Eq.~\ref{eq:p2XW}  and Eq.~\ref{eq:re_var};  see dashed  line  in the
Figure).  In the  simplistic assumption that $\frac{r_{e,g}}{r_{e,K}}$
is  a  good  proxy  for  the  internal colour  gradient  in  ETGs,  the
increasing of KR  slope from $g$ through $K$  would imply that smaller
size ETGs have stronger  (more negative) internal colour gradients than
galaxies with  larger \re.  This point  will be further  analysed in a
forthcoming paper,  studying the  optical--NIR colour gradients  in the
SPIDER sample (see also~  \citealt{LdC09}) and their correlations with
galaxy properties.

The slope of the MLSB fit  can be compared to that obtained from LBM03
for a  sample of ETGs in  clusters at intermediate  redshifts, from $z
\sim 0$ to $z \sim 0.64$. Using  the MLSB fit, LBM03 found $p_2 = 2.92
\pm 0.08$ in V-band restframe.  This should be compared with the value
of $p_2=3.44 \pm  0.04$ we obtain for the SPIDER sample  in the g band
(see   Tab.~\ref{tab:KR_cof}),  which  matches   approximately  V-band
restframe.   The slope  of LBM03  is significantly  flatter,  by $\sim
15\%$, that that  we find here. One should  notice that LBM03 selected
ETGs by a cut in the  Sersic index $n$ ($n>2$), while ETGs are defined
here    according   to    several   photometric    and   spectroscopic
criteria. Moreover, ETGs  in the SPIDER sample reside  in a wide range
of  environments, while  ETGs in  LBM03 mostly  belong to  rich galaxy
clusters.   Both  these issues  might  be  responsible  for the  above
difference of KR slope values.

\begin{table*}
\small
\begin{minipage}{160mm}
\centering
\caption{Coefficients of the Kormendy relation in $grizYJHK$.}
\begin{tabular}{c|c|c|c|c|c|c}
\hline
band & $p_1$ & $p_2$ & $s_{_{KR}}$ & $p_1$ & $p_2$ & $s_{_{KR}}$ \\
& \multicolumn{3}{c|}{MLSB fit} & \multicolumn{3}{c|}{MLSO fit} \\
\hline
g & $ 19.16 \pm  0.04$ & $  3.44 \pm  0.04$ & $ 0.126 \pm 0.002$ & $ 18.92 \pm  0.02$ & $  
3.68 \pm  0.02$ & $ 0.115 \pm 0.001$ \\
r & $ 18.16 \pm  0.02$ & $  3.55 \pm  0.02$ & $ 0.120 \pm 0.002$ & $ 18.02 \pm  0.02$ & $  
3.72 \pm  0.03$ & $ 0.114 \pm 0.001$ \\
i & $ 17.74 \pm  0.02$ & $  3.60 \pm  0.02$ & $ 0.122 \pm 0.002$ & $ 17.60 \pm  0.02$ & $  
3.74 \pm  0.02$ & $ 0.117 \pm 0.002$ \\
z & $ 17.42 \pm  0.02$ & $  3.61 \pm  0.02$ & $ 0.121 \pm 0.002$ & $ 17.29 \pm  0.02$ & $  
3.73 \pm  0.03$ & $ 0.116 \pm 0.002$ \\
Y & $ 16.59 \pm  0.02$ & $  3.67 \pm  0.02$ & $ 0.125 \pm 0.002$ & $ 16.39 \pm  0.02$ & $  
3.90 \pm  0.03$ & $ 0.117 \pm 0.001$ \\
J & $ 16.03 \pm  0.02$ & $  3.72 \pm  0.02$ & $ 0.126 \pm 0.002$ & $ 15.84 \pm  0.02$ & $  
3.95 \pm  0.02$ & $ 0.117 \pm 0.002$ \\
H & $ 15.31 \pm  0.02$ & $  3.77 \pm  0.02$ & $ 0.126 \pm 0.002$ & $ 15.13 \pm  0.02$ & $  
3.99 \pm  0.03$ & $ 0.119 \pm 0.001$ \\
K & $ 14.91 \pm  0.02$ & $  3.80 \pm  0.02$ & $ 0.128 \pm 0.002$ & $ 14.70 \pm  0.02$ & $  
4.04 \pm  0.03$ & $ 0.119 \pm 0.001$ \\
\hline
  \end{tabular}
\label{tab:KR_cof}
\end{minipage}
\end{table*}

\section{The Faber--Jackson relation}
\label{sec:fj}
\begin{figure*}
\begin{center}
\includegraphics[width=160mm]{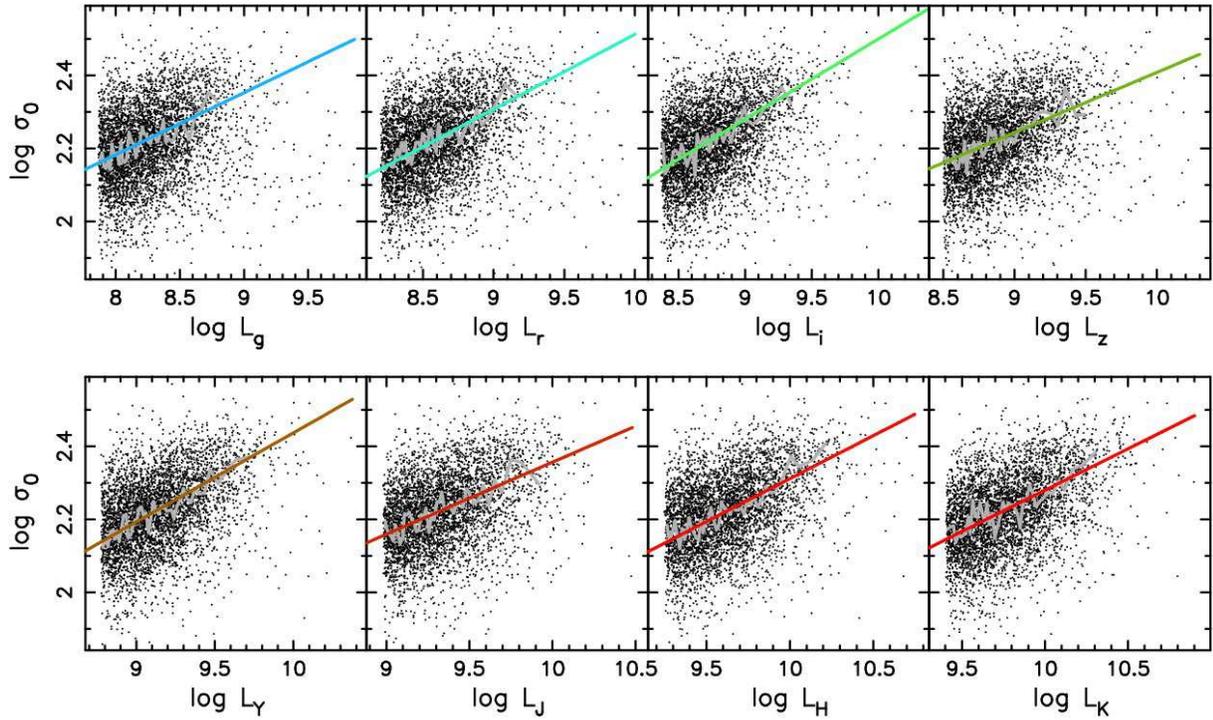}
\caption{The  Faber-Jackson   relation  of  ETGs   in  the  $grizYJHK$
  wavebands (from left  to right and top to  bottom).  For each panel,
  the grey curve is obtained by binning the data with respect to $\log
  L$, with  each including the  same number ($N=40$) of  galaxies. For
  each  bin,  the  bi-weight  peak  of  the  \ls  \,  distribution  is
  computed.  Coloured  lines show  the  orthogonal  fit  to the  binned
  data.~\label{fig:FJ_grizYJHK}}
\end{center}
\end{figure*}

 We write the Faber-Jackson (hereafter FJ) relation as:
\begin{equation}
 \log \sigma_0 = \lambda_0 + \lambda_1 (\log L + 0.4 X)
\label{eq:FJ}
\end{equation}
where \fja  \, and \fjb \, are  the offset and slope  of the relation,
and   X  is   the   magnitude   limit  in   a   given  waveband   (see
Tab.~\ref{tab:FP_samples}).    According   to   this   notation,   the
coefficient \fja \, is the \ls \, value predicted from the FJ relation
for galaxies of magnitude $X$.  The galaxy luminosity, $L$, is defined
as  $10^{-0.4  \times {^{0.07}M}}$,  where  $^{0.07}M$  is the  2DPHOT
absolute  magnitude  in  the  given  band.  In  order  to  derive  the
coefficients \fja \, and \fjb  \, we use the colour-selected samples of
ETGs (Sec.~\ref{sec:FPsamples}).  Fig.~\ref{fig:FJ_grizYJHK} plots the
distributions of ETGs in the \ls \, vs. $\log L$ diagrams. Each sample
is binned in \ls, and the peak value of the $\log L$ distribution in a
given  bin is computed  by the  bi-weight statistics~\citep{Beers:90}.
Since all  colour-selected samples are magnitude  complete, the binning
procedure produces unbiased estimates of the average $\log L$ value as
a function of \ls. The binned values  of $\log L$ vs.  \ls \, are then
fitted with  an orthogonal least--squares fitting  procedure. For each
band,  the fit  is  performed over  a  fixed luminosity  range of  one
decade, with $-0.4 X \le \log  L \le -0.4 X+1$.  Uncertainties on \fja
\, and \fjb \, are estimated by $N=500$ bootstrap iterations, shifting
each time  the $\log L$ binned  values according to  their error bars.
The values of \fja \, and  \fjb \, in $grizYJHK$ bands are reported in
Tab.~\ref{tab:FJ_cof}, along with the  \ls \, scatter of the relation,
$\sigma_{_{FJ}}$,       and       its      intrinsic       dispersion,
$\sigma_{_{FJ}}^{i}$. The  scatter is estimated as  follows.  For each
bootstrap  iteration, we  calculate the  rms of  the \ls  \, residuals
through the  median absolute deviation  estimator. The mean  value and
the  standard  deviation  of   the  rms  values  among  the  different
iterations provide the $\sigma_{_{FJ}}$,  and its error. The intrinsic
scatter is computed by a similar procedure, subtracting in quadrature,
for each iteration, the amount  of dispersion due to the uncertainties
on  $\log  L$  and  \ls  \,  from the  rms  values.   Considering  the
uncertainties, the slopes of the FJ relations are consistent among the
different  wavebands, with  the mean  value  of \fjb  \, amounting  to
$0.198   \pm   0.007$.   Using   the   magnitude-   rather  than   the
colour-selected samples of ETGs, this  result does not change, with the
value of \fjb  \, varying from $0.192 \pm 0.018$  to $0.209 \pm 0.018$
in r band, and from $0.220 \pm  0.023$ to $0.216 \pm 0.032$ in K band.
Using STARLIGHT \ls \, values  would also not change significantly the
\fjb \, values, with the mean value of \fjb \, varying from $0.198 \pm
0.007$  to  $0.187  \pm   0.007$.  For  what  concerns  the  intrinsic
dispersion  around the  FJ relation,  it smoothly  decreases  by $\sim
0.008$~dex from $g$  through $K$, with a value  of $\sim 0.091$~dex in
the optical and $\sim0.083$~dex in K band.  Fixing the slope of the FJ
relation in all wavebands to  the average value of \fjb$=0.198$, would
make  this  amount  of   variation  to  be  $0.017$~dex,  rather  than
$0.008$~dex.     Subtracting    in    quadrature   the    values    of
$\sigma_{_{FJ}}^{i}$ between  the g- and K-bands, one  obtains a value
of $\sim 0.037$~dex (i.e. $\sim 9\%$ in $\sigma_0$).

The slope value of the r-band  FJ relation is close, but flatter, than
that  of $0.25$  reported by~\citet{BERN07}  (see their  eq.~2).  This
difference can  be explained  by the fact  that we use  Sersic (rather
than  de Vaucouleurs)  total magnitudes  and by  the  small systematic
effect in SDSS model magnitudes (see~\papdata).  As shown in~\papdata,
both  effects  make  2DPHOT  total  magnitudes to  be  shifted  toward
brighter values with respect to  SDSS model magnitudes.  The amount of
shift is larger for bright than faint galaxies, producing a flatter FJ
relation.  { The difference might  also be related to the fact that
  the  slope of  the  FJ relation  seems  to change  according to  the
  magnitude    range     where    galaxies    are     selected    (see
  e.g.~\citealt{MG:05})}.   The slope  value of  the  K-band relation,
\fjb$\sim0.23$, is fully consistent  with the value of $0.24$ reported
by~\citet{PdCD98a}.   For what concerns  the intrinsic  dispersion, we
find  a value of  $\sigma_{_{FJ}}^{i} \sim  0.09$~dex in  the optical,
while~\citet{BERN07}, find a smaller value of $\sim 0.07$~dex.

\begin{table*}
\small
\begin{minipage}{160mm}
\centering
\caption{Coefficients of the Faber-Jackson relation in $grizYJHK$.}
\begin{tabular}{c|c|c|c|c}
\hline
 band & $\lambda_0$ & $\lambda_1$ & $\sigma_{_{FJ}}$ & $\sigma_{_{FJ}}^{i}$\\
\hline
$ g $ & $    2.158 \pm    0.008 $ & $    0.172 \pm    0.018 $ & $    0.097
\pm    0.002 $ & $    0.091 \pm    0.002 $ \\
$ r $ & $    2.151 \pm    0.008 $ & $    0.192 \pm    0.018 $ & $    0.096
\pm    0.002 $ & $    0.090 \pm    0.002 $ \\
$ i $ & $    2.155 \pm    0.008 $ & $    0.185 \pm    0.016 $ & $    0.093
\pm    0.002 $ & $    0.087 \pm    0.002 $ \\
$ z $ & $    2.158 \pm    0.008 $ & $    0.172 \pm    0.018 $ & $    0.097
\pm    0.002 $ & $    0.091 \pm    0.002 $ \\
$ Y $ & $    2.144 \pm    0.007 $ & $    0.217 \pm    0.016 $ & $    0.094
\pm    0.002 $ & $    0.087 \pm    0.009 $ \\
$ J $ & $    2.168 \pm    0.008 $ & $    0.194 \pm    0.022 $ & $    0.091
\pm    0.002 $ & $    0.084 \pm    0.002 $ \\
$ H $ & $    2.140 \pm    0.008 $ & $    0.233 \pm    0.018 $ & $    0.091
\pm    0.002 $ & $    0.084 \pm    0.002 $ \\
$ K $ & $    2.143 \pm    0.009 $ & $    0.220 \pm    0.023 $ & $    0.090
\pm    0.002 $ & $    0.083 \pm    0.002 $ \\
\hline
\end{tabular}
\end{minipage}
\label{tab:FJ_cof}
\end{table*}

\section{FP slopes}
\label{sec:fp}

\begin{table*}
\small
\begin{minipage}{160mm}
\centering
\caption{Coefficients of the FP in $grizYJHK$ from the orthogonal fit for the
magnitude-selected sample of ETGs.}
\begin{tabular}{c|c|c|c|c|c}
\hline
  band & $a$ & $b$ & $c$ & $s_{r_e}$ & $s^{i}_{r_e}$ \\
\hline
$ g $ & $ 1.384 \pm 0.024$ & $ 0.315 \pm 0.001$ & $-9.164 \pm 0.079$ & $ 0.125 \pm
0.002$ & $ 0.095 \pm 0.003$ \\
$ r $ & $ 1.390 \pm 0.018$ & $ 0.314 \pm 0.001$ & $-8.867 \pm 0.058$ & $ 0.112 \pm
0.002$ & $ 0.082 \pm 0.002$ \\
$ i $ & $ 1.426 \pm 0.016$ & $ 0.312 \pm 0.001$ & $-8.789 \pm 0.053$ & $ 0.110 \pm
0.002$ & $ 0.079 \pm 0.002$ \\
$ z $ & $ 1.418 \pm 0.021$ & $ 0.317 \pm 0.001$ & $-8.771 \pm 0.072$ & $ 0.111 \pm
0.002$ & $ 0.079 \pm 0.003$ \\
$ Y $ & $ 1.467 \pm 0.019$ & $ 0.314 \pm 0.001$ & $-8.557 \pm 0.058$ & $ 0.107 \pm
0.002$ & $ 0.081 \pm 0.002$ \\
$ J $ & $ 1.530 \pm 0.017$ & $ 0.318 \pm 0.001$ & $-8.600 \pm 0.060$ & $ 0.111 \pm
0.001$ & $ 0.083 \pm 0.002$ \\
$ H $ & $ 1.560 \pm 0.021$ & $ 0.318 \pm 0.002$ & $-8.447 \pm 0.077$ & $ 0.117 \pm
0.002$ & $ 0.087 \pm 0.003$ \\
$ K $ & $ 1.552 \pm 0.021$ & $ 0.316 \pm 0.002$ & $-8.270 \pm 0.076$ & $ 0.118 \pm
0.002$ & $ 0.089 \pm 0.002$ \\
\hline
  \end{tabular}
\label{tab:FP_orth_coef}
\end{minipage}
\end{table*}

\subsection{Variation from $g$ through $K$}
\label{sec:fp_slopes}
\begin{figure*}
\begin{center}
\includegraphics[width=160mm]{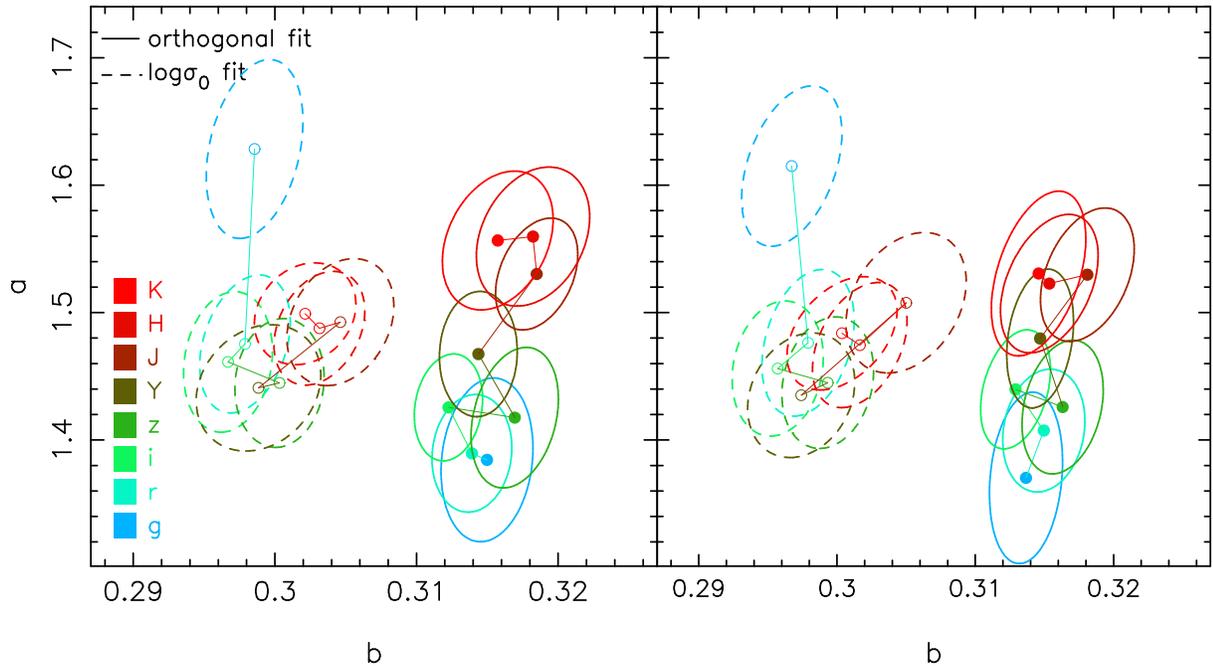}
\caption{The \ls \, slope, $''a''$, of  the FP is plotted against the \mie
  \, slope $''b''$. The left panel shows the case where the same sample of
  ETGs is  used to  derive the  FP in all  wavebands, while  the right
  panel  exhibits the  results  obtained {  for the  colour-selected
    samples (see Sec.~\ref{sec:FPsamples})}.  In each panel, different
  colours denote different wavebands as shown in the lower--left corner
  of  the figure. Filled  and empty  symbols mark  the results  of the
  orthogonal  and \ls  \, fits,  respectively, with  dashed  and solid
  ellipses       corresponding      to       1$\sigma$      confidence
  contours.~\label{fig:FP_grizYJHK}}
\end{center}
\end{figure*}

Due to the large sample size and the wide wavelength baseline provided
by SDSS+UKIDSS,  we can  establish the waveband  dependence of  the FP
with  unprecedented  accuracy.   Fig.~\ref{fig:FP_grizYJHK} plots  the
slopes of the  FP in different wavebands, obtained  for the magnitude-
and colour-selected  subsamples of ETGs  (Sec.~\ref{sec:FPsamples}). In
each  case, we  show the  results of  both the  \ls \,  and orthogonal
regression procedures.

The slopes of  the orthogonal fit are corrected  for the magnitude cut
bias as described in Sec.~\ref{sec:sel_effects}.  In the $r$ band, the
2DPHOT magnitude limit of the magnitude- and colour-selected samples of
ETGs are $-20.55$ and  $-20.60$, respectively. We convert these values
to model  magnitude limits in r-band  at redshift $z  \sim 0.025$, and
then  estimate the  corresponding  correction factors  on $''a''$  and
$''b''$ from  the polynomial curves in  Fig.~\ref{fig:bias}.  Since we
have  selected either the  same sample  of ETGs  at all  wavebands, or
ETG's  samples  with  {\it  equivalent} magnitude  limits  {  (i.e.
  colour-selected samples, see Sec.~\ref{sec:FPsamples})}, we apply the
same correction  factors to all the  $grizYJHK$ wavebands.  Therefore,
although the values  of FP coefficients in a given  band depend on the
correction factors,  their relative variation from $g$  through $K$ is
essentially independent of them. For  the \ls \, fitting method, which
is     not     affected     from     the    magnitude     cut     (see
Sec.~\ref{sec:sel_effects}),  no  correction  is  applied.   For  both
fitting methods, the slopes are  also corrected for the (small) effect
of      correlated       errors      on      effective      parameters
(Sec.~\ref{sec:errors_bias}),   using   the   correction  factors   in
Tab.~\ref{tab:corr_bias_errors}. For  the magnitude-selected sample of
ETG,   the   corrected   values   of   FP   slopes   are   listed   in
Tabs.~\ref{tab:FP_orth_coef}    and~\ref{tab:FP_ls_coef}    for    the
orthogonal  and  \ls   \,  regression  procedures,  respectively.   In
Tab.~\ref{tab:FP_orth_coef}  $''a''$ and $''b''$  are the  slopes, and
$''c''$ and $s_{r_e}$ are the offset and the \lre \, dispersion of the
FP.   Error  bars  denote  $1\sigma$ standard  errors.   The  quantity
$s^{i}_{r_e}$  is  the  intrinsic  dispersion of  the  relation  along
\lre.  In  Tab.~\ref{tab:FP_ls_coef},  $''a''$  and  $''b''$  are  the
slopes,  and $''c''$  and $s_{r_e}$  are the  offset and  the  \lre \,
dispersion of  the FP.  Error  bars denote $1\sigma$  standard errors.
The quantities $''c''$  and $s_{r_e}'$ are the offset  and the scatter
of the FP  re-measured by fixing the values of  $''a''$ and $''b''$ in
all wavebands. The quantity $\delta_{r_e}$ is the amount of dispersion
in the \lre \, direction around the plane due to measurement errors on
effective parameters  and velocity dispersion,  while $s^{i}_{r_e}$ is
the  intrinsic scatter  of the  FP  along the  \lre \,  axis. For  the
colour-selected  samples, the  coefficients are  very similar  to those
obtained  for the  magnitude-selected  samples, and  are not  reported
here.  These tables show how small the statistical uncertainties on FP
slopes are, amounting to only  a few percent in all wavebands.  Notice
that the large number of ETGs  makes the NIR FP coefficients to have a
much better accuracy than any previous study.

\begin{figure}
\begin{center}
\includegraphics[width=80mm]{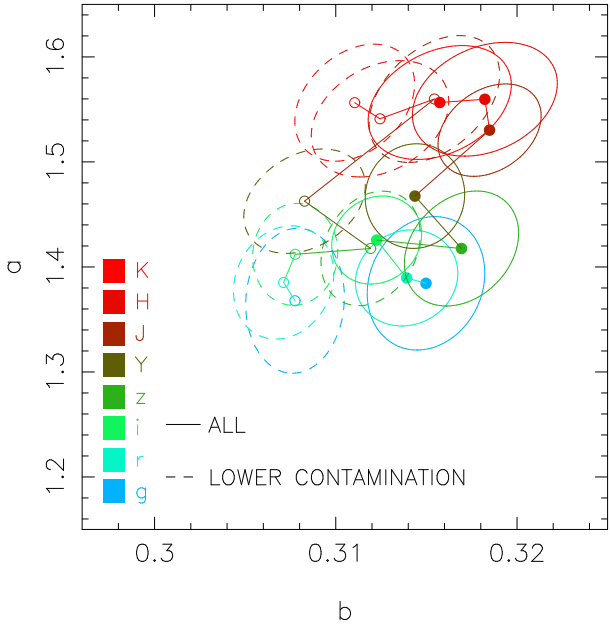}
\caption{Effect of  limiting the analysis  to the sample of  ETGs with
  lower   contamination   from   galaxies  with   residual   disc-like
  morphological  features (see  paper  I).  Filled  circles and  solid
  ellipses refer  to the results  of the orthogonal  fitting procedure
  for the magnitude-limited sample  of ETGs. Dashed ellipses and empty
  circles are  those obtained for the sample  with lower contamination
  (see the  text). Ellipses denote 2~$\sigma$ error  contours.  The FP
  coefficients turn out to be consistent between the two cases, from g
  through K.  ~\label{fig:FP_grizYJHK_lowcont}}
\end{center}
\end{figure}

Both the  magnitude- and colour-selected  samples of ETGs  exhibit very
similar trends  in Fig.~\ref{fig:FP_grizYJHK}. For the \ls  \, fit, we
do not see any systematic variation of the FP with waveband.  From $r$
through  $K$,  the values  of  $''a''$  are  consistent at  less  than
2$\sigma$.  In  g-band, the \ls  \, slope is  larger than that  in the
other bands. The difference between $g$ and $r$-band values of $''a''$
is   significant  at   $\sim  3   \sigma$,  after   the  corresponding
uncertainties  are   taken  into  account~\footnote{To   estimate  the
  significance level, we  add in quadrature the errors  on $''a''$ for
  the  two wavebands,  assuming  they can  be  treated as  independent
  uncertainties.}.  For what concerns the coefficient $''b''$, all the
values  are   very  consistent.   On  the   contrary,  the  orthogonal
regression  exhibits a  clear, though  small, variation  of  the slope
$''a''$ from  $g$ through $K$. The  value of $''a''$ is  found to vary
from $\sim  1.38$ in  $g$ to $\sim  1.55$ in  $K$, implying a  $12 \%$
variation  across the $grizYJHK$  wavebands.  The  coefficient $''b''$
does not  change with  waveband. We analysed  if these results  can be
affected  by  the (small)  contamination  of  the  SPIDER sample  from
early-type spirals. In paper I,  we showed that the contamination from
such  systems  is expected  to  be $\sim  13\%$.   We  also defined  a
subsample   of   ETGs   with    a   lower   contamination   of   $\sim
5\%$. Fig.~\ref{fig:FP_grizYJHK_lowcont} compares the FP slopes of the
magnitude-selected  sample  with  those  obtained  by  selecting  only
galaxies in  the lower contamination subsample. The  values of $''a''$
are fully  consistent between  the two cases  in all  wavebands, while
there is  only a marginally  significant ($\sim$2~$\sigma$) difference
in $''b''$.

\begin{table*}
\small
\begin{minipage}{180mm}
\centering
\caption{Coefficients of the FP in $grizYJHK$ from the \ls \, fit for the
magnitude-selected sample of ETGs.}
\begin{tabular}{c|c|c|c|c|c|c|c|c}
\hline
  band & $a$ & $b$ & $c$ & $s_{r_e}$ & $c'$& $s'_{r_e}$& $\delta_{r_e}$& $s^{i}_{r_e}$ \\
\hline
$ g $ & $ 1.615 \pm 0.032$& $ 0.297 \pm 0.002$& $-9.275 \pm 0.095$& $ 0.135 \pm
0.002$& $-9.080 \pm 0.002$& $ 0.128 \pm 0.002$& $ 0.080 \pm 0.001$& $ 0.100
\pm 0.002$\\
$ r $ & $ 1.476 \pm 0.029$& $ 0.298 \pm 0.002$& $-8.726 \pm 0.083$& $ 0.112 \pm
0.001$& $-8.813 \pm 0.002$& $ 0.115 \pm 0.001$& $ 0.075 \pm 0.001$& $ 0.087
\pm 0.002$\\
$ i $ & $ 1.456 \pm 0.027$& $ 0.296 \pm 0.002$& $-8.517 \pm 0.074$& $ 0.107 \pm
0.001$& $-8.694 \pm 0.002$& $ 0.111 \pm 0.001$& $ 0.075 \pm 0.001$& $ 0.082
\pm 0.002$\\
$ z $ & $ 1.445 \pm 0.026$& $ 0.299 \pm 0.002$& $-8.477 \pm 0.073$& $ 0.104 \pm
0.001$& $-8.605 \pm 0.002$& $ 0.108 \pm 0.001$& $ 0.075 \pm 0.001$& $ 0.078
\pm 0.002$\\
$ Y $ & $ 1.435 \pm 0.025$& $ 0.297 \pm 0.002$& $-8.164 \pm 0.073$& $ 0.099 \pm
0.001$& $-8.353 \pm 0.002$& $ 0.105 \pm 0.001$& $ 0.066 \pm 0.001$& $ 0.081
\pm 0.002$\\
$ J $ & $ 1.508 \pm 0.028$& $ 0.305 \pm 0.002$& $-8.308 \pm 0.085$& $ 0.103 \pm
0.001$& $-8.195 \pm 0.002$& $ 0.102 \pm 0.001$& $ 0.062 \pm 0.001$& $ 0.081
\pm 0.002$\\
$ H $ & $ 1.474 \pm 0.025$& $ 0.302 \pm 0.002$& $-7.966 \pm 0.074$& $ 0.105 \pm
0.001$& $-7.991 \pm 0.002$& $ 0.106 \pm 0.001$& $ 0.068 \pm 0.001$& $ 0.082
\pm 0.002$\\
$ K $ & $ 1.484 \pm 0.023$& $ 0.300 \pm 0.002$& $-7.844 \pm 0.072$& $ 0.106 \pm
0.001$& $-7.872 \pm 0.002$& $ 0.107 \pm 0.001$& $ 0.067 \pm 0.001$& $ 0.082
\pm 0.002$\\
\hline
\end{tabular}
\label{tab:FP_ls_coef}
\end{minipage}
\end{table*}

The values  of $''a''$ and $''b''$  in Tab.~\ref{tab:FP_orth_coef} can
be  compared  with those  obtained  from  previous  studies using  the
orthogonal fitting  procedure. The r-band value of  $''a''$ is consistent,
at $2  \sigma$, with that  of $a=1.49 \pm  0.05$ found by  BER03b, and
with the value of $a  \sim 1.434$ reported by~\citet{HB09}.  The value
of $''a''$  is larger, at the  $2 \sigma$ level, than  that of $a=1.24
\pm 0.07$  found by JFK96.  As noticed by  BER03b, the origin  of such
difference is  still not understood,  although one may notice  that it
further  reduces  when considering  the  value of  $''a''=1.31\pm0.07$
found from JFK96  for ETGs in the Coma cluster.  For what concerns the
coefficient $''b''$ of  the FP, its value in r  band ($\sim 0.314$) is
consistent with that  of $0.328\pm0.008$ found by JFK96,  and with the
value  of $\sim 0.316$  from~\citet{HB09}. On  the other  hand, BER03b
report a  somewhat lower value  of $''b''=0.300 \pm 0.004$.   For what
concerns the  NIR FP, the  values of the  slopes can be  compared with
those obtained  from~\citet{PDdC98b}, who found  $''a''=1.53 \pm 0.08$
and  $''b''=0.316\pm0.012$, still very  consistent with  our findings.
The  finding that $''b''$  does not  change with  waveband is  in full
agreement with what already suggested by~\citet{PDdC98b}.

The  above results on  the waveband  dependence of  the FP  extend the
findings of~LBM08,  who derived the  FP in the  $r$ and $K$  bands for
$1,400$ ETGs selected  with similar criteria as in  the present study.
For  the  orthogonal  fit,  LBM08  obtain $''a''=1.42  \pm  0.05$  and
$''b''=0.305 \pm  0.003$ in  $r$ band, and  $''a''=1.53 \pm  0.04$ and
$''b''=0.308 \pm 0.003$ in $K$  band.  The values of $''a''$ are fully
consistent   with  those   reported   in  Tabs.~\ref{tab:FP_orth_coef}
and~\ref{tab:FP_ls_coef}, while the values  of $''b''$ are smaller, at
$2.5~\sigma$,  than those we  find here.   This (small)  difference is
likely explained  by the  different correction procedure  adopted here
with respect to that of LBM08.  In agreement with LBM08, we find that,
when  considering the \ls  \, fit,  one does  not see  any significant
variation of  FP slopes with waveband.  When  comparing the orthogonal
fit results in  $r$ and $K$ bands, LBM08 found a  variation of only $8
\pm  4\%$   (see  the  values  reported  in   their  table~1).   Here,
considering   the   $r$   and   $K$   band  values   of   $''a''$   in
Tab.~\ref{tab:FP_orth_coef}, we find a variation of $11 \pm 2 \%$. The
variation is even smaller, amounting to $\sim 8.5\%$, when considering
the  colour-selected samples  of  ETGs.  Both  values, are  consistent,
within the uncertainties, with those found by LBM08.

\subsection{Dependence on velocity dispersion estimates and 
magnitude range}
\label{sec:fp_slopes_sig_mag}
\begin{figure}
\begin{center}
\includegraphics[width=80mm]{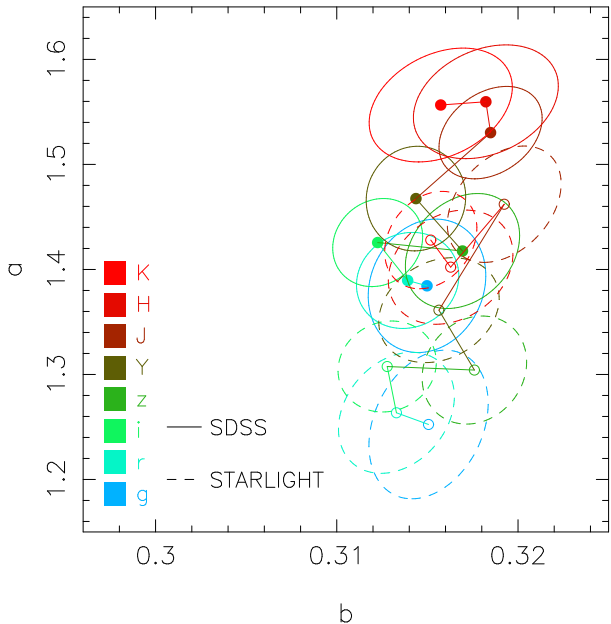}
\caption{Effect of changing the  method to derive velocity dispersions
  on FP slopes.  Filled and empty  symbols are the values of FP slopes
  obtained  by   using  SDSS-DR6  and   STARLIGHT  $\sigma_0$  values,
  respectively, for the magnitude-selected  sample of ETGs.  Solid and
  dashed ellipses plot 2~$\sigma$  confidence contours for the samples
  with  SDSS-DR6  and  STARLIGHT velocity  dispersions,  respectively.
  Different    colours    mark     different    wavebands,    as    in
  Fig.~\ref{fig:FP_grizYJHK}. Notice that scales and labelling are the
  same                              as                              in
  Fig.~\ref{fig:FP_grizYJHK_lowcont}.~\label{fig:FP_SDSS_SL}}
\end{center}
\end{figure} 
As   described  in~\papdata,   two  alternative   velocity  dispersion
estimates are available for the entire sample of ETGs, those retrieved
from  SDSS-DR6  and  the new  values  we  have  measured by  means  of
STARLIGHT~\citep{CID05}.   Fig.~\ref{fig:FP_SDSS_SL}  compares the  FP
slopes we derive  in the different wavebands when  using either one or
the other set of $\sigma_0$  values. Although we find a good agreement
among STARLIGHT and SDSS-DR6  $\sigma_0$ values (see~\papdata), the FP
slopes slightly  change when using either  one or the  other source of
$\sigma_0$'s. In  particular, the  value of $''a''$  is sistematically
smaller for STARLIGHT with respect to SDSS-DR6. Averaging over all the
wavebands, the difference  amounts to $\sim -9\%$. We  notice that the
r-band  value  of $a=1.26\pm  0.03$  from  the STARLIGHT  $\sigma_0$'s
matches  exactly   the  value  of  $''a''$  obtained   by  JFK96  (see
Sec.~\ref{sec:fp}), implying that the method to measure the $\sigma_0$
might be one main driver of the differences in FP coefficients between
BER03b and JFK96. Notice also that the value of $''b''$ is essentially
independent of the velocity dispersion estimates.
\begin{figure}
\begin{center}
\includegraphics[width=75mm]{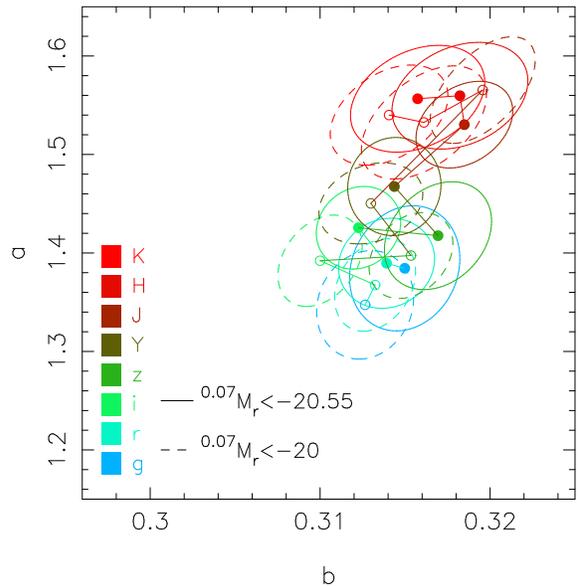}
\caption{Same  as Fig.~\ref{fig:FP_grizYJHK_lowcont}  but  showing the
  effect  of   using  the  entire  SPIDER  sample,   rather  than  the
  magnitude-selected sample of ETGs, on the waveband dependence of the
  FP.   Notice  that   scales  and  labelling  are  the   same  as  in
  Fig.~\ref{fig:FP_grizYJHK_lowcont}.~\label{fig:FP_grizYJHK_spiderall}}
\end{center}
\end{figure} 
In order to analyze if the  waveband dependence of the FP is sensitive
\begin{figure}
\begin{center}
\includegraphics[width=75mm]{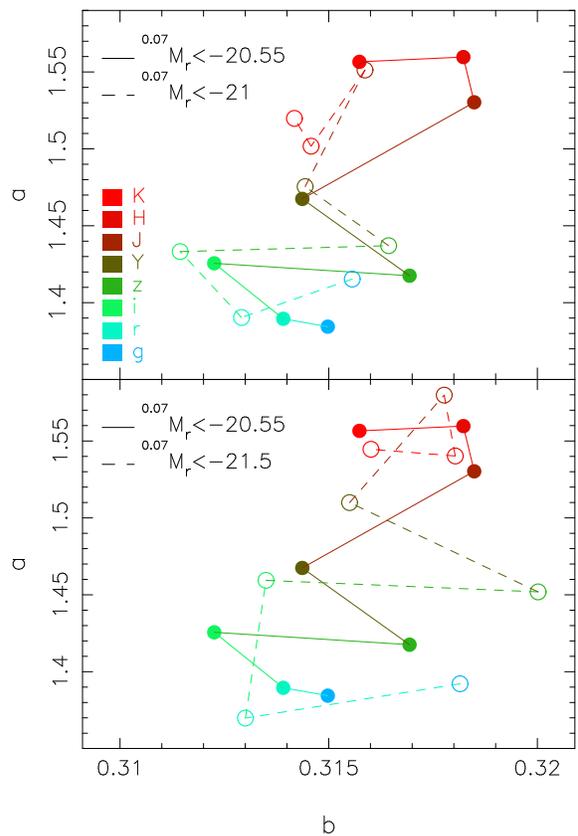}
\caption{Effect  of  changing  the  magnitude limit  on  the  waveband
  dependence  of the  FP.  The  upper and  lower panels  plot  as open
  circles,  connected by dashed  segments, the  FP slopes  $''a''$ and
  $''b''$  obtained  when   applying  different  magnitude  limits  of
  \mr=$-21$ and  $-21.5$, respectively.  In both panels,  the results,
  obtained for  the entire magnitude-selected sample,  are also shown,
  for  comparison, as  filled  circles, connected  by solid  segments.
  Different    colours     mark    different    wavebands,     as    in
  Fig.~\ref{fig:FP_grizYJHK}.~\label{fig:FP_grizYJHK_mag}}
\end{center}
\end{figure} 
to the magnitude range where ETGs are selected, we proceed as follows.
First, we  select all the ETGs  in the SPIDER  sample, with photometry
available in  $grizYJHK$ and reduced  $\chi^2$ smaller than  $3$. This
selection is  the same  as for the  magnitude-selected sample  of ETGs
(Sec.~\ref{sec:data}),  but  without  applying  any magnitude  cut  in
r-band.      The    sample     consists    of     $4,981$    galaxies.
Fig.~\ref{fig:FP_grizYJHK_spiderall}  compares the  FP  slopes of  the
magnitude-selected sample  of ETGs with those obtained  for the entire
sample.  

The  slopes of the FP  are fully consistent  in all wavebands
between the  two cases.  We  also define two subsamples  consisting of
all  the  ETGs with  available  photometry  in  $grizYJHK$ and  r-band
magnitude brighter  that \mr$=-21$ and  \mr$=-21.5$, respectively.  We
exclude galaxies whose  Sersic model fit gives an  high $\chi^2$ value
($>3$).  These \mr$=-21$  and \mr$=-21.5$ subsamples include $N=3,411$
and $N=2,091$  galaxies, respectively.  Fig.~\ref{fig:FP_grizYJHK_mag}
compares the slopes  of the FP obtained for these  two samples, by the
orthogonal  fitting procedure,  with those  obtained for  ETGs  in the
magnitude  range of  \mr$\le  -20.55$.   In order  to  allow a  direct
comparison  of  the  amounts  of  variation with  waveband,  for  both
subsamples, the best-fitted values of $''a''$ and $''b''$ are rescaled
to  match the  values of  $''a''$ and  $''b''$ in  the r-band  for the
magnitude-selected sample. The figure  clearly shows that the waveband
dependence  of  the FP  is  essentially  the  same regardless  of  the
magnitude  range.   For \mr$\le  -21$,  the  variation  of $''a''$  is
smaller,  but consistent  within the  errors, with  that  obtained for
\mr$\le-20.55$. For  \mr$\le -21.5$, the trend of  $''a''$ vs. $''b''$
matches very well that obtained  for the entire sample.  In all cases,
the \ls \, coefficient increases from $g$ through $K$, while the value
of $''b''$ is independent of waveband.

\subsection{Dependence on galaxy parameters}
\label{sec:fp_gal_pars}
The  FP relation  and its  dependence  on waveband  might change  when
selecting samples of ETGs  with different properties.  To analyze this
aspect, we split the  magnitude-selected sample according to the value
of different galaxy parameters, i.e. the axis ratio, $b/a$, the Sersic
index,  $n$, the  $r-K$ colour  index, and  the  average discy/boxiness
parameter, $a_4$. We utilize the values  of $b/a$, and $n$, in the $r$
band,  while for  $a_4$, we  adopt its  median value  among  the $gri$
wavebands  (see~\papdata).  The  $r-K$ colour  is computed  from 2DPHOT
total magnitudes.

\begin{figure*}
\begin{center}
\includegraphics[width=120mm]{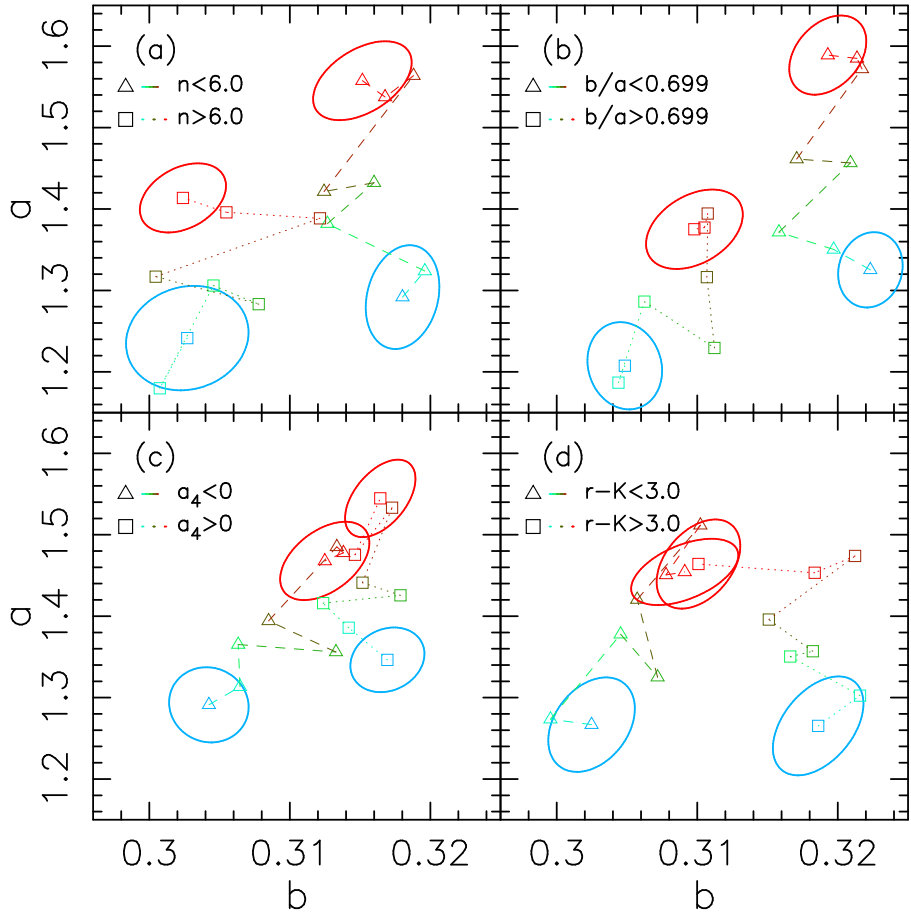}
\caption{FP slopes, from $g$  through $K$, for different subsamples of
  ETGs.   Each panel  shows the  FP slopes  obtained by  splitting the
  magnitude-selected sample  of ETGs  in two subsamples,  according to
  the Sersic index (panel a),  the axis ratio (panel b), the isophotal
  parameter,  $a_4$  (panel c),  and  the  $r-K$  colour index  (panel
  d). For  each quantity,  the two  bins are defined  as shown  in the
  lower-right corner  of the corresponding panel.   The slope's values
  are  plotted  with  different  symbols  and  are  connected  through
  different  line types  for the  two  bins, as  shown in  lower-right
  corner  of  each  plot.  Different wavebands  are  represented  with
  different  colours,   as  in  Fig.~\ref{fig:FP_grizYJHK}.   Ellipses
  denote 1$\sigma$  confidence contours  for $''a''$ and  $''b''$.  To
  make the  plot more clear, only  the ellipses in  $g$- and $K-$bands
  are shown.  Notice  that scales and labelling of  each panel are the
  same                              as                              in
  Fig.~\ref{fig:FP_grizYJHK_lowcont}.~\label{fig:FP_grizYJHK_pars_om}}
\end{center}
\end{figure*}

Fig.~\ref{fig:FP_grizYJHK_pars_om} plots  the slope $''a''$  of the FP
as a function of $''b''$.   The slope's values are those obtained from
the orthogonal  fit, applying the  same correction factors as  for the
entire sample (Sec.~\ref{sec:fp_slopes}).  Each panel corresponds to a
given  parameter  $p$:  $a_4$,   $b/a$,  $n$,  and  $r-K$.   For  each
parameter, the  magnitude-selected sample of  ETGs is splitted  in two
subsamples, having values  of $p$ either lower or  higher than a given
cut value, $p_c$.  For $p=b/a$, $n$,  and $r-K$, we set $p_c$ equal to
the median value of the distribution of $p$ values.  The median values
are   $p_c=0.699,  6.0$  and   $3.0$  for   $b/a$,  $n$,   and  $r-K$,
respectively. For $a_4$, we divide the sample into discy ($a_4>0$) and
boxy ($a_4<0$) galaxies.  Notice that, for a given parameter, galaxies
in  the   two  subsamples  can  populate  different   regions  of  the
\lre-\mie-\ls \, space. For  instance, because of the luminosity--size
relation and  the KR, galaxies  with higher Sersic index  are brighter
and  tend  to  have  higher  values of  \mie.   This  {\it  geometric}
difference might  produce spurious  differences in FP  coefficients. A
trivial example  of this geometric effect is  the magnitude selection:
the  bias on  FP coefficients  changes  for samples  of ETGs  spanning
different luminosity ranges (Sec.~\ref{sec:sel_effects}).  In order to
minimize  any geometric  difference,  for a  given  parameter $p$,  we
extract the two subsamples of ETGs by constraining their distributions
in magnitude and \mie \, to be the same (see App.~\ref{app:match_dist}
for details).
 
Fig.~\ref{fig:FP_grizYJHK_pars_om} shows  that the waveband dependence
of the  FP is similar  for all subsamples,  i.e. the value  of $''b''$
tends to be constant while  the coefficient $''a''$ increases by $\sim
15\%$  from   $g$  through  $K$.    However,  the  FP   slopes  change
significantly  for samples  of  ETGs with  different properties.   The
differences can be summarized as follows.
\begin{description}
\item[-] Galaxies with  higher $n$ have a lower  value of $''b''$; the
  value of $''a''$ in the NIR is smaller for the subsample with $n>6$,
  while in the optical both subsamples have consistent $''a''$.
\item[-] The FP  of round galaxies (higher $b/a$)  is more tilted (smaller $''a''$) than
  that of galaxies  with low $b/a$. The difference  is more pronounced
  in the NIR than in the optical.
\item[-] For $a_4$ and $r-K$, one can notice a different behaviour. In
  the NIR, the  FP slopes of the two  subsamples are fully consistent,
  while  in the  optical,  there  is a  detectable  difference in  the
  coefficient  $''b''$.  Boxy and blue (i.e.  $r-K<3$) galaxies  tend to
  have lower $''b''$.
\end{description}
{ We remark that all these trends remain essentially unchanged when
  replacing  SDSS-DR6 with  STARLIGHT velocity  dispersions,  with the
  exception that  $''a''$ is slightly lower for  STARLIGHT relative to
  SDSS $\sigma_0$ values (see Sec.~\ref{sec:fp_slopes_sig_mag}).}

\begin{figure*}
\begin{center}
\includegraphics[width=100mm]{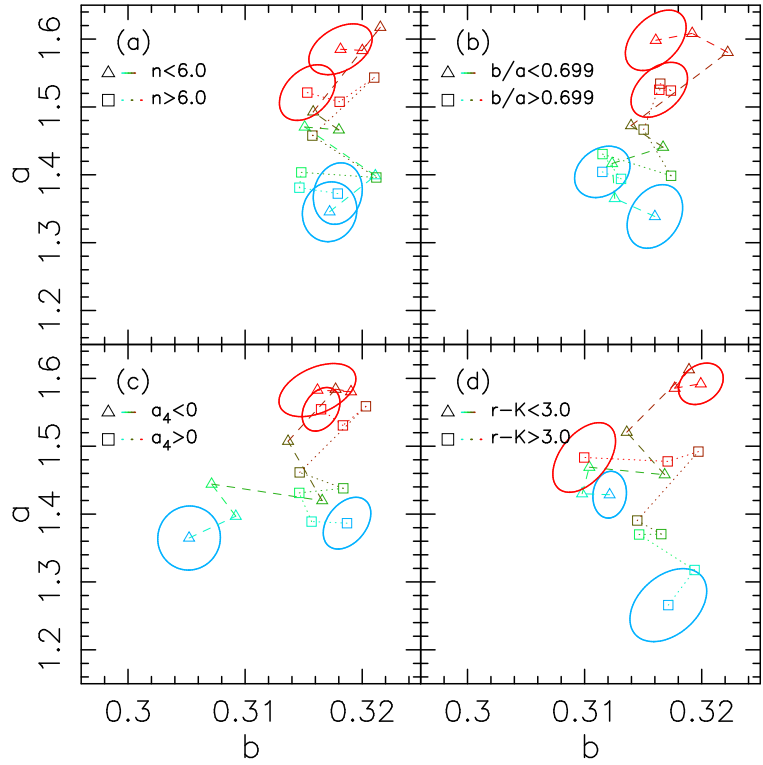}
\caption{The same as in Fig.~\ref{fig:FP_grizYJHK_pars_om}, but without
matching the distributions in magnitude and mean surface brightness of
subsamples in the two bins of a given quantity.
~\label{fig:FP_grizYJHK_pars}}
\end{center}
\end{figure*}

Fig.~\ref{fig:FP_grizYJHK_pars}  shows  the  FP  slopes  obtained  for
different  subsamples  as  in Fig.~\ref{fig:FP_grizYJHK_pars_om},  but
without imposing  the constraint that,  for a given quantity,  the two
subsamples  consist  of  galaxies   with  the  same  distributions  in
magnitude  and \mie.   No  difference would  have  been detected  with
respect to  $n$ and  $b/a$, while a  (spurious) difference in  the NIR
value of $''a''$ between red  and blue galaxies would have been found.
The     comparison      of     Fig.~\ref{fig:FP_grizYJHK_pars}     and
Fig.~\ref{fig:FP_grizYJHK_pars_om}  proves that accounting  for purely
{\it  geometric} differences  in  the  space of  FP  parameters is  of
paramount  importance to  correctly analyze  the scaling  relations of
different galaxy samples.

\section{The edge- and face-on projections of the FP}
\label{sec:fp_proj}
{ So far,  we have analysed the waveband dependence  of the FP, and
  that of the  FJ and KR. Since  the FJ and KR are  projections of the
  FP, we expect  their waveband dependence to be  connected to that of
  the  distribution  of  galaxies   in  the  FP.   We  establish  this
  connection by analysing the edge- and face-on projections of the FP.

\begin{figure*}
\begin{center}
\includegraphics[width=150mm]{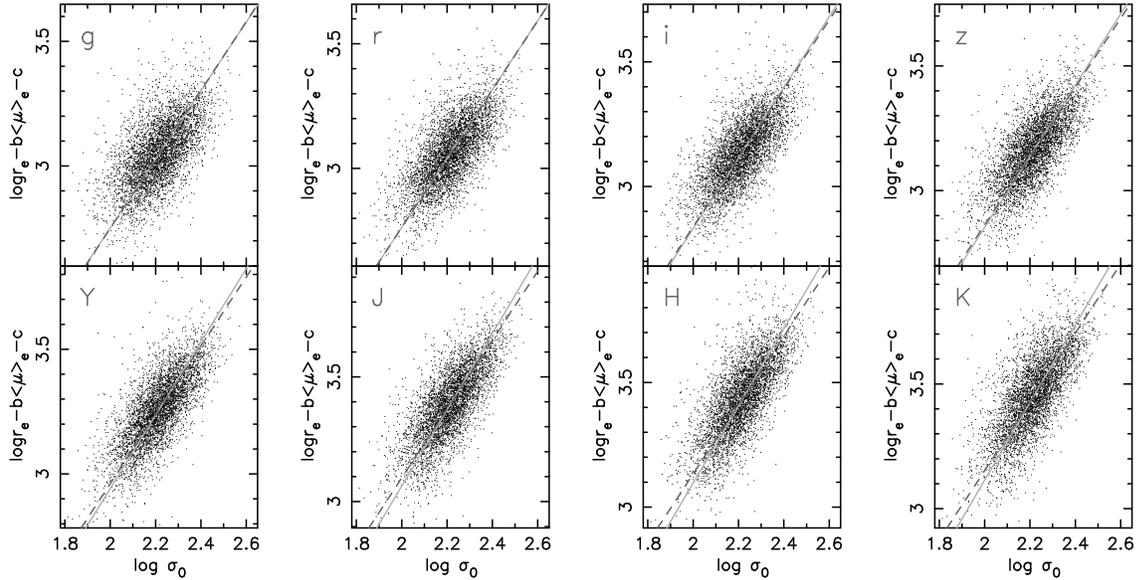}
\caption{Short     edge-on    projection     of     the    FP     (see
  Sec.~\ref{sec:fp_proj}), where the photometric quantity entering the
  FP,  $log r_e  -  b  < \!   \mu  \!  >_e$,  is  plotted against  the
  spectroscopic quantity \ls. Different panels correspond to different
  passbands,  from  $g$  (upper-left)  through $K$  (lower-right),  as
  indicated  in the  upper-left corner  of  each plot.   In the  short
  edge-on  projection, the  FP projects  into a  line, having  a slope
  equal  to  the  FP   coefficient  $''a''$.   For  each  panel,  this
  projection  is shown  by the  solid  light-grey line.   In order  to
  emphasize the waveband dependence of  $''a''$, in each panel we plot
  as  a  reference,  with  a  dashed dark-grey  line,  the  r-band  FP
  projection.  Notice  that the value of $''b''$,  defining the y-axis
  variable  changes among  different panels,  according to  the values
  reported in Tab.~\ref{tab:FP_orth_coef}. { Notice that for a more
    direct comparison of the FP projection in different wavebands, the
    lengths   of  the   x-   and   y-axes  are   the   same  for   all
    panels. }~\label{fig:FP_edgeon_grizYJHK}}
\end{center}
\end{figure*}

Fig.~\ref{fig:FP_edgeon_grizYJHK}  presents the so-called  {\it short}
edge-on  projection  of the  FP,  from  $g$  through $K$,  namely  the
combination of effective structural parameters, $\log r_e - b < \! \mu
\! >_e$, as  a function of \ls.  This corresponds to  the FP along the
   {\it  shortest}  axis,   whose  slope  is  equal  to   its  \ls  \,
   coefficient,         $''a''$.          Each        panel         in
   Fig.~\ref{fig:FP_edgeon_grizYJHK}, for a  given passband, shows the
   FP  obtained from the  orthogonal fitting  method (solid  line), as
   well as the r-band fitted FP (dashed line). From these plots we can
   see the  increasing of  $''a''$ from the  optical through  the NIR.
   Comparing the solid and dashed  lines we see that the increasing is
   quite small  (see Sec.~\ref{sec:fp_slopes}).  The  observed scatter
   in the  edge-on projection decreases  from the optical  through the
   NIR,  as it  can be  attested from  the values  of the  FP  \lre \,
   dispersion,            $s_{r_e}$,            reported            in
   Tab.~\ref{tab:FP_ls_coef}~\footnote{  We actually  refer  to the
     scatter  values reported  in  Tab.~\ref{tab:FP_ls_coef} (and  not
     those reported in Tab.~\ref{tab:FP_orth_coef})  as for the \ls \,
     fitting procedure the FP  slopes do not change significantly with
     waveband,  allowing  a  meaningful  comparison of  the  $s_{r_e}$
     values from $g$ through $K$.}.

\begin{figure*}
\begin{center}
\includegraphics[width=140mm]{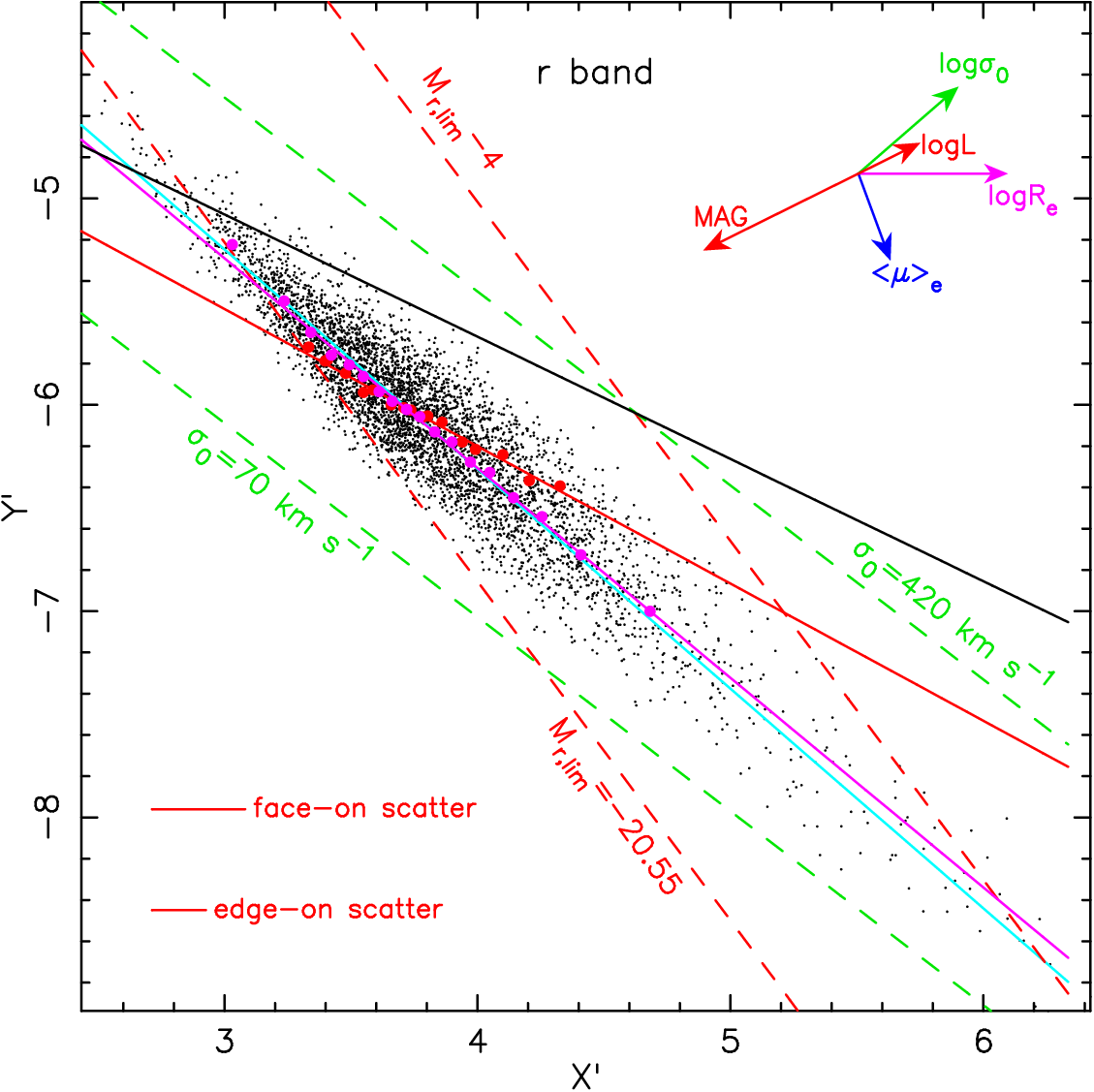}
\caption{Face-on projection  of the  FP in r  band. The  projection is
  such that the x-axis variable,  $X'$, is proportional to \lre, while
  the  y-axis  variable,  $Y'$,  is  proportional to  $-$\mie  \,  and
  \ls. The  arrows in  the upper-right corner  of the plot  denote the
  directions  where the quantities,  $MAG$, \ls,  $\log L$,  \mie, and
  \lre \, increase,  where $MAG$ is total galaxy  magnitude, and $\log
  L$  is logarithmic luminosity  in r  band.  The  size of  the arrows
  amounts  to  $0.5$~dex, $1  \,  mag  \,  arcsec^{-2}$, $3$~mag,  and
  $0.5$~dex  for \ls, \mie,  $MAG$, and  \lre, respectively.   The red
  dashed   lines   correspond   to   the   r-band   magnitude   limit,
  $M_{r,lim}=-20.55$, and a  bright-end limit four magnitudes brighter
  than $M_{r,lim}$,  as shown by the corresponding  labels.  The green
  dashed  lines correspond  to the  \ls \,  lower and  upper selection
  limits of  $70$ and $420 \, km  \, s^{-1}$.  The solid  cyan line is
  the MLSO best-fitting relation to  the data (see the text). { The
    thick solid  black line marks  the exclusion zone, $Y'=-0.59  X' +
    const$, as  originally defined by~\citealt{BBF:92}.  The intercept
    of this  line has  been arbitrarily normalized  to mark  the upper
    envelope of  the face-on projection of  the FP. The  presence of a
    few points above  the line is likely due  to measurement errors on
    FP  variables.}   The magenta  and  red  circles  are obtained  by
  binning the data with respect to \lre \, and $\log L$, respectively,
  computing  the median  values of  $X'$ and  $Y'$ in  each  bin.  The
  magenta and red solid lines are the best-fitting lines to the binned
  data-points.  The size  of the 2~$\sigma$ scatter around  the fit of
  the face-on projection  (cyan line) is given by  the long segment in
  the lower-left corner of the plot.  Notice how this segment is about
  twice larger than the corresponding \lre \, scatter of the FP, given
  by the short segment at lower-left.  ~\label{fig:FP_faceon_r} }
\end{center}
\end{figure*}

In order  to represent the FP  face-on projection, we  follow the same
formalism as  in~\citet{GLB:93} (hereafter GLB93).  We  project the FP
into a  plane defined  by two orthogonal  directions, one of  which is
perpendicular to the \lre \, axis. The axes of the projection are:
\begin{eqnarray}
X' & = & \left( x_0 \log r_e + b' \log < \! I \! >_e + a \log \sigma_0 \right)/\sqrt{x_0 \times (1 + x_0)}  \label{eq:xp} \\
Y' & = & \left( a \log  < \! I \! >_e - b' \log \sigma_0 \right)/\sqrt{x_0}, \label{eq:yp}
\end{eqnarray}
where $b'=-b \times 2.5$, $x_0 = a^2 + (b')^2$, and $< \! I \! >_e$ is
the mean  surface brightness in flux units,  with $< \! \mu  \!  >_e =
-2.5 \log <  \! I \! >_e$.  From the  FP equation and Eq.~\ref{eq:xp},
it   follows    that   $X'$   is   simply    proportional   to   \lre.
Fig.~\ref{fig:FP_faceon_r}  shows  the  distribution  of ETGs  on  the
face-on projection of the FP  in r-band, together with the \lre, \mie,
and \ls,  directions, as  well as the  directions of  increasing total
magnitude, $MAG$, and logarithmic luminosity, $\log L$, on the face-on
FP.   The dashed  lines  in  the plot  illustrate  the $\sigma_0$  and
magnitude  selection limits of  the sample  (Sec.~\ref{sec:data}).  As
already  noticed  in  previous  studies  {  (e.g.~\citealt{BBF:92};
  GLB93; JFK96)},  ETGs are  confined in a  {\it small} region  of the
face-on projection,  only partly  due to selection  effects.  Galaxies
populate a diamond-shaped region, limited at low $X'$ by the magnitude
limit  of the  sample, $M_{r,lim}=-20.55$,  and at  high $X'$,  by the
bright-end knee of the galaxy luminosity function, i.e.  the fact that
there are  no galaxies  brighter than a  magnitude threshold  of about
$M_{r,lim}-4$.  Notice that the $\sigma_0$ selections (see paper I and
Sec.~\ref{sec:data}) do  not affect the  shape of the  distribution in
the face-on  projection, as  all galaxies lie  well within  the region
defined by these additional cuts (dashed green lines in the figure).

Since the  \lre \, and \mie  \, directions form  almost a $90^{\circ}$
angle on the FP (see the blue and magenta arrows in the upper-right of
Fig.~\ref{fig:FP_faceon_r}),  the  KR  is essentially  reflecting  the
face-on  distribution,  as already  noticed  by  GLB93.   In order  to
establish this  connection, we perform an  orthogonal least-squares fit
of the  diamond-shaped region, accounting for  the magnitude selection
in  the   $X'$--$Y'$  plane  by   the  MLSO  fitting   procedure  (see
Sec.~\ref{sec:kr}). The relation is:
\begin{equation}
Y' = const + A' \times X',
\label{eq:FP_faceon}
\end{equation}
where $A'$ is the slope, and $const$  is an offset. For the r band, we
obtain a best-fitting value of  $A'=-1.08 \pm 0.01$. Since the \lre \,
and \mie  \, directions are approximately orthogonal,  the fitted line
is  very similar  to what  we would  obtain by  binning the  data with
respect to \lre \, and take the median values of $X'$ and $Y'$ in each
of those  bins. The result of  this binning procedure is  shown by the
magenta circles in Fig.~\ref{fig:FP_faceon_r}. The magenta line is the
best fit  of the binned  data-points, with a slope  of -1.01$\pm$0.02,
very close to the MLSO  fit reported above.  The 2~$\sigma$ scatter of
the  MLSO fit,  along  the $X'$,  is  displayed by  a  segment in  the
lower-left  of Fig.~\ref{fig:FP_faceon_r},  with  the shorter  segment
corresponding  to the  2~$\sigma$ \lre  \, dispersion  of the  FP seen
edge-on (Tab.~\ref{tab:FP_orth_coef}). The  scatter around the edge-on
FP  is about  twice smaller  than that  of the  face-on FP  as already
noticed by GLB93 implying that the FP is more like a {\it band} rather
than  a  plane,  in  the  \lre,  \mie,  \ls  \,  space.   We  can  use
Eq.~\ref{eq:FP_faceon}  to  eliminate  \ls  \, from  the  FP  equation
(Eq.~\ref{eq:FP}).  This  leads to a  linear relation between  \mie \,
and  \lre, similar to  Eq.~\ref{eq:KR}, i.e.   the KR,  whose expected
slope is:
\begin{equation}
p'_2=\frac{b}{x_0} - a \times A' \times \frac{\sqrt{1+x_0}}{0.4 x_0}.
\label{eq:krp}
\end{equation}
Inserting    the    r-band    value    of   the    FP    slope    from
Tab.~\ref{tab:FP_orth_coef} and the best-fitting value of $A'$ in this
equation, we obtain  $p'_2=3.56 \pm 0.03$, in good  agreement with the
KR  r-band slope,  $p_2  \sim 3.55$,  obtained  by the  MLSB fit  (see
Tab.~\ref{tab:KR_cof}).

As far as the FJ relation, we notice that the $\log L$ and \ls \, axes
form  a small  angle  on the  FP,  and are  almost  orthogonal to  the
direction of  the long  diagonal of the  diamond-shaped region.   As a
consequence, the best-fitting line  of the face-on distribution cannot
be directly connected to the FJ, as we do for the KR.  In fact, the FJ
relation almost  coincides with the (short) edge-on  projection of the
FP (e.q.~GLB93).   In order  to overcome this  problem and  relate the
face-on distribution and the FJ relation, we bin the data with respect
to $\log  L$ and then  compute the median  values of $X'$ and  $Y'$ in
each  bin.   This   binning  procedure  allows  us  to   look  at  the
distribution of galaxies in different luminosity bins, in the same way
as for the FJ relation.  The $\log L$-binned points are plotted as red
circles  in   Fig.~\ref{fig:FP_faceon_r}.   The  corresponding  linear
best-fitting is  displayed as a red  line. The slope value  of the red
line amounts to $A'_L=-0.72 \pm 0.03$. Notice how the MLSO fit and the
red  line  differ  significantly.    Replacing  $A'$  with  $A'_L$  in
Eq.~\ref{eq:FP_faceon}, and combining  the resulting equation with the
FP, we obtain  a linear relation between $\log L$  and \ls, similar to
the FJ equation (Eq.~\ref{eq:FJ}), with an expected slope of
\begin{equation}
\lambda'_1 = \frac{a-b' \times A'_L \times \sqrt{1+x_0}}{a \times A'_L \times \sqrt{1+x_0} + b' + 2 x_0}.
\label{eq:fjp}
\end{equation}
Inserting the value of $A'_L$  and the FP coefficients in this equation we
obtain $\lambda'_1=0.14  \pm 0.01$,  very close to  the measured
slope of the  FJ relation in r band ($\lambda_1 =  0.19 \pm 0.02$, see
Tab.~\ref{tab:FJ_cof}). The difference between $\lambda_1$
and $\lambda'_1$ does not reflect any inconsistency in the data, but
just the fact that the  $\log L$ and  \ls \, directions form  a small
angle on  the FP, and hence  it is not straightforward  to connect the
distribution  of galaxies  on the  face-on projection  to that  on the
$\log L$--\ls \, plane. Eq.~\ref{eq:fjp}  is used here as an empirical
tool to  analyse the  dependence of the  FJ relation on  waveband (see
below).

\begin{figure*}
\begin{center}
\includegraphics[width=130mm]{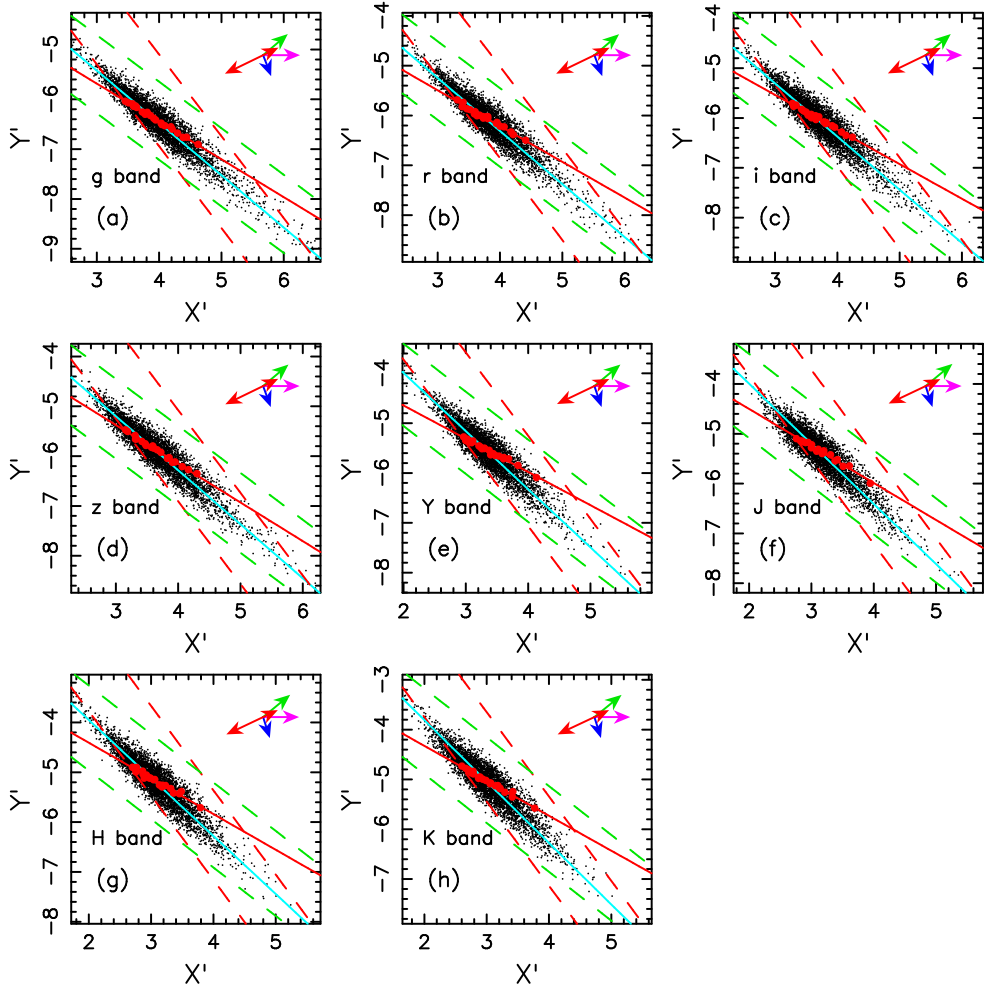}
\caption{The  same as  Fig.~\ref{fig:FP_faceon_r}  for the  $grizYJHK$
  wavebands,  from left  to  right,  and top  to  bottom.  For  better
  displaying     the     plots,     the     internal     labels     of
  Fig.~\ref{fig:FP_faceon_r} are  not shown. Panel (b) is  the same as
  Fig.~\ref{fig:FP_faceon_r}  and  is   repeated  to  allow  a  direct
  comparison with the other  panels (wavebands).  { Notice that for
    a more  direct comparison of  the face-on projection  in different
    wavebands, the  lengths of the x-  and y-axes are the  same in all
    panels. } ~\label{fig:FP_faceon_grizYJHK}}
\end{center}
\end{figure*}

Fig.~\ref{fig:FP_faceon_grizYJHK} shows the face-on projections of the
FP from $g$ through $K$. For  each band, we have performed an MLSO fit
of the data, as  well as a $\log L$-binned fit, in  the same way as we
do  in Fig.~\ref{fig:FP_faceon_r}.  For  each band,  the corresponding
slopes,  $A'$  and $A'_L$,  are  reported in  Tab.~\ref{tab:cof_proj},
together with the predicted slopes of the KR, $p'_2$, and FJ relation,
$\lambda'_1$,  from Eqs.~\ref{eq:krp}  and~\ref{eq:fjp}, respectively.
From  Tab.~\ref{tab:cof_proj} we  see  that  the slope  of  the KR  is
expected to increase with waveband,  in agreement with what we measure
(Sec.~\ref{sec:kr}).  This can be seen directly from Eq.~\ref{eq:krp},
as $''b''$ does  not change significantly with waveband,  and the same
holds for  the term $\sqrt{1+X_0}/X_0$.  It follows  that the waveband
dependence  of the  KR slope  is  driven by  the term  $-a \times  A'$
(second  term  of  Eq.~\ref{eq:krp}).   From  the values  of  $A'$  in
Tab.~\ref{tab:cof_proj},  we see that  $-A'$ increases  with waveband,
i.e.  the MLSO fitted line steepens  with waveband, in the same way as
$''a''$ does (Sec.~\ref{sec:fp_slopes}).   Therefore, the variation of
the KR from $g$ through $K$  is connected to the variation of both the
slope, $''a''$,  and the face-on  projection of the FP  with waveband.
The steepening of  the MLSO fit from $g$ through  $K$ can be explained
by  the variation of  optical to  NIR radii  along the  ETG's sequence
(Sec.~\ref{sec:kr}), and considering the fact that $X'$ is essentially
proportional to \lre.

For   what   concerns  the   FJ   relation,   the   slope  listed   in
Tab.~\ref{tab:cof_proj}  does  not change  with  wavelength, which  is
consistent   with   the   results  presented   in   Sec.~\ref{sec:fj}.
Eq.~\ref{eq:fjp}  explains the  reason for  this behaviour.   First, we
notice that the FP slope, $''a''$, appears both in the upper and lower
part of the second  term of Eq.~\ref{eq:fjp}.  Therefore, the waveband
dependence  of  $''a''$  does   not  affect  the  $\lambda'_1$  value.
Moreover,  from   Tab.~\ref{tab:cof_proj},  we  see   that  the  $\log
L$-binned slope of the FP face-on distribution is {\it independent} of
waveband, making  the value of $\lambda'_1$ constant  from $g$ through
$K$.  In other words, the  face-on distribution of the FP changes with
waveband  in  a  complex  way,  so  that  the  long  diagonal  of  the
diamond-shape region steepens with waveband, while the $\log L$-binned
envelope  of the  distribution  does not  change  with waveband.   The
former  effect,  together  with  the  waveband  variation  of  the  FP
coefficient, $''a''$,  determines a dependence of the  KR on waveband,
while the latter is consistent  with the FJ relation not changing from
$g$ through $K$.

\begin{table}
\centering
\small
\begin{minipage}{90mm}
\caption{Slopes of the face-on projection  of the FP, $A'$ and $A'_L$,
  and  {\it predicted} slopes of the FJ and KR, $\lambda'_1$ and $p'_2$.}
\begin{tabular}{c|c|c|c|c}
\hline
 band & $A' (fit)$ & $A'_L$ & $p_2'$ & $\lambda_1'$ \\
\hline
$g$ &  $ -1.05 \pm    0.01$ & $ -0.77 \pm    0.03$ &  $ 3.48 \pm    0.02$ & $  0.11 \pm    0.01$ \\ 
$r$ &  $ -1.08 \pm    0.01$ & $ -0.72 \pm    0.03$ &  $ 3.56 \pm    0.03$ & $  0.14 \pm    0.01$ \\ 
$i$ &  $ -1.07 \pm    0.01$ & $ -0.69 \pm    0.03$ &  $ 3.51 \pm    0.03$ & $  0.15 \pm    0.01$ \\ 
$z$ &  $ -1.09 \pm    0.01$ & $ -0.78 \pm    0.04$ &  $ 3.54 \pm    0.03$ & $  0.10 \pm    0.02$ \\ 
$Y$ &  $ -1.14 \pm    0.02$ & $ -0.67 \pm    0.04$ &  $ 3.64 \pm    0.05$ & $  0.16 \pm    0.02$ \\ 
$J$ &  $ -1.18 \pm    0.02$ & $ -0.68 \pm    0.03$ &  $ 3.66 \pm    0.06$ & $  0.14 \pm    0.02$ \\ 
$H$ &  $ -1.17 \pm    0.02$ & $ -0.73 \pm    0.04$ &  $ 3.67 \pm    0.06$ & $  0.13 \pm    0.01$ \\ 
$K$ &  $ -1.24 \pm    0.01$ & $ -0.70 \pm    0.04$ &  $ 3.83 \pm    0.04$ & $  0.14 \pm    0.01$ \\ 
\hline
  \end{tabular}
\label{tab:cof_proj}
\end{minipage}
\end{table}

}

\section{Stellar populations along the FP}
\label{sec:SP_slopes}
\begin{figure}
\begin{center}
\includegraphics[width=80mm]{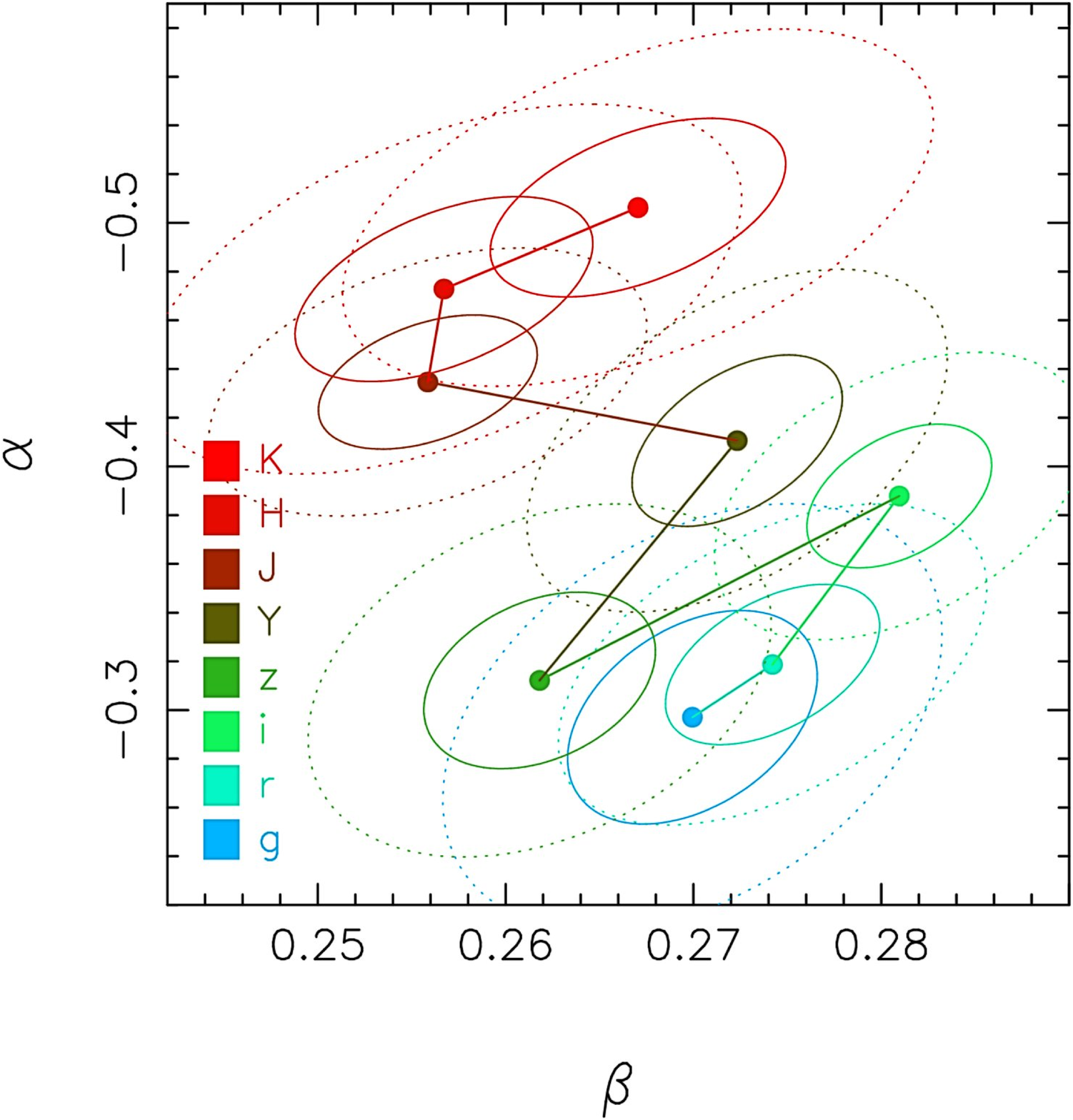}
\caption{  Slopes,   $\alpha$  and   $\beta$,  of  the   power-law  of
  $\frac{M}{L}$ as a function of  $\sigma_0$ and $M$ in the $grizYJHK$
  wavebands.  Solid and dotted  ellipses denote $1\sigma$ and $2\sigma$
  confidence contours,  respectively, as implied  by the uncertainties
  on                FP                coefficients                (see
  Tab.~\ref{tab:FP_orth_coef}).~\label{fig:alfa_beta_grizYJHK}}
\end{center}
\end{figure}

Under  the  homology  assumption,  one  can combine  the  FP  relation
(Eq.~\ref{eq:FP}) with the virial theorem
\begin{equation}
\sigma_0^2 \propto \frac{M}{L} < \! I \! >_e r_e,
\label{eq:VT}
\end{equation}
and parametrize the mass-to-light  ratio, $\frac{M}{L}$, as a function
of two  variables out of $M$, $L$,  $\sigma_0$, $r_e$, and $<  \! I \!
>_e$ (Djorgovski, de Carvalho, \&  Han~1988). Here, we denote as $< \!
I \! >_e$ the mean surface  brightness within $r_e$ in flux units.  In
order  to analyze  how stellar  population parameters  vary  along the
sequence of  ETGs, it  is convenient to  parametrize such  sequence by
means  of  variables  that   are  independent  of  stellar  population
parameters.   To  this effect,  we  consider  the  quantities $M$  and
$\sigma_0$, and write
\begin{equation}
 \frac{M}{L} \propto M^{\beta_x} \sigma_0^{\alpha_x},
\label{eq:ML}
\end{equation}
where   the  index  $x$   runs  over   all  the   available  wavebands
($x=grizYJHK$).    Using  Eq.~\ref{eq:FP}  and   Eq.~\ref{eq:VT},  one
obtains the following expressions for $\alpha_x$ and $\beta_x$:
\begin{eqnarray}
 \alpha_x & = & 4 - 0.4 \left( \frac{a_x+2}{b_x}
\right) \label{eq:alfabeta1}\\
 \beta_x & = & \frac{0.4}{b_x}-1, \label{eq:alfabeta2}
\end{eqnarray}
where $a_x$ and $b_x$ are the values  of the \ls \, and \mie \, slopes
of the FP  in the waveband $x$.  These equations  imply that, at fixed
$\sigma_0$  the  variation  of  the  $M/L$  with  mass  is  completely
determined by the coefficient $''b''$  of the FP.  On the contrary, at
fixed  $M$,  the  variation  of  $M/L$  with  velocity  dispersion  is
determined  by both  the values  of  $''a''$ and  $''b''$. Hence,  the
result  that  $''b''$  does  not  change from  $g$  through  $K$  (see
Sec.~\ref{sec:fp}) implies  that, at  fixed $\sigma_0$, the  change of
$M/L$  with mass  is independent  of  waveband. On  the contrary,  the
dependence of  $M/L$ with $\sigma_0$  (at fixed $M$) changes  from $g$
through $K$. This  is shown in Fig.~\ref{fig:alfa_beta_grizYJHK} where
we  plot  the  values  of  $\alpha$  and  $\beta$  in  the  $grizYJHK$
wavebands.   For each  band, we  calculate $\alpha$  and  $\beta$ from
Eqs.~\ref{eq:alfabeta1}    and~\ref{eq:alfabeta2},   using    the   FP
coefficients  obtained  by   the  orthogonal  fitting  procedure  (see
Tab.~\ref{tab:FP_orth_coef}). The  reason for adopting  the FP slope's
values from  the orthogonal regression  is discussed in  Sec.~9.1.  As
expected, the  value of $\beta$ is constant,  while $\alpha$ increases
from  $g$ through  $K$.  Although  the variation  of $''a''$  from $g$
through $K$ amounts to only $\sim 12\%$, the corresponding increase in
the $\alpha$ value is significant, amounting to $\sim 70\%$.

\begin{figure}
\begin{center}
\includegraphics[width=80mm]{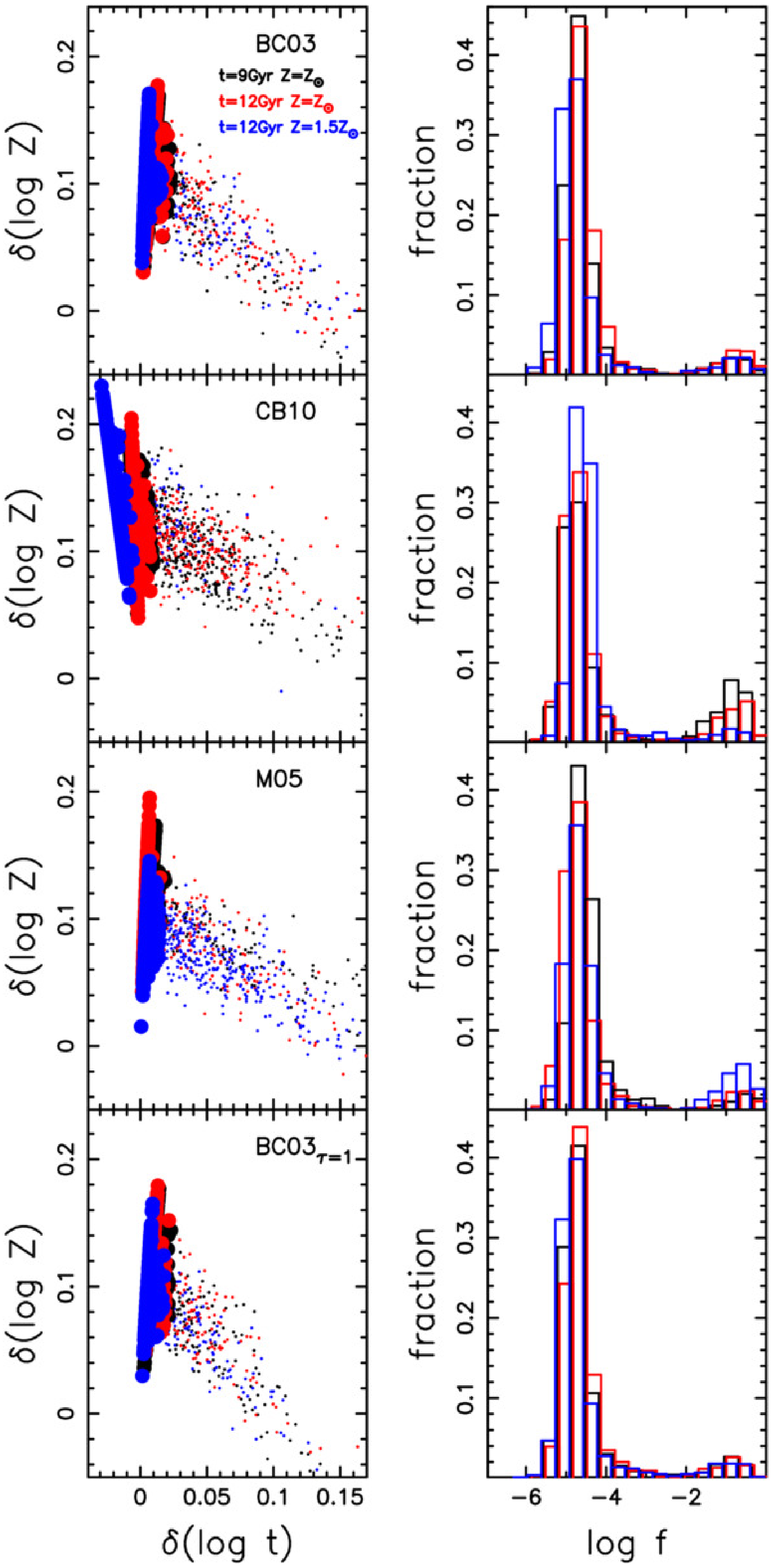}
\caption{Best-fitting values of  \dt, \dz \, and $f$  values. The left
  panels  plot  \dt  \,  vs.    \dz  for  the  BC03,  CB10,  M05,  and
  $BC03_{\tau=1}$ SSP  models (from top to  bottom).  Different colours
  denote the  different combinations  of $t$ and  $Z$ as shown  in the
  upper--right corner of  the top panel. For a  given panel and colour,
  the  different points correspond  to the  solutions obtained  in the
  $2000$  minimisation  iterations.   The solutions  corresponding  to
  $f<0.1$ are plotted  with larger symbols. The right  panels plot the
  corresponding  distributions of $f$  values. To  make the  plot more
  clear,  small  shifts  (of  $\pm$0.05)  have  been  applied  to  the
  histograms with different colours.~\label{fig:age_zeta}}
\end{center}
\end{figure} 

One  can also  notice that  the values  of $\alpha$  and  $\beta$ have
opposite sign.   Since the value of  $\alpha$ is negative,  at a given
mass,  the \ml  \,  is a  decreasing  function of  $\sigma_0$. On  the
contrary, for fixed $\sigma_0$, the \ml \, increases with $M$. { In
  order to characterise the overall variation of $M/L$ along the ETG's
  sequence, parametrized in  terms of galaxy mass, we  have to project
  Eq.~\ref{eq:ML}, i.e.  the FP itself, into the $M/L$--$M$ plane.  To
  this effect, we  can take advantage of a  specific projection of the
  FP,  such as  the FJ  relation, i.e.   the fact  that  luminosity is
  proportional to $\sigma_0$. In this approach, the FJ relation is not
  providing any extra information $wrt$  the FP itself, but is used as
  an empirical tool to project  Eq.~\ref{eq:ML} into an $M/L$ vs.  $M$
  power-law~\footnote{  A  different  approach  would  be  that  of
    measuring directly dynamical  mass from the data, by  means of the
    virial theorem.  This approach  (see e.g.~JFK96) relies on a given
    galaxy model to translate $\sigma_0$ and $r_e$ into $M$, and hence
    implies  several assumptions about,  for instance,  the dark-matter
    component of ETGs.   This analysis is currently under  way for the
    SPIDER   sample,  and   will   be  presented   in  a   forthcoming
    contribution.  For the present study, we adopt a model-independent
    approach, using only the information provided by the FP.}.}  Using
Eq.~\ref{eq:FJ} to replace $\sigma_0$ with $L$ in Eq.~\ref{eq:ML}, for
a given waveband $X$, we obtain:
\begin{equation}
 M/L \propto M^{\gamma_x},
\label{eq:ml_m}
\end{equation}
where
\begin{equation}
\gamma_x = \frac{\left( \beta_x + \alpha_x \cdot \lambda
\right) }{ \left(  1 + \alpha_x \cdot \lambda \right) },
\label{eq:gamma}
\end{equation}
and  $\lambda =  0.198  \pm 0.007$  is  the average  slope  of the  FJ
relation (Sec.~\ref{sec:fj}).   We point out that using  the values of
$\lambda$ we have measured for  each waveband (see Tab.~5) rather than
the  average $\lambda$  value  would  not change  at  all the  results
presented here.  {  The values of $\gamma$, derived  with the above
  procedure,  are reported in  Tab.~\ref{tab:gamma} (col.~2)  for each
  waveband}.  The $\gamma$ has a  positive value in all wavebands, and
tends to slightly  decrease, by $20 \pm 5 \%$, from  $g$ to $K$.  This
variation  can  be  interpreted  as  a change  of  stellar  population
properties along the sequence of  ETGs.  To this effect, following the
same  approach  of~LBM08,  we  assume that  also  the  stellar
mass-to-light ratio of ETGs, \mls, is a power--law of $M$:
\begin{equation}
 M_{\ast}/L \propto M^{\gamma^{\ast}_x}.
\label{eq:mls_m}
\end{equation}
Eq.~\ref{eq:ML} can then be written as 
\begin{equation}
 M/L \propto M^{\gamma'+\gamma^{\ast}_x},
\label{eq:ml_m_gamma}
\end{equation}
where  $\gamma'=\gamma_x-\gamma^{\ast}_x$  defines  how the  ratio  of
stellar  to  total mass  changes  along  the  mass sequence  of  ETGs,
$M_{\ast}/M  \propto  M^{-\gamma'}$  and  thus  it is  assumed  to  be
independent     of     waveband.      Introducing    the     parameter
$f=\gamma_K^{\ast}/\gamma_K$, which defines the fraction of the K-band
slope of the $M/L$ vs. $M$ relation due to stellar population effects,
we obtain the following system of equations:
\begin{equation}
 \left( \frac{1-f}{f} \right) \gamma_K^\star + \gamma_x^\star = 
\gamma_x. \label{eq:gamma_sp}
\end{equation}
We note that $f$ can vary  between $0$ and $1$.  For $f=0$, the K-band
tilt  is independent  of stellar  populations ($\gamma^{\ast}_{K}=0$),
while for $f=1$  the tilt is entirely explained  by stellar population
effects ($\gamma^{\ast}_K=\gamma_K$  and $\gamma'=0$).  The quantities
$\gamma^{\ast}_x$ depend  on how stellar  population properties change
along the  mass sequence of ETGs.  Considering only the  age, $t$, and
the metallicity, $Z$, one can write:
\begin{equation}
 \gamma^{\ast}_x = \frac{\delta (\log M_{\ast}/L)}{\delta(\log M)}= c_{t_x}
\cdot
\frac{\delta (\log t)}{\delta (\log M)} + c_{Z_x} \frac{\delta (\log
Z)}{\delta (\log M)}
\label{eq:gammas}
\end{equation}
where the quantities \dt \, and \dz \, are the logarithmic differences
of age  and metallicity  between more and  less massive  galaxies (per
decade    in   mass),    while    $c_{t_x}=\frac{\partial   \log    \!
  M_{\ast}/L_x}{\partial \log t}$  and $c_{Z_x}=\frac{\partial \log \!
  M_{\ast}/L_x}{\partial   \log  Z}$   are  the   partial  logarithmic
derivatives of $M_{\ast}/L$ (in the  waveband $x$) with respect to $t$
and $Z$. Deriving the  coefficients $\gamma_x$ from the slope's values
of       the      FP       in       the      different       wavebands
(Eqs.~\ref{eq:alfabeta1},~\ref{eq:alfabeta2}, and~\ref{eq:gamma}), and
inserting the expression of $\gamma^{\ast}_x$ from Eq.~\ref{eq:gammas}
into Eq.~\ref{eq:gamma_sp}, we obtain a system of eight equations, one
for each  of the $grizYJHK$  wavebands, in the three  unknowns $\delta
(\log  t)$, $\delta  (\log  Z)$, and  $f$.  We solved  this system  by
minimizing the sum of relative residuals:
\begin{equation}
 \chi^2 = \sum_x \left[ \frac{ (1-f)/f \cdot \gamma_K^\star +
\gamma_x^\star  - \gamma_x }{\gamma_x} \right]^2.
\label{eq:chi2}
\end{equation}
We  estimated the  quantities $c_{\rm  t,x}$ and  $c_{\rm  Z,x}$ using
simple  stellar  population   (SSP)  models  from  different  sources:
~\citet{BrC03}  (BC03),  \citet{M05} (M05),  and  Charlot and  Bruzual
(2010,  in preparation; CB10).   These models  are based  on different
synthesis techniques and have different  IMFs.  The M05 model uses the
fuel consumption approach instead  of the isochronal synthesis of BC03
and   CB10.    The   CB10    code   implements   a   new   AGB   phase
treatment~\citep{MG07}.  The  IMFs are: Scalo  (BC03), Chabrier (M05),
and  Salpeter  (CB10). Moreover,  we  also  used  a composite  stellar
population  model from  BC03  having exponential  star formation  rate
(SFR) with e-folding time of $\tau=1$~Gyr (hereafter $BC03_{\tau=1}$).
The models are  folded with the $grizYJHK$ throughput  curves, and the
\ml \, values  computed for different values of $t$  and $Z$. In order
to evaluate  the impact of changing  $t$ and $Z$,  we considered three
cases, with (i) an age of  $9$~Gyr and solar metalicity, (ii) an older
age of  $12$~Gyr and solar metallicity,  and (iii) an  age of $12$~Gyr
and  super--solar metallicity  ($Z=1.5 Z_{\odot}$).   The minimization
was  performed $2000$  times  for each  kind  of model,  and for  each
combination  of  $t$  and  $Z$  values,  shifting  each  time  the  FP
coefficients    according    to    the   corresponding    (correlated)
uncertainties.   For  each iteration,  we  found  that  all the  eight
equations were solved with an accuracy better than $10\%$.

Fig.~\ref{fig:age_zeta}  plots  \dt  \,  vs.   \dz,  as  well  as  the
distribution of  $f$ values obtained in  each case for  all the $2000$
iterations. For almost  all solutions, the $f$ is  very close to zero,
implying that  the tilt of  the NIR  FP is not  due to a  variation of
stellar population properties of ETGs  with mass. For instance, in the
case $t=9Gyr$ and solar metallicity, the percentages of solutions with
$f<0.05$ amounts to  $94\%$, $82\%$, $95\%$, and $94\%$  for the BC03,
CB10, M05, amd  BC03$_{\tau=1}$ models, respectively. Considering only
the solutions with $f<0.05$, we estimated the mean value of \dt \, and
\dz.   The  mean  values,  and the  corresponding  uncertainties,  are
reported in Tab.~\ref{tab:age_zeta}.  The uncertainties were estimated
by the standard deviation of the \dt \, and \dz \, values obtained for
a given  model and for a given  combination of $t$ and  $Z$.  The mean
values do not depend significantly  on either the model or the adopted
values of  $t$ and  $Z$.  On average,  \dt \,  is very close  to zero,
while the  \dz \, mean value  amounts to $\sim  0.1$~dex. This implies
that  ETGs  have  syncronous  luminosity-weighted ages,  with  an  age
variation smaller  than a  few percent per  decade in mass,  while the
metallicity variation  per mass decade  amounts to $\sim 23  \%$. {
  These results remain unchanged when using STARLIGHT rather than SDSS
  velocity dispersions. Inserting the values of FP coefficients and FJ
  slope as  estimated by STARLIGHT (rather than  SDSS) $\sigma_0$'s in
  Eq.~\ref{eq:gamma}, the values of the $\gamma$'s increase on average
  by  $\sim 12  \%$,  as shown  by  comparing the  values in  column~3
  (STARLIGHT)  and  column~2   (SDSS)  of  Tab.~\ref{tab:gamma}.   For
  instance,  the value of  $\gamma_g$ changes  from $0.224$  (SDSS) to
  $0.251$  (STARLIGHT),  while in  K  band  the  $\gamma$ varies  from
  $0.186$  (SDSS)  to $0.218$  (STARLIGHT).   Applying the  procedure
  described above to  estimate $f$, $\delta( \log t )  $ and $ \delta(
  \log Z)$, we  find that, for BC03 SSP models with  an age of $9$~Gyr
  and solar metallicity, the  corresponding distribution of $f$ values
  is  still strongly  peaked around  zero,  while the  mean values  of
  $\delta( \log  t ) $ and $  \delta( \log Z)$ amount  to $\sim 0.008$
  and $\sim 0.100$, respectively,  fully consistent with what obtained
  from SDSS $\sigma_0$'s  ($\delta( \log t ) \sim  0.013$ and $\delta(
  \log  Z) \sim  0.105$, see  Tab.~\ref{tab:age_zeta}).  As  a further
  test,  we estimated  the $\gamma$'s  by not  using the  FJ relation.
  Combining Eq.~\ref{eq:mls_m} with the  virial theorem, we obtain the
  equation:
\begin{equation}
\log r_{\rm e} \propto  2 \times \frac{1-\gamma_x}{1+\gamma_x} \log \sigma_{\rm 0} + \frac{0.4}{1+\gamma_x} < \! \mu \! >_{\rm e},
\label{eq:ml_m_fp}
\end{equation}
which  reduces to  the  FP for  $a_x=2 (1-\gamma_x)/(1+\gamma_x)$  and
$b_x=0.4/(1+\gamma_x)$.  For each waveband, we estimate the $\gamma_x$
by minimising the expression:
\begin{equation}
\chi^2 = \left( a_x-2 \frac{1-\gamma_x}{ 1 + \gamma_x } \right)^2/(\delta a_x)^2 + \left( b_x-\frac{0.4}{1+\gamma_x} \right)^2/(\delta b_x)^2,
\label{eq:chi2_gamma}
\end{equation}
where    $a_x$   and    $b_x$   are    the   FP    coefficients   from
Tab.~\ref{tab:FP_orth_coef}, and $\delta a_x$ and $\delta b_x$ are the
errors  on $a_x$ and  $b_x$.  The corresponding  values of
$\gamma_x$ are reported  in Tab.~\ref{tab:gamma} (col.~4). On average,
the  values  of $\gamma_x$  tend  to  increase, $wrt$ to  those  from
Eq.~\ref{eq:gamma}, by $\sim 13\%$.  Even in this case, this variation
does not  impact at all  the above conclusions,  i.e. the $f$  is zero,
while the mean  values of $\delta( \log  t ) $ and $  \delta( \log Z)$
amount to about $0.01$ and $0.1$~dex, respectively.}


\begin{table}
\centering
\small
\begin{minipage}{80mm}
\caption{{  Slopes of  the \ml  \, vs.  mass relation,  obtained by
    projecting  the FP  through  the FJ  relation  (col.~2), by  using
    STARLIGHT (rather than SDSS)  $\sigma_0$'s (col.~3), and fitting the
    $\gamma$ to the FP coefficients  rather than using the FJ relation
    (col.~4).}}
\begin{tabular}{c|c|c|c}
\hline
 waveband & \multicolumn{3}{c}{$\gamma$}\\
   & ${\rm SDSS} \, \sigma_0$'s & STARLIGHT $\sigma_0$'s & $\alpha$--$\beta$ fit\\
  (1) & (2) & (3) & (4) \\
\hline
g & $    0.224 \pm    0.008 $ & $0.251 \pm 0.009$ & $0.249 \pm 0.008$ \\
r & $    0.225 \pm    0.006 $ & $0.253 \pm 0.007$ & $0.248 \pm 0.007$ \\
i & $    0.221 \pm    0.006 $ & $0.247 \pm 0.006$ & $0.254 \pm 0.007$ \\
z & $    0.213 \pm    0.008 $ & $0.236 \pm 0.006$ & $0.233 \pm 0.007$ \\
Y & $    0.208 \pm    0.007 $ & $0.230 \pm 0.007$ & $0.227 \pm 0.007$ \\
J & $    0.186 \pm    0.007 $ & $0.202 \pm 0.008$ & $0.215 \pm 0.008$ \\
H & $    0.180 \pm    0.009 $ & $0.221 \pm 0.007$ & $0.208 \pm 0.007$ \\
K & $    0.186 \pm    0.009 $ & $0.218 \pm 0.007$ & $0.214 \pm 0.008$ \\
\hline
  \end{tabular}
\label{tab:gamma}
\end{minipage}
\end{table}

\begin{table}
\centering
\small
\begin{minipage}{80mm}
\caption{Age and metallicity differences per decade of galaxy mass.}
\begin{tabular}{c|c|c|c|c}
\hline
 $ MODEL $ & $t \, (Gyr)$ & $Z/Z_{\odot}$ & $ \delta( \log t )
$ & $ \delta( \log Z) $ \\
\hline
 $ BC03          $ & $12$ & $1$ & $  0.013 \pm  0.021 $ & $  0.104 \pm
0.026 $ \\
 $ BC03          $ & $9$  & $1$ & $  0.013 \pm  0.017 $ & $  0.105 \pm 
0.025 $ \\
 $ BC03          $ & $12$ & $1.5$ & $  0.004 \pm  0.001 $ & $  0.106 \pm 
0.019 $ \\
 $ CB10          $ & $12$ & $1$ & $  0.003 \pm  0.021 $ & $  0.121 \pm 
0.022 $ \\
 $ CB10          $ & $9$  & $1$ & $  0.005 \pm  0.025 $ & $  0.121 \pm 
0.026 $ \\
 $ CB10          $ & $12$ & $1.5$ & $ -0.019 \pm  0.004 $ & $  0.150 \pm 
0.027 $ \\
 $ M05           $ & $12$ & $1$ & $  0.008 \pm  0.018 $ & $  0.112 \pm 
0.025 $ \\
 $ M05           $ & $9$  & $1$ & $  0.012 \pm  0.023 $ & $  0.108 \pm 
0.022 $ \\
 $ M05           $ & $12$ & $1.5$ & $  0.005 \pm  0.001 $ & $  0.094 \pm 
0.017 $ \\
 $ BC03_{\tau=1} $ & $12$ & $1$ & $  0.011 \pm  0.012 $ & $  0.107 \pm 
0.023 $ \\
 $ BC03_{\tau=1} $ & $9$  & $1$ & $  0.012 \pm  0.014 $ & $  0.107 \pm 
0.025 $ \\
 $ BC03_{\tau=1} $ & $12$ & $1.5$ & $  0.006 \pm  0.001 $ & $  0.100 \pm 
0.018 $ \\
\hline
\end{tabular}
\label{tab:age_zeta}
\end{minipage}
\end{table}

\section{Discussion}
\label{sec:disc}

\subsection{The fit of the FP in different wavebands}
One  of  the crucial  aspects  of the  present  study  is the  fitting
procedure used to  obtain the coefficients of the FP  and how they are
affected by different systematic effects. Different fitting techniques
produce  different  estimates of  FP  coefficients,  and  may lead  to
erroneous  results  when  comparing  the FP  relations  obtained  with
different samples (LBC00;  \citealt{Saglia01, BER03b}).  To avoid this
problem, we adopt  the same fitting method for  all the ETG subsamples
we  analyze.  Selection effects  and  correlated  errors on  effective
parameters can be taken into account analytically under the assumption
that  the  FP  variables  are  normally  distributed~\citep{Saglia01}.
Although~\citet{BER03b}  showed that the  joint distribution  of \lre,
\ls,  and   galaxy  magnitude  is  relatively  well   described  by  a
multivariate  Gaussian,  this  might  not  necessarely  be  true  when
effective parameters are derived by the Sersic (2DPHOT) rather than de
Vaucouleurs  (Photo)  model.  These  two  pipelines yield  significant
differences   in  \lre  \,   and  magnitudes   (see~\papdata).   These
differences depend on  galaxy magnitude, and may be  partly due to the
sky  overestimation problem  affecting the  SDSS Photo  parameters. We
have adopted a non- parametric approach, first fitting the FP relation
and then correcting the  slopes for different systematic effects using
extensive  Monte-Carlo simulations. We  find that  the main  source of
bias  on   the  FP  slopes   is  the  magnitude  cut.    In  agreement
with~\citet{HB09}, we show that for  the orthogonal fit this cut leads
to  underestimating  the FP  coefficients,  with  the effect  becoming
negligible  only at  faint  magnitude limits  ($M_r  \sim -18.5$,  see
Fig.~\ref{fig:bias}).  The effect is negligible when we use the \ls \,
fitting  method.  As  shown by~LBC00,  minimizing the  $\log \sigma_0$
residuals leads to a \ls \, slope of the FP systematically higher than
that obtained by  other fitting techniques (see also  JFK96).  We also
find  that the coefficient  $''a''$ of  the FP  in the  optical (SDSS)
wavebands  are systematically  larger when  we  use the  \ls \  method
compared to  results obtained  with the orthogonal  fitting procedure.
In r-band the difference amounts to $\sim 6\%$. On the other hand, the
coefficient  $''b''$ turns out  to be  systematically lower,  by $\sim
5\%$,  for  the~\ls  \,  method (see  Tabs.~\ref{tab:FP_ls_coef}  and~
\ref{tab:FP_orth_coef}).  Another important result we find is that the
difference produced  by different  fitting method depends  on waveband
(see also~LBM08). The FP coefficients  do not change with the waveband
when using  the \ls  \, method,  while they smoothly  vary, by  $ \sim
12\%$, from $g$ through $K$ when using the orthogonal method. This can
be explained by the fact that  the \ls \, regression minimizes the rms
of  residuals in  the  perpendicular direction  to  the \lre--\mie  \,
plane,  and  hence  it  is   less  sensitive  to  differences  in  the
distribution  of galaxies in  that plane,  like those  among effective
parameters measured  in different  wavebands. The problem  of deriving
the  best  fitting coefficients  of  correlations among  astrophysical
quantities has been  addressed by~\citet{ISO90}.  They concluded that,
in case one aims to study the underlying functional relation among the
variables,   regression   procedures   treating  all   the   variables
symmetrically,  like the  orthogonal method,  should be  adopted.  For
this  reason,  we  have  analyzed  the implications  of  the  waveband
dependence  of   the  FP  adopting  the  results   of  the  orthogonal
regression.

\subsection{Variation of $r_{OPT} / r_{NIR}$ with galaxy radius}
{ In the present study, we find that the slope of the KR exhibits a
  small systematic variation with  waveband, steepening by $\sim 10\%$
  from $g$ through $K$.} This  variation may be explained as the ratio
of optical to NIR effective  radii decreasing for galaxies with larger
\re, namely, while  smaller size ETGs have, on  average, optical radii
larger than the  NIR ones, the most massive  galaxies have $r_{_{OPT}}
\sim r_{_{NIR}}$.  In the assumption that $r_{_{OPT}} / r_{_{NIR}}$
is a  proxy for the  internal colour gradient  of an ETG,  this finding
implies that the  stellar populations of the most  massive ETGs have a
more  homogeneous  spatial  distribution  inside  the  galaxies,  i.e.
flatter  radial gradients, than  less massive  systems.  \citet{SPF09}
found that the relation  between the internal metallicity gradient and
mass in early-type systems is bimodal, with a sharp transition at $M_B
\sim -19$.  This magnitude  corresponds approximately to the lower cut
applied to the SPIDER sample (paper I).  For $M_B > -19$, ETGs exhibit
a tight correlation between  the metallicity gradient and either mass,
luminosity, or \ls.  Brighter galaxies tend to have steeper gradients,
as  expected by  the lower  efficiency of  feedback processes  in less
bound  (massive)  systems~\citep{Larson:74}.   At higher  mass,  colour
gradients exhibit a larger scatter, with no sharp dependence on galaxy
mass.  It is currently not  clear how the results of \citet{SPF09} can
be reconciled with the variation  in the ratio of effective radii with
radius we find  here. In fact, colour gradients  are also determined by
the  change in  the profile  shape (i.e.   the Sersic  index), besides
radius,  with waveband.   Moreover, both  metallicity and  (small) age
gradients can combine to produce the observed internal colour gradients
of ETGs  (see~\citealt{LdC09}).  The trend  of $r_{_{OPT}}/r_{_{NIR}}$
with    $r_{_{NIR}}$   is   consistent    with   a    recent   finding
by~\citet{Roche:09},  who analyzed  how the  ratio of  effective radii
measured in  $g$ and  $r$ (using SDSS)  correlate with  several galaxy
properties, for  different families of ETGs (normal  E/S0 galaxies and
BCGs). Although limited to the optical regime, they find that the mean
ratio  of radii  measured in  $g$ and  $r$ become  flatter  for larger
galaxies       (Fig.~\ref{fig:reg_rek}).       The       trend      of
$r_{_{OPT}}/r_{_{NIR}}$ can be  explained by the increasing importance
of dissipationless  mergers in the formation of  more massive galaxies
with  galaxy  mass. Indeed,  dry  mergers  are  expected to  wash  out
internal   differences    of   stellar   population    properties   in
galaxies~\citep{White:80, diMatteo:09}.   A major role  of dry mergers
in  the  formation of  massive  ETGs has  also  been  suggested, in  a
theoretical framework, by~\citet{Naab:06} and \citet{deLucia:06}.

\subsection{The FP from $g$ through $K$}
LBM08  derived the  FP relation  in the  $r$ (SDSS)  and  $K$ (UKIDSS)
wavebands, showing that  the FP slopes exhibit only  a small variation
with waveband, and  that this variation is degenerate  with respect to
(i)  the gradients  of stellar  population properties  (i.e.   age and
metallicity)  with galaxy  mass, $\delta(\log  t)/\delta(\log  M)$ and
$\delta(\log Z)/\delta(\log M)$, and (ii) the fraction of the FP tilt,
$f$, which  is caused by stellar  populations. One main  result of the
present study is  that using the $grizYJHK$ coefficients  of the FP we
are  able  to  break   this  degeneracy.   The  resulting  probability
distribution of $f$  is sharply peaked around zero,  implying that the
tilt of  the FP  in the NIR  is not  due to stellar  populations. This
result is in agreement with  that of~Trujillo et al.~(2004), who found
that the slope  of the $M/L$ vs. luminosity relation  in K-band can be
entirely    due   to    structural   non-homology    of    ETGs   (see
also~\citealt{BCC97, GrC97}). In B band,  they found that a minor, but
still significant fraction (one-quarter) of the tilt is due to stellar
populations.  The  results of~Trujillo  et  al.~(2004) contrast  those
of~\citet{BBT07}, who argued that the  tilt is more likely caused by a
variation  of  the  dark   matter  content  with  mass,  with  stellar
populations  playing  a  minor  role,  which  fully  agrees  with  our
finding. Recently, ~\citet{Jun:08} have  derived the FP relation for a
sample of fifty-six ETGs in the visible (V), NIR (K), and MIR (Spitzer
IRAC)  wavelengths and  concluded that  the  slope $''a''$  of the  FP
increases with  the waveband.   However, the uncertainties  (see their
tab.~2) seem  to be still too  large to conclude  if $''a''$ increases
even further in the MIR wavebands.

Spectroscopic studies  of stellar  population properties in  ETGs have
found  that the (luminosity-weighted)  age of  ETGs tends  to increase
along the galaxy sequence, as parametrized in terms of either velocity
dispersion  or  stellar  and  dynamical  mass  (e.g.~  \citealt{THO05,
  Gallazzi:06}).   The  ages  are  usually  estimated  comparing  line
spectral  indices  with   the  expectations  from  stellar  population
models. In particular, \citet{Gallazzi:06} found that the slope of the
$\log  t$  vs. $\log  M$  relation is  $0.115  \pm  0.056$ (see  their
tab.~4). This value is estimated for a sample of ETGs with a dynamical
mass $M  \widetilde{>} 10^{10}  M_{\odot}$, with a  limiting magnitude
comparable  to  that  we  adopt  here  in this  work.   The  value  of
$\delta(\log   t)  /   \delta(\log  M)$   from~\citet{Gallazzi:06}  is
significantly  larger  than  what   we  obtain  here,  although  still
marginally            consistent           within           2-$\sigma$
(Tab.~\ref{tab:age_zeta}). Moreover, we have  to consider that age and
metallicity  values from  spectroscopic  studies always  refer to  the
central galaxy region. Aperture  corrections are based on measurements
of line spectral  indices for small samples of ETGs at  $z \sim 0$ and
apply  only   to  a  relatively  small  radial   range,  with  $R<R_e$
(see~\citealt{JORG:97}).   \citet{Gallazzi:06}   adopt   a   different
approach and instead  of correcting the indices, test  how the stellar
population parameters  vary with  redshift, up to  $z \sim  0.12$, for
galaxies with similar physical properties (e.g.  dynamical mass).  The
main drawback  of this  approach is that  it relies on  the assumption
that  spectral  indices  and   their  gradients  do  not  evolve  with
redshift.  Considering the redshift  range ($z<0.12$),  large galaxies
are still observed  only in a radial region  of $R \widetilde{<} R_e$.
The  values  of   $\delta(\log  t)/\delta(\log  M)$  and  $\delta(\log
Z)/\delta(\log M)$ we  obtain from the FP analysis  describe the total
stellar  population content  of  ETGs, as  the photometric  parameters
entering the FP are defined in terms of the total galaxy luminosity of
the  2D Sersic  model.   The information  encoded  in the  FP is  more
similar to that provided by the colour-magnitude relation, where galaxy
colours are usually measured within a larger aperture than that sampled
by spectroscopic  studies.  In fact,  in agreement with  our findings,
\citet{Kod98}  showed that  the  small redshift  evolution  of the  CM
relation implies that (i) all  the (luminous) ETGs are equally old and
(ii)  more massive  galaxies are  more  metal rich  than less  massive
systems.

In the  framework of the SAURON  project, for a  sample of twenty-five
ETGs, \citet{Cap06} found  that the variation of the  dynamical \ml \,
is  well correlated  with the  H$_\beta$ line-strength,  implying that
most of the tilt of the FP { (i.e. the deviation of FP coefficients
  from the Virial Theorem expectation under the assumption of homology
  and constant $M/L$)} is indeed  due to galaxy age varying with mass.
This  result   apparently  contrasts  with   findings  of~Trujillo  et
al.~(2004) and~\citet{BBT07}, and with our results, where both $f$ and
$\delta(\log t)/\delta(\log M)$ are consistent with zero.  However, as
also  noticed  by~LBM08, 68\%  of  the  galaxies in  the~\citet{Cap06}
sample  are  fast  rotators  and  20\% have  low  velocity  dispersion
($\sigma$=60-85 km  s $^{-1}$).  \citet{ZGZ06}  and~\citet{DOF08} have
found that the  FP of spheroidal systems depends  on the covered range
in  mass and  velocity dispersion  { (see  also~\citealt{GG:08} and
  references  therein)},  with   the  tilt  becoming  larger  (smaller
$''a''$) for galaxies in the low $\sigma_0$ regime.  ~\citet{Jeong:09}
derived the NUV and FUV FP of thirty-four ETGs from the SAURON sample.
They showed that  the tilt is significantly affected  by residual star
formation in  ETGs, mostly found at  low $\sigma_0$ ($\widetilde{<}100
\, km  \, s^{-1}$). Hence,  the above mentioned disagreement  with the
findings of~\citet{Cap06} might be explained by the different range of
velocity dispersion and different  selection criteria of both samples.
It    is   important   to    remember,   as    we   have    shown   in
Sec.~\ref{sec:fp_gal_pars}, that  different subsamples of  ETGs do not
share the same FP relations.  When binning the SPIDER sample according
to  Sersic index  and axis  ratio, we  find that  the tilt  of  the FP
becomes larger  (i.e. the slopes  of the FP  decrease) by a  small but
detectable amount for galaxies with  higher $n$ and larger $b/a$, with
the effect being mainly due to a difference in the $''b''$ coefficient
of the FP.  The result  for $n$ is consistent with~ \citet{DOF08}, who
found  that in the  optical regime  the $''b''$  coefficient decreases
significantly  as  the  Sersic   index  increases,  while  $''a''$  is
constant.   However,  one  should  notice that~\citet{DOF08}  did  not
account  for the  fact that  galaxies in  different bins  of  $n$ have
different distributions in  the space of the FP  variables, and, as we
show  in  Sec.~ \ref{sec:fp_gal_pars},  this  might  prevent a  proper
comparison of FP coefficients.  The  fact that the variation of the FP
tilt among galaxies  with different $n$ and $b/a$  is similar from $g$
through $K$ suggests that it  is more related to differences of galaxy
properties (structural  and dynamical), rather than  to differences in
the galaxy stellar population content.~\citet{Kelson00} derived the FP
of  $56$  ellipticals,  lenticulars,  and early-type  spirals  in  the
cluster  environment at  redshift  $z \sim  0.3$.   In agreement  with
JFK96, they found  that the FPs of Es and  S0s have consistent slopes.
They also  found that the FP  of early-type spirals has  a larger tilt
(smaller  $''a''$) with  respect  to that  of  ETGs, likely  due to  a
variation  of  the luminosity-weighted  age  with  galaxy mass.   This
result  might  explain what  we  find  when  binning the  SPIDER  ETGs
according to  their optical--NIR colours and  the discy/boxy parameter
$a_4$.   Galaxies with  bluer  colours and  more pronounced  disc-like
isophotes  tend  to  have  a  more  tilted FP  (mainly  because  of  a
smaller $''b''$), with  this effect smoothly  disappearing from
$g$ through $K$.

\begin{figure}
\begin{center}
\includegraphics[width=80mm]{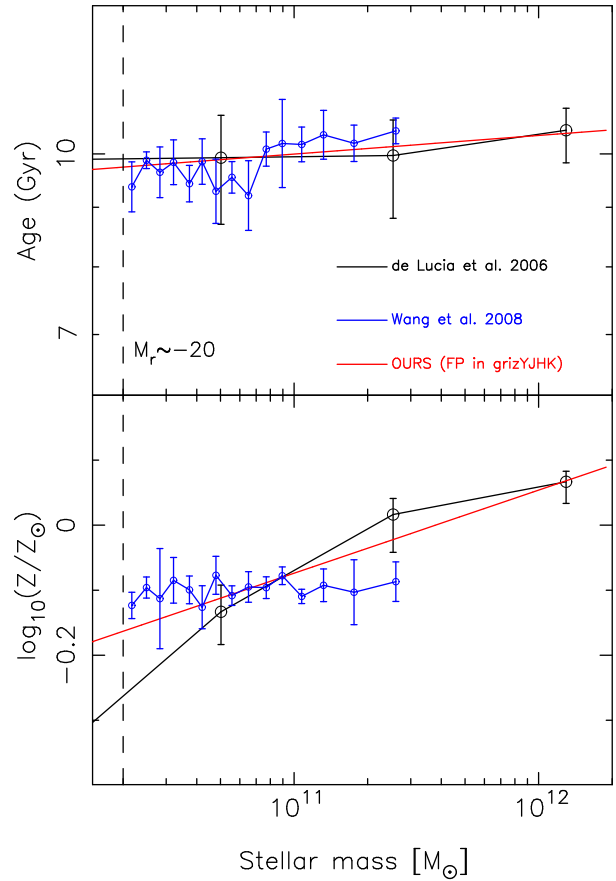}
\caption{Comparison  of  the  variation   of  age  (upper  panel)  and
  metallicity (lower panel) with stellar mass from the grizYJHK FP and
  the predictions of semi-analytical models of galaxy formation. Black
  circles  and error  bars are  the same  as in  fig.~6  of~deL06, and
  represent  median  values of  luminosity--weighted  age and  stellar
  metallicities of  model elliptical galaxies.  Black  error bars link
  the  upper  and  lower   quartiles  of  the  age  and  metallicity
  distribution in a given bin  of stellar mass. The magnitude limit of
  the SPIDER  sample corresponds  to a stellar  mass limit of  $\sim 2
  \times 10^{10}$~$M_\odot$, marked in the plot by the vertical dashed
  line. The blue  circles are the peak values  of the distributions of
  luminosity--weighted  ages  and   stellar  metallicities  for  model
  elliptical   galaxies   from   the  updated   semi-analytical   model
  of~\citet{Wang:08},  where  the  WMAP3  cosmology (rather  WMAP1  as
  in~deL06) is adopted. The peak  values are computed by the bi-weight
  estimator.   Error  bars  denote  $2\sigma$  uncertainties  on  peak
  values.   Model ellipticals  are selected  as those  objects  in the
  semi-analytical  model with  a stellar  mass fraction  in  the bulge
  larger than  $80\%$, and colour index $g-r>0.5$  (consistent with the
  distribution  of  ETG's colours  for  the  SPIDER  sample, see  paper
  I). Ages refer to redshift $z=0$, for both models. The red lines are
  the result of the analysis of Sec.~\ref{sec:SP_slopes}. Their offset
  is arbitrarily  chosen to  match the models,  while the  slope's are
  obtained from the values of $\delta  (\log t) / \delta (\log M)$ and
  $   \delta    (\log   Z)   /   \delta   (\log    M)$   reported   in
  Tab.~\ref{tab:age_zeta}  for  the BC03  model,  with $t=12$~Gyr  and
  $Z/Z_\odot=1$~\label{fig:FP_SAMs}.  }
\end{center}
\end{figure}

\subsection{Comparison to semi-analytical models of galaxy formation}
\label{sec:SAMs}
Explaining the  stellar population properties  of ETGs is  a lingering
problem  for  theories of  galaxy  formation  and  evolution.  In  the
hierarchical scheme of galaxy formation, larger systems assemble their
mass at later times. Hence, if star formation closely follows the mass
assembling,  one would naively  expect more  massive galaxies  to have
younger stellar populations, in  evident disagreement with (i) the red
colours and  old stellar  populations characterizing the  massive ETGs,
and (ii)  the observed bimodality  of galaxies in  the colour-magnitude
diagram~\citep{Strateva:01}.     As   shown   by~\citet{Kauffmann:96},
semianalytical models (SAMs) of  galaxy formation in the CDM framework
can  account for  point (i)  because more  massive systems  are indeed
those  forming stars  at higher  redshift. For  massive  galaxies, the
galaxy bimodality  is also reproduced, by preventing  cooling flows in
the centre  of dark-matter haloes.  This is achieved with  some ad-hoc
recipe, like  the feedback  from AGNs~\citep{Croton:06}. As  a result,
both the luminosity-weighted age and metallicity of ETGs increase from
lower  to  higher mass  systems  (see fig.~6  of~\citealt{deLucia:06},
hereafter deL06).  Since the  $grizYJHK$ FP sets strong constraints on
the variations  of age and  metallicity with galaxy mass,  the natural
question is if the amount of such variations can be accomodated in the
framework   of    current   models   of    galaxy   formation.    {
  Fig.~\ref{fig:FP_SAMs} compares the logarithmic variation of age and
  metallicity per decade  in stellar mass, $\delta(\log t)/\delta(\log
  M_{\star})$  and $\delta(\log  Z)/\delta(\log  M_{\star})$, that  we
  infer   from  the   FP~\footnote{  The   quantities  $\delta(\log
    t)/\delta(\log   M_{\star})$   and   $\delta(\log   Z)/\delta(\log
    M_{\star})$   are  computed  from   the  values   of  $\delta(\log
    t)/\delta(\log M)$ and $\delta(\log Z)/\delta(\log M)$ reported in
    Tab.~\ref{tab:age_zeta},  and  the   relation  $\delta(\log  M)  =
    \delta(\log  M_{\star})/(1-\gamma_K)$, that  holds for  $f=0$ (see
    Sec.~\ref{sec:SP_slopes}).   } (Sec.~\ref{sec:SP_slopes})  and the
  expectation from  SAMs.}  The plot  shows the variation of  the mean
luminosity--weighted age  and stellar  metallicities as a  function of
galaxy stellar mass, $M_{\star}$, of model elliptical galaxies for the
SAM of~deL06  (black circles), and that  of~\citet{Wang:08}, where the
latter   model    has   been   updated   according    to   the   WMAP3
cosmology. Interestingly,  we see that current SAMs  are actually able
to match  the results obtained  from the analysis  of the FP  from $g$
through $K$.  Massive ETGs ($>M_{\star} \sim {2 \times 10^{10}}$) have
essentially  coeval stellar  populations, with  more  massive galaxies
being  slightly more  metal rich,  by a  difference in  metallicity of
$\sim 0.1$~dex per decade in mass, than less massive systems.

In  a forthcoming  paper, we  will  continue the  analysis of  scaling
relations of  ETGs, by  presenting the dependence  of the FP  from $g$
through $K$  as a function  of the environment where  galaxies reside,
and discussing the implications for current models of galaxy formation
and evolution.

\section{Summary}
\label{sec:summary}
{ In this contribution, we present a thorough analysis of the FP of
  ETGs using  a homogeneous dataset  obtained in two  wide-sky surveys
  (SDSS-DR7 and UKIDSS-LAS). As far as the FP derivation is concerned,
  we discuss  fitting procedure, bias  due to selection  effects, bias
  due  to correlated  errors  on $r_e$  and  \mie, and  how to  obtain
  meaningful  FP coefficients.  Below we  summarise some  of  the main
  findings of this paper:

\noindent 1  - We examine the  KR for all the  wavebands available and
find a smooth increase in slope  from $g$ ($\sim 3.44\pm 0.04$) to $K$
($\sim 3.80\pm  0.02$), while the  scatter seems to be  independent of
the waveband. Although the KR is just a projection of the FP relation,
these results serve as a benchmark  at the nearby Universe and will be
essential for studies  of ETGs at high redshift,  for which not always
large  samples exist  to  probe  the FP.   In  agreement the  waveband
variation of the  KR slope, we find that the  ratio of effective radii
measured in $g$ to  that measured in $K$, ($\frac{r_{e,g}}{r_{e,K}}$),
decreases as $r_{e,K}$ increases.


\noindent  2 -  We  measure the  waveband  dependence of  the FP  with
unprecedented accuracy.   The trends  of the FP  coefficients, $''a''$
and  $''b''$  (see  Eq.~\ref{eq:FP}),   with  waveband  are  all  very
consistent   regardless   of    the   sample   used   (magnitude-   or
colour-selected).  When  using the $\log  \sigma_{\rm 0}$ fit,  we find
that $''a''$ is consistent, within 2-$\sigma$, from $r$ to $K$ and for
$g$  band  $''a''$  differs  significantly by  $\sim$3$\sigma$,  while
$''b''$  is all very  consistent. Using  the orthogonal  fit, however,
$''a''$ significantly varies by 12\%  from $g$ through $K$ and $''b''$
does not change at all.

\noindent 3 -  The analysis of the face-on  and edge-on projections of
the FP indicate,  first of all, consistency with  the results obtained
when examining the FJ and KR. Moreover, the scatter around the edge-on
projection  is  about  twice  smaller  than  that  of  the  face-on's,
indicating that the FP is more like a band rather than a plane.

\noindent  4 -  We test  the  sensitivity of  the FP  solution to  the
velocity dispersion measurement used, $\log \sigma_{\rm 0}$(STARLIGHT)
versus   $\log   \sigma_{\rm   0}$(SDSS-DR6).   Although   these   two
measurements  agree   remarkably  well,   the  value  of   $''a''$  is
systematically  smaller  when  using  the STARLIGHT  values  of  $\log
\sigma_{\rm    0}$,   while   $''b''$    is   insensitive    to   both
measurements. Also, we find that  the waveband dependence of the FP is
the same regardless of the magnitude range used in the analysis.

\noindent 5 - The sample analysed is formed by ETGs covering a certain
domain  in galaxy properties,  like axis  ratio ($b/a$),  Sersic index
($n$), $r-K$ colour, and $a_{4}$.  The FP slopes vary significantly for
ETGs with different properties in  the following way: ETGs with larger
$n$ have lower $''b''$; $''a''$ is smaller  in the NIR for the $n > 6$
subsample, and in the  optical both subsamples have similar $''a''$'s;
The FP  of round  galaxies has smaller  $''a''$ (and  smaller $''b''$)
than the  FP obtained for  lower $b/a$ ETGs  - the difference  is more
evident in  the NIR. Also, boxy  and bluer ($r-K$) ETGs  exhibit an FP
with  lower $''b''$,  with  this difference  disappearing  in the  NIR
wavebands.

\noindent 6 - Finally, we  show that current Semi Analytical Models of
galaxy formation match the results  here obtained from the analysis of
the FP tilt  from $g$ through $K$. This analysis  implies that the NIR
tilt of  the FP is not  due to stellar populations:  massive ETGs have
coeval stellar populations, and are  more metal rich than less massive
systems. This is  one of the crucial points of  the FP study presented
here.  }

\appendix
\section{The MLSO fit}
\label{app:mlso}
We consider two  random variables, $X$ and $Y$,  related by the linear
model:
\begin{equation}
 Y = p_1 + p_2 X,
\end{equation}
where  $p_1$ and  $p_2$ are  the offset  and slope,  respectively.  We
indicate as $x$ and $y$ the outputs of $X$ and $Y$.  Assuming that the
$y$ values are normally  distributed along the orthogonal direction to
the line, the probability of observing a given $x$ and $y$ pair is:
\begin{equation}
 P(r) d \! r = \left( 2 \pi \sigma_o^2 \right)^{-1/2} \cdot 
                exp\left[-r^2/(2\sigma_o^2)\right] d \! r,
\label{eq:prob_orth}
\end{equation}
where $r$ is the orthogonal residual, $r=(y-p_1-p_2 \cdot x)
\cdot \left( 1 + p_2^2\right)^{-1/2}$, and $\sigma_o$ is the orthogonal
scatter around the relation. In case where a selection cut is applied:
\begin{equation}
 y < c_1 + c_2 x,
\end{equation}
with $c_1$ and $c_2$ assigned constants, Eq.~\ref{eq:prob_orth} 
modifies as follows:
\begin{equation}
 P(r) d \! r = K(p_1,p_2,c_1,c_2; x) \cdot 
                exp\left[-r^2/(2\sigma_o^2)\right] f(y-c_1-c_2 x) d \! r,
\label{eq:prob_orth_sel}
\end{equation}
where the  function $f$ is equal  to one when its  argument is smaller
than  zero, and vanishes  otherwise. The  function $K(p_1,p_2,c_1,c_2;
x)$  is obtained  by the  normalization condition  $\int P(r)  d  \! r
=1$. If no  selection cut is applied ($f=1$  identically), one obtains
$K=\left(   2   \pi  \sigma_o^2   \right)^{-1/2}$,   and  we   recover
Eq.~\ref{eq:prob_orth}.  In general, the $K$ is given by:
\begin{equation}
 K=\left( 2 \pi \sigma_o^2 \right)^{-1/2} \cdot 2 \cdot \left[
1+erf(t)\right]^{-1},
\end{equation}
with  $t= \left[  (c_1-p_1)  + (c2-p_2)  x \right]/(\sqrt{2}  \sigma_o
\sqrt{1+p_2^2})$, and  $erf$ denotes the  error function. For  a given
sample of data-points, the likelyhood, $\it L$, can be written as
\begin{equation}
  L = \sum \frac{r^2}{2 \sigma_o^2} - \sum (\ln K)
\end{equation}
where both sums are performed over the entire dataset.  In the case of
the KR, one  has $y=$\mie \, and $x=$\lre  \, (Sec.~\ref{sec:kr}). The
magnitude cut can be written as $<  \! \mu \! >_e < M_{lim} + 38.56578
+  5  \log  r_e$,  where  $M_{lim}$  is the  magnitude  limit  of  the
sample.  This expression  is  identical to  Eq.~\ref{eq:prob_orth_sel}
provided  that   $c_1=M_{lim}  +  38.56578$  and   $c_2=5$.  The  MLSO
coefficients of  the KR are then  obtained by minimizing  the $L$ with
respect to $p_1$, $p_2$, and $\sigma_o$.

\section{Matching the magnitude and surface brightness distributions of
ETG samples}
\label{app:match_dist}
We consider the  case where a set of $n$  galaxy samples, with running
indices    $i=1$   to   $n$,    are   given.     In   the    case   of
Sec.~\ref{sec:fp_gal_pars},  we have  $n=2$, and  the two  samples are
obtained by splitting the  magnitude-complete sample of ETGs according
to a  given galaxy  parameter $p$.  First,  we select the  sample with
lowest sample size. For such sample, we define the minimum and maximum
values of absolute magnitude, $M_{min}$ and $M_{max}$, and the minimum
and   maximum  values  of   \mie,  \mie$_{,min}$   and  \mie$_{,max}$,
respectively.   We then  construct a  grid in  the  magnitude--\mie \,
plane, over  the rectangular region  from $M_{min}$ to  $M_{max}$, and
\mie$_{,min}$ to \mie$_{,max}$. For a given cell $k$ over the grid, we
count  the   number  of  galaxies   of  each  sample  in   that  cell,
$n_{i,k}$. We  take the minimum  value of $n_{i,k}$, $n_k$,  among all
the given  samples. For  each sample, we  then randomly  extract $n_k$
galaxies  whose magnitude  and \mie  \, values  fall inside  the given
cell.  This  step is  performed for  all the cells  in the  grid.  The
procedure  provides a subsample  of galaxies  from each  input sample,
with all  subsamples having the same  number of galaxies  and the same
absolute magnitude and \mie \,  distributions. { The mean values of
  $M_{min}$   and  $M_{max}$,   among  the   subsamples   analysed  in
  Sec.~\ref{sec:fp_gal_pars},  amount to  about $-24.6$  and $-20.55$,
  respectively,   while   the  mean   values   of  \mie$_{,min}$   and
  \mie$_{,max}$  amount to  about $15.2$  and $27.2$~$mag/arcsec^{2}$.
  The step  sizes in $M$  and \mie \,  are chosen to be  $0.2$~mag and
  $0.2$~$mag/arcsec^{2}$,  respectively.   This  makes the  number  of
  galaxies  in each  cell of  the grid  to be  smaller than  $40$.  We
  verified that either reducing or  increasing the bin size in a given
  direction by  a factor  of two  does not change  at all  the results
  presented in Sec.~\ref{sec:fp_gal_pars}.}

\section*{Acknowledgements}

We thank M.  Capaccioli for  the support provided to this project.  We
thank R.R.   Gal for several suggestions and  comments throughout this
project.   We thank G.  de Lucia  and J.  Wang for  helping us  in the
comparison to semi-analytical models of galaxy formation. { We also
  thank the anonymous referee for  comments which helped us to improve
  the manuscript. } We
have used data  from the 4th data release of  the UKIDSS survey, which
is  described  in detail  in  \citet{War07}.   The  UKIDSS project  is
defined  in~\citet{Law07}.  UKIDSS  uses the  UKIRT Wide  Field Camera
(WFCAM; Casali et  al.~2007).  The photometric system  is described in
Hewett et  al.~(2006), and the  calibration is described  in Hodgkin et
al.~(2009). The pipeline processing and science archive are described
in Irwin et  al.~(2009, in prep) and Hambly et  al.~(2008).  Funding for
the  SDSS  and SDSS-II  has  been provided  by  the  Alfred P.   Sloan
Foundation,  the  Participating  Institutions,  the  National  Science
Foundation, the  U.S.  Department of Energy,  the National Aeronautics
and Space Administration, the  Japanese Monbukagakusho, the Max Planck
Society, and  the Higher Education  Funding Council for  England.  The
SDSS Web  Site is  http://www.sdss.org/.  The SDSS  is managed  by the
Astrophysical Research Consortium  for the Participating Institutions.
The  Participating Institutions  are  the American  Museum of  Natural
History,  Astrophysical   Institute  Potsdam,  University   of  Basel,
University of  Cambridge, Case Western  Reserve University, University
of Chicago,  Drexel University,  Fermilab, the Institute  for Advanced
Study, the  Japan Participation  Group, Johns Hopkins  University, the
Joint  Institute for  Nuclear  Astrophysics, the  Kavli Institute  for
Particle Astrophysics  and Cosmology, the Korean  Scientist Group, the
Chinese Academy of Sciences  (LAMOST), Los Alamos National Laboratory,
the     Max-Planck-Institute     for     Astronomy     (MPIA),     the
Max-Planck-Institute   for  Astrophysics   (MPA),  New   Mexico  State
University,   Ohio  State   University,   University  of   Pittsburgh,
University  of  Portsmouth, Princeton  University,  the United  States
Naval Observatory, and the University of Washington.

\bsp

\label{lastpage}

\end{document}